\documentclass[a4paper,12pt]{article}
\usepackage{test} %

\usepackage{xcolor}
\definecolor{pantone}{RGB}{1, 33, 105}
\usepackage{graphicx}
\usepackage{placeins}
\usepackage{booktabs,caption,etoolbox}
\newif\ifinpapertable
\AtBeginEnvironment{table}{\inpapertabletrue}
\AtBeginEnvironment{table*}{\inpapertabletrue}
\AtBeginEnvironment{tabular}{%
  \ifinpapertable
    \footnotesize\setlength{\tabcolsep}{4pt}%
  \fi}

\usepackage{setspace}
\usepackage{hyperref}
\hypersetup{colorlinks,linkcolor=blue,citecolor=blue,urlcolor=pantone}
\usepackage{doi}

\usepackage[style=ext-authoryear-comp,
sorting=nyt, %
sortcites=false, %
dashed=false, %
maxcitenames=2, %
maxbibnames=99, %
uniquelist=false,
uniquename=false,
giveninits=true, %
useprefix=true, %
natbib, %
date=year %
]{biblatex}

\AtBeginRefsection{\GenRefcontextData{sorting=ynt}}
\AtEveryCite{\localrefcontext[sorting=ynt]}

\DeclareFieldFormat{pages}{#1} %
\renewbibmacro{in:}{\ifentrytype{article}{}{\printtext{\bibstring{in}\intitlepunct}}} %

\DeclareFieldFormat[article,inbook,incollection,inproceedings,patent,thesis,unpublished]{titlecase:title}{\MakeSentenceCase*{#1}} %

\newbibmacro*{atoda:titlelink}[1]{%
  \ifhyperref
    {\iffieldundef{doi}
       {\iffieldundef{eprint}
          {\iffieldundef{url}
             {#1}
             {\href{\thefield{url}}{#1}}}
          {\href{https://arxiv.org/abs/\thefield{eprint}}{#1}}}
       {\href{https://doi.org/\thefield{doi}}{#1}}}
    {#1}}

\DeclareFieldFormat{title}{\usebibmacro{atoda:titlelink}{\mkbibemph{#1}}}
\DeclareFieldFormat
  [article,inbook,incollection,inproceedings,patent,thesis,unpublished]
  {title}{\usebibmacro{atoda:titlelink}{\mkbibquote{#1\isdot}}}

\renewbibmacro*{doi+eprint+url}{}

\usepackage[inline,shortlabels]{enumitem}
\setlist[enumerate,1]{label=(\roman*)}

\usepackage[margin = 1.25in]{geometry}

\title{Does Calibration Pick the Right Model? Likelihood Evidence from Income Risk}
\author{Jiuyi Cheng\thanks{Emory University. Email: \href{mailto:jiuyi.cheng@emory.edu}{jiuyi.cheng@emory.edu}.} \and Alexis Akira Toda\thanks{Department of Economics, Emory University and Research Institute for Economics and Business Administration, Kobe University. Email: \href{mailto:alexis.akira.toda@emory.edu}{alexis.akira.toda@emory.edu}.}}

\numberwithin{equation}{section}

\newcommand{\cL}{\mathcal{L}}
\newcommand{\cM}{\mathcal{M}}
\newcommand{\cN}{\mathcal{N}}
\newcommand{\cQ}{\mathcal{Q}}
\newcommand{\cW}{\mathcal{W}}
\newcommand{\cX}{\mathcal{X}}
\newcommand{\KL}{\operatorname{KL}}

\begin{document}
\maketitle
	
\begin{abstract}
Quantitative macroeconomics typically disciplines models by matching selected moments. We show how finite-state hidden Markov approximations make likelihood estimation and model selection feasible for nonlinear, non-Gaussian latent processes on short panels, without simulation. We illustrate the method with PSID earnings data. The calibrations that best match moments of earnings dynamics fit complete histories worse than a likelihood-estimated process with a Gaussian persistent component. In a standard saving model, calibrations of one process that fit the targets about equally well imply mean assets that differ by a factor of three. Close agreement with selected moments can leave economic predictions largely undetermined.

\medskip

\noindent
\textbf{Keywords:} calibration, hidden Markov model, income risk,
likelihood, misspecification, precautionary saving.

\medskip

\noindent
\textbf{JEL codes:} C13, C52, C63, D15, E21.
\end{abstract}

\section{Introduction}

Quantitative macroeconomics often disciplines models by calibration. A
researcher chooses parameters so that the model reproduces a small set of
selected moments of the data, and then uses the model to study behavior,
welfare, or policy
\citep{KydlandPrescott1982,KydlandPrescott1996}.\footnote{We use
``calibration'' to include procedures that fit a small set of selected
moments, including the simulated method of moments as used in our
application.} Calibration is transparent and directs attention to
features that the researcher considers important. However, agreement
with the targets is evidence only about the targets. Different models,
or different parameters within one model, may reproduce the same targets
and still imply different responses to risk and policy
\citep{HansenHeckman1996,BrowningHansenHeckman1999}. Whether this
concern matters in practice is an empirical question: does matching the
targets pin down the economic answer?

Answering this question requires a criterion that uses more information
than the selected moments. The likelihood is such a criterion. It
evaluates the joint distribution of complete observed histories, so a
model can reproduce selected statistics and still assign low probability
to the histories from which we compute those statistics. The obstacle is
computation. When a model contains a latent (unobserved) state with
nonlinear dynamics or non-Gaussian shocks, the Kalman filter no longer
applies, and the likelihood requires integrating over all possible paths
of the latent state. Particle filters approximate this integral by
simulation \citep{Fernandez-VillaverdeRubio-Ramirez2007}, but the
simulated likelihood contains sampling noise and is not smooth in the
parameters, which complicates optimization. For these reasons,
likelihood comparison remains rare in quantitative work.

In this paper, we show that maximum likelihood estimation of nonlinear,
non-Gaussian latent-state models on short, irregular panels is now
feasible, and that it leads to economic answers different from those of
calibration. In our application, calibrations of a single earnings model
fit the targets about equally well, yet imply mean assets that differ by
a factor of about three, with the same data, preferences, and prices.
The likelihood, by contrast, ranks the models and their parameters by
their fit to complete earnings histories.

We approximate
the continuous latent state by a finite-state Markov chain while leaving
the observations continuous. This approximation turns the model into a
hidden Markov model (HMM), in which an unobserved finite-state chain
generates the observed outcomes \citep{Kitagawa1987,Farmer2021}. For
discrete-time processes, we construct the chain by the conditional moment
matching method of \citet{FarmerToda2017QE}. For a diffusion, we
discretize its generator (the operator that describes the local drift and
variance of the diffusion). The HMM forward recursion sums over all
latent histories by successive matrix multiplications. We integrate
permanent individual effects (time-invariant unobserved differences) by
numerical quadrature. We retain each individual's entry age and use
transition matrices for the actual observation gaps, so irregular
observation dates pose no difficulty. The likelihood requires no
simulation, and standard optimizers estimate all parameters jointly.

We also provide a framework for inference and model comparison under
misspecification. We treat each complete history as one independent
observation and let the number of individuals grow while histories remain
short. Under stated identification and regularity conditions, we give
inference for the parameters and compare non-nested models using
individual log likelihood differences \citep{White1982,Vuong1989}. The
population likelihood ranks models by their Kullback--Leibler divergence (a measure of the discrepancy between two
probability distributions)
from the data distribution. We distinguish inference for the fixed
finite-state approximation from inference for the continuous-state model,
and we state conditions on the numerical approximation error.

We apply the HMM approach to earnings dynamics, a central input into
quantitative household models. We use 28,422 biennial observations of
4,683 men aged 30--60 from the Panel Study of Income Dynamics (PSID),
after removing deterministic earnings effects. We compare four processes
for the persistent component of earnings. Our benchmark is an
Ornstein--Uhlenbeck (OU) process, which is a Gaussian AR(1) at annual
dates. The second is a stationary adaptation of
\citet*{GuvenenKarahanOzkanSong2021} (GKOS), an AR(1) whose innovations
are a mixture of Gaussian distributions. The third is a stationary
adaptation of \citet*{GuvenenMcKayRyan2026} (GMR), an AR(1) with losses
that affect both current earnings and the persistent state. The fourth is
our piecewise linear variance (PLV) diffusion, whose volatility depends
on the persistent state. Each process also includes permanent individual
effects and either Gaussian or Gaussian-mixture transitory shocks, giving
eight specifications. We show that all four stationary processes can
share the entire earnings covariance function
(Proposition~\ref{prop:covariance-equivalence}). Thus second moments
alone cannot separate them. Our application leads to three main results.

First, the likelihood distinguishes the candidate
specifications. Gaussian-mixture transitory shocks improve the log
likelihood by more than 1,500 points in every model. Under this
specification, PLV has the highest likelihood, followed by GKOS, GMR, and
OU. The data favor PLV and GKOS over GMR but do not clearly separate PLV
from GKOS.

Second, close target fit need not extend to complete
earnings histories. We calibrate each process with mixture transitory
shocks to 70 moments of earnings changes and levels, following the
designs of \citet{GuvenenKarahanOzkanSong2021,GuvenenMcKayRyan2026} as closely as our data allow. To prevent cherry-picking, we fixed the
targets and weights before computing any results. The calibrations match
the percentiles of changes and levels closely and lower the moment
objective by about 80 percent relative to the likelihood estimates. Yet
all four calibrations fit the histories worse than every
likelihood-estimated specification with mixture transitory shocks,
including OU.\footnote{An earlier calibration to 18 moments of earnings
changes only, in which GKOS and GMR match the targets within our
tolerance, leads to the same conclusions
(Appendix~\ref{subsec:calibration-changes}).}

Third, the empirical criterion changes both the level
of saving and the ranking of models. We solve a standard optimal saving problem,
holding preferences, prices, mean earnings, and transfers fixed. When we
match the processes to the fitted OU covariance function, PLV moves from
the most wealth to the least, and mean assets fall within 4.2 to 4.8,
compared with 4.8 to 10.5 under likelihood estimation. Calibrated PLV
implies mean assets of 3.76, compared with 10.51 under likelihood estimation.
Among evaluated calibrations whose objectives lie within 0.0092 of the
reported objective, mean assets range from 2.2 to 8.2 for GKOS and from
3.1 to 9.3 for PLV. Thus close target fit can leave the economic answer
largely undetermined.

These results do not mean that the likelihood-preferred model is
correct. Every model we consider is misspecified, and the fitted models
still miss features of the data, including the extreme kurtosis of
earnings changes. We therefore recommend a division of labor: use the
likelihood to compare and select models, and use targeted moments to
diagnose where the selected model fails. The HMM approach applies
whenever the latent state is low-dimensional, so researchers can assess
the fit of observed histories before using a model for economic
analysis. In empirical industrial organization, researchers routinely
estimate and test structural models by likelihood \citep{Rust1987}. The
HMM approach makes the same discipline available for the household models
of quantitative macroeconomics.

\subsection{Related literature}
\label{subsec:related-literature}

Our paper contributes to the literature on the empirical foundations of
quantitative models. \citet{KydlandPrescott1982,KydlandPrescott1996}
develop and defend calibration as a computational experiment.
\citet{HansenHeckman1996} and \citet{BrowningHansenHeckman1999} examine
the empirical support for the parameters and specifications used in
quantitative macroeconomics. \citet{GregorySmith1991} and
\citet{Canova1994} develop simulation-based inference for calibrated
models, and \citet{Watson1993} measures discrepancies in their dynamic
second moments. \citet{HnatkovskaMarmerTang2012} compare misspecified
calibrated models by minimum distance and study the role of the
weighting matrix. \citet{AndrewsGentzkowShapiro2017} measure the local
sensitivity of parameter estimates to estimation moments. Closely related
to our application, \citet{CrawleyHolmTretvoll2026}
show that misspecification makes income-risk estimates depend on the
moments and weights used, and they propose a richer transitory component.
We instead compare models by the likelihood of complete observed
histories and show that close target fit can leave economic predictions
undetermined.

The HMM approximation of continuous-state models has a long history.
\citet{Kitagawa1987} evaluates non-Gaussian state-space likelihoods by
numerical integration, and \citet{LangrockMacDonaldZucchini2012} maximize
the likelihood of fine-grid HMM approximations to nonlinear models.
\citet{Farmer2021} develops the ``discretization filter'' for economic
models, combines HMM likelihood evaluation with moment-matching
discretizations such as \citet{FarmerToda2017QE}, and establishes
likelihood theory for long time series. \citet{Altman2007} develops mixed
HMMs with random effects, and \citet{DelattreGenon-CatalotSamson2012}
study likelihood estimation of diffusions with random effects. We bring
HMM likelihood evaluation and comparison into a short-panel framework
with permanent heterogeneity, nonstationary initialization, and irregular
observation gaps. We give inference under misspecification and separate
the statistical conditions from the conditions on numerical accuracy.

Our illustration builds on the earnings-dynamics literature.
\citet{AbowdCard1989} and \citet{MeghirPistaferri2011} study the
covariance structure of earnings, and
\citet{StoreslettenTelmerYaron2004} use earnings risk in a quantitative
household model. \citet{GewekeKeane2000} estimate PSID earnings dynamics
with normal-mixture shocks by Bayesian methods.
\citet{MeghirPistaferri2004} and \citet{BotosaruSasaki2018} allow
state-dependent variances, and \citet{ArellanoBlundellBonhomme2017} allow
nonlinear persistence. \citet{GuvenenKarahanOzkanSong2021} document non-Gaussian earnings changes in
administrative data, and \citet{GuvenenMcKayRyan2026} propose a tractable process with large
losses and scarring.
\citet{DeNardiFellaPaz-Pardo2020} and \citet{GuvenenOzkanMadera2024}
study the consumption and welfare implications of richer earnings risk.
We use earnings dynamics to demonstrate HMM model selection and measure
how the choice of empirical criterion changes economic predictions.

\section{Hidden Markov likelihood for short panels}
\label{sec:panel-likelihood}

In this section, we describe how we evaluate and compare likelihoods of
short-panel models. We approximate the continuous latent state by a
finite-state Markov chain, which turns the model into a hidden Markov
model (HMM). We let the number of independent individuals grow while
individual histories remain bounded in length. Our inference
accommodates permanent heterogeneity without requiring the latent
process to mix over time. We present the practical steps here and collect the
asymptotic theory in Appendix~\ref{sec:asymptotic-theory}.

\subsection{Short-panel state-space model}
\label{subsec:short-panel-model}

Let $i=1,\dots,I$ index independent and identically distributed individuals.
We observe individual $i$, who has baseline covariates $Z_i$, at dates
$0\le t_{i0}<\dots<t_{iJ_i}\le T$, where $J_i\le\bar J<\infty$. Write
$Y_{ij}$ for the observation at $t_{ij}$ and let
$\Delta_{ij}\coloneqq t_{ij}-t_{i,j-1}>0$ denote the observation gap for
$j\ge1$. The observed design is
$D_i\coloneqq(Z_i,J_i,t_{i0},\dots,t_{iJ_i})$, whose distribution does not
depend on the model parameter. Conditional on $Z_i$, the observation
schedule is independent of the latent path and
measurement shocks.\footnote{If the schedule depends on latent states or
unobserved outcomes, the conditional likelihood below generally requires
a model of the observation process or conditions ensuring ignorability.}
The schedule may be fixed or include nonresponse and dropout. We
condition on $D_i$ throughout.

Let the model parameter $\theta$ belong to a compact set
$\Theta\subset\R^{d_\theta}$, where the parameter dimension $d_\theta$
is fixed. Conditional on $Z_i$, let the latent process $X_i(t)$ take
values in $\cX\subset\R^{d_x}$, where the state dimension $d_x$ is fixed.
Let $\nu_{\theta,0}$ be the state distribution at $t=0$, and let
$K_{\theta,\Delta}(x,\diff x')$ be the time-homogeneous transition kernel
over a gap $\Delta$. The state distribution at the first observation is
$\nu_{\theta,t_{i0}}\coloneqq\nu_{\theta,0}K_{\theta,t_{i0}}$.
Conditional on the latent path and $D_i$, the observations are independent
across dates, and $Y_{ij}$ has a positive density
$g_{\theta,ij}(y\mid X_i(t_{ij}))$.\footnote{For each observed design,
densities are Radon--Nikodym derivatives with respect to a reference
measure that does not depend on $\theta$. The reference measure for the
outcome vector is the product of the measures for its components.
Examples include Lebesgue measure for continuous outcomes and counting
measure for discrete outcomes.}
We suppress the dependence of these laws on $Z_i$ in the notation. We
model any permanent individual effects through a common distribution
governed by $\theta$. There are no unrestricted individual-specific
parameters.

Let $y_{ij}$ denote the observed value of $Y_{ij}$. The exact individual
likelihood is
\begin{equation}
	L_i(\theta)\coloneqq\int_{\cX^{J_i+1}}\nu_{\theta,t_{i0}}(\diff x_0)g_{\theta,i0}(y_{i0}\mid x_0)\prod_{j=1}^{J_i}\set*{
 K_{\theta,\Delta_{ij}}(x_{j-1},\diff x_j)
 g_{\theta,ij}(y_{ij}\mid x_j)}.
 \label{eq:panel-exact-likelihood}
\end{equation}
Each complete history is one independent sampling unit. Since histories
remain short, the influence of initialization does not vanish as
$I\to\infty$.

\subsection{Population objective and identification assumption}
\label{subsec:panel-identification}

Let $\E_0$ denote expectation under the actual distribution of the observed
individual record, which need not belong to the posited model. Define the
exact log likelihood and its population objective by
\begin{equation}
 \ell_i(\theta)\coloneqq\log L_i(\theta),
 \qquad
 \cQ(\theta)\coloneqq\E_0[\ell_i(\theta)].
 \label{eq:panel-population-objective}
\end{equation}
We assume that $\cQ$ is finite and continuous on $\Theta$ and has a unique
maximizer $\theta_0$. The state-space representation alone does not
establish identification.

To interpret this condition, let $f_\theta(\cdot\mid D)$ denote the
conditional density of the outcome vector for a given design $D$, as
\eqref{eq:panel-exact-likelihood} specifies. Its value at individual
$i$'s observations is $L_i(\theta)$. When the actual conditional law is
absolutely continuous with respect to the same reference measure, denote
its density by $f(\cdot\mid D)$. For two densities
$p$ and $q$, write $\KL(p\parallel q)$ for the
Kullback--Leibler divergence, the expectation under $p$ of
$\log(p/q)$. If $f(\cdot\mid D_i)=f_{\theta^*}(\cdot\mid D_i)$ for
almost every design $D_i$ at some $\theta^*\in\Theta$, then
\begin{equation}
 \cQ(\theta^*)-\cQ(\theta)
 =\E_0[
 \KL(
 f_{\theta^*}(\cdot\mid D_i) \parallel
 f_\theta(\cdot\mid D_i))]\ge0.
 \label{eq:panel-conditional-KL}
\end{equation}
Under the uniqueness assumption, $\theta_0=\theta^*$. Under
misspecification, $f$ need not be one of the modeled densities. If the
expected divergences below are finite, then
\begin{multline}
 \cQ(\theta_0)-\cQ(\theta)\\
=\E_0[\KL(f(\cdot\mid D_i)
       \parallel f_\theta(\cdot\mid D_i))-\KL(f(\cdot\mid D_i)
       \parallel f_{\theta_0}(\cdot\mid D_i))]\ge0.
 \label{eq:panel-pseudo-true-KL}
\end{multline}
Thus $\theta_0$ minimizes the expected conditional KL divergence from
the actual outcome distribution over the posited model.

\subsection{Discretized likelihood and estimator}

To evaluate the exact likelihood in \eqref{eq:panel-exact-likelihood}, we
integrate over the latent state at every observation date. In two important
cases, we can do so recursively: we apply the Kalman filter to a linear
Gaussian state-space model \citep{Kalman1960}, or the hidden Markov forward
algorithm to a finite latent state space
\citep{BaumPetrie1966,Hamilton1989}.  With a continuous latent state and
nonlinear or non-Gaussian dynamics, these successive integrations generally
have no closed-form recursion. Evaluating them numerically for every
individual and trial parameter value can therefore be costly, even when
individual panels are short. Thus, following \citet{Kitagawa1987},
\citet{LangrockMacDonaldZucchini2012}, and \citet{Farmer2021}, we approximate the latent process by a finite-state
Markov chain. We then evaluate the resulting likelihood by matrix
multiplication, retaining the specified observation density.

For an $N$-point grid
$\cX_N\coloneqq\set{x_{1,N},\dots,x_{N,N}}$, let
$\nu_{\theta,N,i}$ be the column vector of initial state probabilities
approximating $\nu_{\theta,t_{i0}}$, and let $P_{\theta,N}(\Delta)$ be a
row-stochastic transition matrix approximating $K_{\theta,\Delta}$.
Define $\iota_N$ as the $N$-vector of ones and
\begin{equation*}
 G_{\theta,N,ij}\coloneqq
 \operatorname{diag}
 \set{g_{\theta,ij}(y_{ij}\mid x_{n,N})}_{n=1}^N.
\end{equation*}
The approximate likelihood is
\begin{equation}
	L_{i,N}(\theta)
	\coloneqq\nu_{\theta,N,i}^\top
 G_{\theta,N,i0}
 \set*{\prod_{j=1}^{J_i}
 P_{\theta,N}(\Delta_{ij})G_{\theta,N,ij}}\iota_N,
 \label{eq:panel-discrete-likelihood}
\end{equation}
where we order the matrix product chronologically.
We evaluate the transition matrix at each observed gap $\Delta_{ij}$,
so the observation dates need not be equally spaced.

We evaluate \eqref{eq:panel-discrete-likelihood} with the standard forward
recursion for a hidden Markov model. We write $\nu_{\theta,i}$,
$P_\theta(\Delta)$, and $G_{\theta,ij}$, suppressing their dependence on
$N$. Let $\xi_{ij\mid j-1}$ and $\xi_{ij\mid j}$ denote
the vectors of state probabilities before and after observing $y_{ij}$.
Initialize $\xi_{i0\mid-1}\coloneqq\nu_{\theta,i}$. At $j=0$, and subsequently
after each prediction step, update according to
\begin{subequations}\label{eq:panel-filter}
\begin{align}
	c_{ij}(\theta)
	&\coloneqq\iota_N^\top G_{\theta,ij}\xi_{ij\mid j-1},
 \label{eq:panel-filter-density}\\
 \xi_{ij\mid j}
 &\coloneqq
 \frac{G_{\theta,ij}\xi_{ij\mid j-1}}{c_{ij}(\theta)}.
 \label{eq:panel-filter-update}
\end{align}
\end{subequations}
For $j=1,\dots,J_i$, the prediction step is
\begin{equation}
	\xi_{ij\mid j-1}
	\coloneqq P_\theta(\Delta_{ij})^\top\xi_{i,j-1\mid j-1}.
 \label{eq:panel-filter-prediction}
\end{equation}
Here $c_{ij}(\theta)$ is the conditional density of $y_{ij}$ given the
earlier observations and $D_i$ under the discretized model. Consequently,
\begin{equation}
 L_{i,N}(\theta)=\prod_{j=0}^{J_i}c_{ij}(\theta),
 \qquad
 \ell_{i,N}(\theta)\coloneqq\log L_{i,N}(\theta)
 =\sum_{j=0}^{J_i}\log c_{ij}(\theta).
 \label{eq:panel-filter-log-likelihood}
\end{equation}
Normalizing at every update keeps the filtered probabilities well scaled
and avoids multiplying many small likelihood contributions directly.
We maximize the sample log likelihood without assuming that the model
is correctly specified. We define the discretized sample objective and
quasi-maximum likelihood estimator by
\begin{equation}
 \widehat \cQ_{I,N}(\theta) \coloneqq\frac{1}{I}\sum_{i=1}^I\ell_{i,N}(\theta), \qquad
 \widehat\theta_{I,N} \in\argmax_{\theta\in\Theta}\widehat \cQ_{I,N}(\theta).
 \label{eq:panel-discrete-estimator}
\end{equation}

\subsection{Inference}
\label{subsec:panel-inference}

For a fixed grid, define the discretized population objective and its
maximizer by
\begin{equation*}
 \cQ_N(\theta)\coloneqq\E_0[\ell_{i,N}(\theta)],
 \qquad
 \theta_N\in
 \argmax_{\theta\in\Theta}\cQ_N(\theta).
\end{equation*}
We call $\theta_N$ the pseudo-true value: the parameter that maximizes
the discretized population criterion. Discretization can shift the
maximizer from $\theta_0$ even when the continuous-state model is correct.
Define the individual score (gradient) and Hessian (matrix of second
derivatives) of the discretized log likelihood by
\begin{equation}
 s_{i,N}(\theta)\coloneqq\nabla\ell_{i,N}(\theta),
 \qquad
 H_{i,N}(\theta)\coloneqq\nabla^2\ell_{i,N}(\theta).
 \label{eq:panel-discrete-score-hessian}
\end{equation}

Appendix~\ref{sec:asymptotic-theory} gives the asymptotic theory as
$I\to\infty$. Two results matter for practice. First, at a fixed grid,
$\widehat\theta_{I,N}$ is an $M$-estimator (an estimator that maximizes
a sample average). Under standard regularity conditions, it is consistent
for $\theta_N$ and asymptotically normal with the sandwich covariance
$A_N^{-1}B_NA_N^{-1}/I$, where
$A_N\coloneqq-\E_0[H_{i,N}(\theta_N)]$ and
$B_N\coloneqq\E_0[s_{i,N}(\theta_N)s_{i,N}(\theta_N)^\top]$
(Propositions~\ref{prop:panel-fixed-grid-consistency} and
\ref{prop:panel-fixed-grid-normality}). Second, suppose the grid becomes
finer as $I$ grows and the discretization bias in the expected score
vanishes faster than $I^{-1/2}$. Then the same estimator is consistent
and asymptotically normal around the continuous-state target $\theta_0$
(Propositions~\ref{prop:panel-refining-grid} and
\ref{prop:panel-continuous-normality}).

The algorithm below gives standard errors for $\theta_N$. Inference for
$\theta_0$ additionally
requires numerical approximation error to be small relative to sampling
error. We therefore recommend checking numerical accuracy by refining
the grid and any numerical integration rules used to evaluate the
likelihood.

\begin{alg}[HMM quasi-maximum likelihood and standard errors]
\label{alg:panel-qmle-inference}
Given the panel data, a grid size $N$, and a parameter space $\Theta$, proceed
as follows.
\begin{enumerate}
	\item\label{item:panel-qmle-components}
	For each trial value $\theta\in\Theta$, construct the initial
	distribution $\nu_{\theta,N,i}$, the observation matrices
	$G_{\theta,N,ij}$, and the transition matrix
	$P_{\theta,N}(\Delta)$ for every distinct observation gap in the sample.

	\item\label{item:panel-qmle-likelihood}
	For each individual, use the normalized forward recursion
	\eqref{eq:panel-filter}--\eqref{eq:panel-filter-prediction} to
	compute the conditional densities $c_{ij}(\theta)$ and the individual log
	likelihood $\ell_{i,N}(\theta)$ in
	\eqref{eq:panel-filter-log-likelihood}. Sum these contributions across
	individuals.

	\item\label{item:panel-qmle-estimator}
	Maximize the sample criterion to obtain the quasi-maximum likelihood
	estimator $\widehat\theta_{I,N}$ in
	\eqref{eq:panel-discrete-estimator}. Use multiple starting values when the
	criterion may have local maxima. A consistent approximate maximizer
	has the same first-order limit if its average score is
	$o_{\mathrm p}(I^{-1/2})$.

	\item\label{item:panel-qmle-derivatives}
	Compute the individual scores $s_{i,N}(\widehat\theta_{I,N})$ and Hessians
	$H_{i,N}(\widehat\theta_{I,N})$ at the estimate. Use analytical derivatives, automatic
	differentiation, or stable numerical differences. Form
	\begin{equation*}
		\widehat A_{I,N}
		\coloneqq-\frac{1}{I}\sum_{i=1}^I
		H_{i,N}(\widehat\theta_{I,N}),
		\qquad
		\widehat B_{I,N}
		\coloneqq\frac{1}{I}\sum_{i=1}^I
		s_{i,N}(\widehat\theta_{I,N})
		s_{i,N}(\widehat\theta_{I,N})^\top.
	\end{equation*}

	\item\label{item:panel-qmle-covariance}
	Estimate the covariance matrix of $\widehat\theta_{I,N}$ by the
	individual-clustered sandwich
	\begin{equation}
		\widehat V_{I,N}^{\mathrm{cl}}
		\coloneqq\frac{1}{I}\widehat A_{I,N}^{-1}
		\widehat B_{I,N}\widehat A_{I,N}^{-1}.
		\label{eq:panel-sample-sandwich}
	\end{equation}
	The standard error of component $k$ is
	$[\widehat V_{I,N}^{\mathrm{cl}}]_{kk}^{1/2}$. When the information
	identity applies, the conventional inverse-information covariance estimator
	is $\widehat V_{I,N}^{\mathrm{ML}}\coloneqq I^{-1}\widehat A_{I,N}^{-1}$.

	\item\label{item:panel-qmle-delta-method}
	If the optimization uses an unconstrained parameter $\vartheta$ and a
	differentiable map $\theta=h(\vartheta)$, carry out
	Steps~\ref{item:panel-qmle-estimator}--\ref{item:panel-qmle-covariance} in
	$\vartheta$ and use the delta method,
	$\widehat V_{\theta}\coloneqq D\widehat V_{\vartheta}D^\top$, where
	$D\coloneqq\partial h(\widehat\vartheta)/\partial\vartheta^\top$. Construct confidence
	intervals from the resulting standard errors.
\end{enumerate}
\end{alg}

\subsection{Comparing misspecified models}
\label{subsec:model-comparison-theory}

Our main use of the likelihood is to compare competing models that are all
misspecified. Let $A$ and $B$ denote two models with population
objectives $\cQ^A$ and $\cQ^B$ and pseudo-true values $\theta_0^A$ and
$\theta_0^B$, defined as in \eqref{eq:panel-population-objective}. By
\eqref{eq:panel-pseudo-true-KL}, the difference
$\cQ^A(\theta_0^A)-\cQ^B(\theta_0^B)$ equals the difference between the
expected KL divergences from the actual outcome distribution to model $B$
and to model $A$. Thus the model with the higher population objective is
closer to the data in KL divergence.

Under the regularity conditions of \citet{Vuong1989}, we use individual
log likelihood differences to compare the models. Let $\widehat\theta^A$
and $\widehat\theta^B$ be the two
estimates, and define the individual log likelihood differences
$d_i\coloneqq\ell_i^A(\widehat\theta^A)-\ell_i^B(\widehat\theta^B)$, their
sample mean $\bar d$, and their sample variance $\widehat\sigma_d^2$. When
the two models are strictly non-nested and equally close to the data, the
statistic $\sqrt I\bar d/\widehat\sigma_d$ is asymptotically standard
normal. Large positive values favor model $A$, and large negative values
favor model $B$. The statistic uses only the individual log likelihoods
that Algorithm~\ref{alg:panel-qmle-inference} already computes. At a fixed
grid, it compares the two discretized models. It also compares the
continuous-state models when the approximation error in each total log
likelihood is small relative to $\sqrt I\widehat\sigma_d$.

\section{PSID data and the Gaussian benchmark}
\label{sec:PSID-data}

We illustrate likelihood-based model comparison with earnings dynamics,
a central input into quantitative models of consumption, saving, and
inequality. We use the Panel Study of Income Dynamics (PSID) because it is
widely used and readily available, allowing
readers to reproduce the exercise without access to restricted
administrative records. We compare models for reported earnings, which
include survey noise. 
We describe the data and estimate a benchmark with Gaussian persistent
and transitory shocks by exact maximum likelihood. We examine how
permanent heterogeneity and initial dispersion affect the estimates.

\subsection{Data and deterministic adjustment}
\label{subsec:data-adjustment}

We use 14 biennial PSID interviews from 1997 through 2023, which report
annual labor earnings for income years 1996, 1998, \ldots, 2022. We
follow individuals using their 1968 family identifier and person number,
selecting annual labor earnings for their current role as Reference Person
or spouse/partner. We compute each individual's birth year as
the rounded median of interview year minus reported age. We use this birth
year to define income-year age and measure elapsed time from age 30.

The baseline sample contains men ages 30--60 in the income year who are
current Reference Persons, spouses, or partners. We require positive, nonmissing annual
labor earnings and nonmissing completed education, as well as at least
two usable observations per person, but not a balanced panel. We keep only
the SRC core sample (1968 family identifiers 1--2,930). This restriction
drops the Census low-income oversample and the immigrant samples added
from 1997, so that the unweighted sample does not overrepresent
low-income households. Following \citet{GuvenenKarahanOzkanSong2021}, we
also require annual labor earnings of at least 13 weeks of 40 hours at
half the federal minimum wage ($\$1{,}885$ for income years 2010--2022),
which excludes observations with minimal attachment to the labor market.
Table~\ref{tab:PSID-sample-flow} reports the restrictions. The final panel
contains 28,422 observations for 4,683 men.

\begin{table}[!htb]
 \centering
 \caption{PSID sample construction.}
 \label{tab:PSID-sample-flow}
 \begin{tabular}{lrr}
  \toprule
  Restriction & Observations & Individuals \\
  \midrule
  Selected person--wave records & $649{,}768$ & $46{,}412$ \\
  Male & $319{,}102$ & $22{,}793$ \\
  Current Reference Person or spouse/partner & $81{,}394$ & $12{,}531$ \\
  Income-year age 30--60 & $52{,}416$ & $9{,}644$ \\
  Positive, nonmissing labor income & $45{,}894$ & $9{,}131$ \\
  Nonmissing completed education & $44{,}371$ & $8{,}862$ \\
  SRC core sample & $29{,}533$ & $5{,}567$ \\
  Earnings at least the minimum threshold & $29{,}278$ & $5{,}539$ \\
  At least two usable observations & $28{,}422$ & $4{,}683$ \\
  \bottomrule
 \end{tabular}
 \caption*{\footnotesize\textit{Note:} The table reports the sample
 restrictions that we apply successively to the 14 PSID interviews from 1997 through 2023. The
 minimum threshold is 520 hours at half the federal minimum wage, averaged
 over the income year.}
\end{table}

Apart from the minimum threshold, we do not trim earnings or use
income-quality flags, and we do not require positive work weeks or hours.
We construct the sample and estimate the models without weights.
The threshold excludes only 0.9 percent of the SRC observations, but these
observations have extreme log earnings and dominate the tails of earnings
changes.\footnote{Without the threshold, the excess kurtosis of two-year
changes in the SRC sample is $32.70$ instead of $10.82$. If we also
include the Census and immigrant samples, the corresponding values are
$26.07$ and $9.11$.}

Before estimating the earnings dynamics model, we remove predictable age--education--year variation from annual log
earnings. Let $Y_{it}>0$ be the selected annual labor earnings and let
$G_{it}\in\set{1,2,3,4}$ denote completed education: less than high school,
high school, some college, or BA or more. Following \citet{KarahanOzkan2013}, we use an education-specific cubic age
profile and additive income-year effects. With centered and scaled age
$z_{it}\coloneqq(a_{it}-45)/10$, we estimate by ordinary least squares
\begin{equation}
 \log Y_{it}
 =\delta_t
 +\sum_{g=1}^4\mathbf{1}\set{G_{it}=g}
 \sum_{p=0}^3\gamma_{gp}z_{it}^p
 +e_{it},
 \label{eq:PSID-residualization}
\end{equation}
where $\delta_t$ is an income-year effect, which we normalize to zero in 1996.\footnote{We use
education-specific cubics for parsimony: replacing them with
separate indicators for each combination of exact age and education group
increases the number of coefficients from 29 to 137, but raises the
first-stage $R^2$ only from $0.1890$ to
$0.1917$.} We use the OLS
residual $u_{it}\coloneqq\widehat e_{it}$ as our outcome, omitting individual
fixed effects so that it retains the persistent heterogeneity modeled
below. The year effects absorb common price-level changes, so deflating
earnings by a year-specific price index leaves the residuals unchanged
apart from numerical rounding.

\subsection{Gaussian benchmark}
\label{subsec:OU-benchmark}

We use a Gaussian persistent--transitory model as a natural benchmark,
given its widespread use in the earnings-dynamics literature
\citep{AbowdCard1989,MeghirPistaferri2011}. We first assess how
permanent heterogeneity and initial dispersion affect estimated persistence.

At elapsed age $t$ since 30, the persistent component follows the Ornstein--Uhlenbeck (OU) process
\begin{equation}
  \diff X_i(t)=-\kappa X_i(t)\diff t+\sigma\diff B_i(t),
  \label{eq:OU-dynamics}
\end{equation}
where $\kappa>0$ controls the speed of mean reversion, $\sigma>0$ is the
instantaneous volatility of persistent shocks, and $B_i$ are independent
standard Brownian motions. We assume $X_i(0)\sim\cN(0,\sigma_0^2)$, with
initial standard deviation $\sigma_0>0$. Persistence over an observation
gap $\Delta>0$ is $\e^{-\kappa\Delta}$. The stationary standard
deviation is $\omega\coloneqq\sigma/\sqrt{2\kappa}$; estimating $\sigma_0$ freely
allows dispersion at age 30 to differ from this long-run value.

In specification~(1) of Table~\ref{tab:OU-benchmarks}, we omit
permanent effects and write
\begin{equation}
  u_{ij}=X_i(t_{ij})+\epsilon_{ij},
  \qquad
  \epsilon_{ij}\sim\cN(0,\sigma_\epsilon^2).
  \label{eq:OU-gaussian-observation}
\end{equation}
The transitory shocks are independent across people and dates and
independent of the initial state and Brownian motions. They capture both
genuine short-lived earnings changes and reporting error, which is
substantial in survey earnings data \citep{BoundBrownDuncanRodgers1994}. We apply the Kalman
filter to evaluate the Gaussian levels likelihood exactly, using the actual
observation gaps. We obtain $\widehat\kappa=0.039$, implying a mean-reversion
half-life of about 18 years.

\begin{table}[!htb]
 \centering
 \caption{OU estimates under alternative permanent effects and initial conditions.}
 \label{tab:OU-benchmarks}
 \begin{tabular}{lrrrrr}
  \toprule
  Specification & \multicolumn{1}{c}{(1)} & \multicolumn{1}{c}{(2)} & \multicolumn{1}{c}{(3)} & \multicolumn{1}{c}{(4)} & \multicolumn{1}{c}{(5)} \\
  Permanent effect & \multicolumn{1}{c}{None} & \multicolumn{1}{c}{Unrestricted} & \multicolumn{1}{c}{Gaussian} & \multicolumn{1}{c}{Gaussian} & \multicolumn{1}{c}{Normal--Laplace} \\
  Initial dispersion & \multicolumn{1}{c}{Free} & \multicolumn{1}{c}{Free} & \multicolumn{1}{c}{Stationary} & \multicolumn{1}{c}{Free} & \multicolumn{1}{c}{Free} \\
  Likelihood data & \multicolumn{1}{c}{Levels} & \multicolumn{1}{c}{Differences} & \multicolumn{1}{c}{Levels} & \multicolumn{1}{c}{Levels} & \multicolumn{1}{c}{Levels} \\
  \midrule
  $\widehat\kappa$           & 0.039 & 0.055 & 0.105 & 0.056 & 0.080 \\
  & $(0.004)$ & $(0.009)$ & $(0.016)$ & $(0.012)$ & $(0.010)$ \\
  $\widehat\sigma$           & 0.223 & 0.247 & 0.259 & 0.235 & 0.247 \\
  & $(0.008)$ & $(0.010)$ & $(0.013)$ & $(0.012)$ & $(0.011)$ \\
  $\widehat\sigma_0$         & 0.531 & $0.000^{\mathrm b}$ & 0.566 & 0.417 & 0.213 \\
  & $(0.015)$ &  & $(0.025)$ & $(0.055)$ & $(0.052)$ \\
  $\widehat\sigma_G$   &  &  & 0.399 & 0.323 & $0.000^{\mathrm b}$ \\
  &  &  & $(0.035)$ & $(0.065)$ &  \\
  $\widehat\sigma_\epsilon$ & 0.327 & 0.310 & 0.304 & 0.320 & 0.314 \\
  & $(0.009)$ & $(0.009)$ & $(0.011)$ & $(0.010)$ & $(0.010)$ \\
  $\widehat a$               &  &  &  &  & 3.807 \\
  &  &  &  &  & $(0.228)$ \\
  $\widehat b$               &  &  &  &  & 2.790 \\
  &  &  &  &  & $(0.105)$ \\
  Log likelihood & $-22{,}694.30$ &  & $-22{,}793.08$ & $-22{,}690.28$ & $-22{,}595.35$ \\
  AIC & $45{,}396.60$ &  & $45{,}594.15$ & $45{,}390.56$ & $45{,}204.70$ \\
  \bottomrule
 \end{tabular}
 \caption*{\footnotesize\textit{Note:} All specifications use Gaussian OU
 state dynamics and the same underlying PSID residual panel. ``Free'' means
 we estimate the age-30 standard deviation $\sigma_0$; ``Stationary''
 imposes $\sigma_0=\omega=\sigma/\sqrt{2\kappa}$. In (2), we eliminate
 unrestricted permanent effects by differencing and retain the serial
 correlation of the differences. We omit its log likelihood and AIC because
 they are not comparable with those for levels. In (5), we use the centered
 normal--Laplace permanent effect defined in \S\ref{subsec:permanent-heterogeneity}, with tail rates $a,b$.
 This seven-parameter specification nests (4) as $a,b\to\infty$.
 Superscript $\mathrm b$ marks estimates at the zero boundary, which have no
 regular standard error. Parentheses report individual-clustered sandwich
 standard errors; the standard error of $\widehat\sigma_0$ in (3) uses the
 delta method.}
\end{table}

Such persistence could partly reflect permanent differences in earnings
levels rather than slowly decaying shocks. To assess
this possibility, we introduce a permanent individual effect $\alpha_i$ and modify \eqref{eq:OU-gaussian-observation} as
\begin{equation}
 u_{ij}=\alpha_i+X_i(t_{ij})+\epsilon_{ij}.
 \label{eq:OU-observation}
\end{equation}
We assess permanent heterogeneity in two ways. In (2), we eliminate
unrestricted $\alpha_i$ by first differencing and maximize the exact joint
Gaussian likelihood of the differences. In (3)--(4), we assume an
independent Gaussian intercept $\alpha_i\sim\cN(0,\sigma_G^2)$ and
estimate in levels, imposing stationary initialization ($\sigma_0=\omega$) in (3) and
estimating $\sigma_0$ freely in (4).

With free $\sigma_0$, we obtain $\widehat\kappa=0.055$ in (2) and
$0.056$ in (4), somewhat faster mean reversion than in (1). In (2), the
estimated initial dispersion is at its zero boundary. In (4), we gain only
$4.02$ in log likelihood from adding a Gaussian intercept.\footnote{At
$\kappa=0$, the Gaussian initial state and permanent effect are
indistinguishable, with their variances entering the likelihood only
through $\sigma_0^2+\sigma_G^2$. High persistence can make them close
substitutes over finite histories. In fact, using the estimates in (4), we have
$\sqrt{\sigma_0^2+\sigma_G^2}=0.528$, close to $\sigma_0=0.531$
in (1).} When we impose
stationarity in (3), we obtain $\widehat\kappa=0.105$, with a log
likelihood $98.78$ below (1). We therefore retain free initial dispersion in the
extensions below. In specification~(5), introduced next, we allow non-Gaussian permanent
heterogeneity to accommodate the distribution of earnings levels.

\subsection{Earnings levels and permanent heterogeneity}
\label{subsec:permanent-heterogeneity}

With Gaussian initial conditions and permanent heterogeneity, the OU
benchmark implies a Gaussian distribution of adjusted log earnings at each
fixed age. To accommodate the power-law patterns documented in the
income distribution literature \citep{Toda2012JEBO}, we consider the
normal--Laplace distribution \citep{ReedJorgensen2004}. It is the
convolution of a Gaussian variable and an asymmetric Laplace variable.
Exponentiating a normal--Laplace variable gives the double Pareto-lognormal
distribution, whose upper and lower tails obey power laws.

\begin{figure}[!htb]
 \centering
 \includegraphics[width=\textwidth]{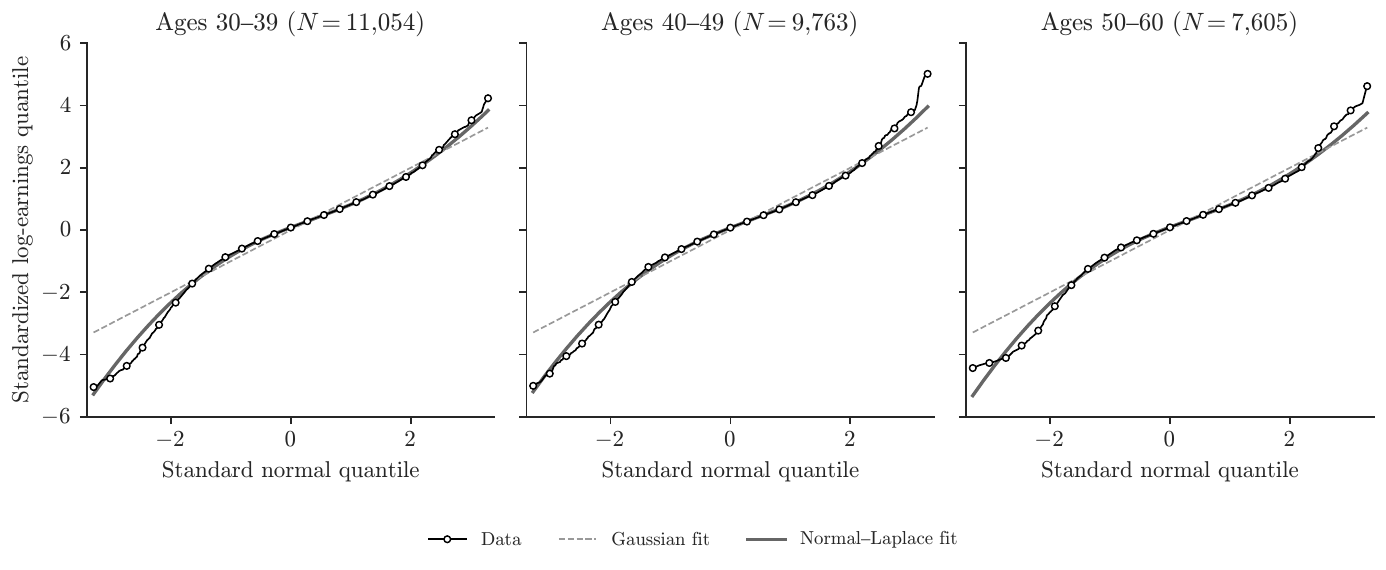}
 \caption{Standardized log-earnings quantiles by age.}
 \caption*{\footnotesize\textit{Note:} The horizontal axis gives standard
 normal quantiles. Black circles show empirical quantiles and solid gray curves
 show quantiles of the normal--Laplace distribution that we fit by maximum
 likelihood within each age group. We standardize both using the empirical
 mean and standard deviation (with divisor $n_g$, the number of
 observations in age group $g$). The dashed
 gray 45-degree line gives the fitted Gaussian quantiles under the same
 standardization. Probabilities range from $0.0005$ to $0.9995$; all panels
 use common axis limits.}
 \label{fig:gaussian-QQ-by-age}
\end{figure}

We fit both distributions separately within broad age groups and compare
their quantile--quantile (Q--Q) plots in Figure~\ref{fig:gaussian-QQ-by-age}. The normal--Laplace
fit improves most clearly in the lower tail and is also closer to the
data between standardized log-earnings quantiles $-2$ and $2$.
Discrepancies remain at the most extreme quantiles. These are descriptive
marginal fits: pooling ages with different variances can produce a
non-Gaussian mixture even under Gaussian OU dynamics. The plots do not by
themselves establish Pareto-tail asymptotics or identify the source of
non-Gaussianity.

We assess permanent heterogeneity as a potential source of non-Gaussian earnings
levels using the panel likelihood. In specification~(5) of Table~\ref{tab:OU-benchmarks}, we add an
asymmetric Laplace component to the Gaussian intercept in (4), retaining
Gaussian persistent and transitory shocks:
\begin{subequations}
 \label{eq:normal-laplace-effect}
\begin{align}
& \alpha_i
 \coloneqq G_i+V_{a,i}-V_{b,i}-1/a+1/b, \\
&G_i\sim\cN(0,\sigma_G^2),\quad
 V_{a,i}\sim\operatorname{Exp}(a),\quad
 V_{b,i}\sim\operatorname{Exp}(b),
\end{align}
\end{subequations}
where $G_i,V_{a,i},V_{b,i}$ are mutually independent and independent of the
OU process and transitory shocks. The exponential distributions use rates
$a,b>0$, which govern the upper and lower tails, respectively. Centering
ensures $\E[\alpha_i]=0$, and
$\var(\alpha_i)=\sigma_G^2+a^{-2}+b^{-2}$. At each fixed age, adjusted
log earnings then have a normal--Laplace distribution. This specification
assigns the non-Gaussian tails to permanent heterogeneity while preserving
Gaussian persistent innovations. It nests specification~(4) as
$a,b\to\infty$.

In (5), we obtain a log-likelihood gain of $94.93$ over (4) with two
additional parameters, lowering AIC by $185.87$.\footnote{We evaluate the
panel likelihood exactly: conditional on the asymmetric-Laplace component
of $\alpha_i$, the model is Gaussian, and the remaining integral has a
closed-form expression. See Appendix~\ref{sec:OU-normal-laplace-likelihood}.}
The Gaussian intercept component is at its zero boundary, and
$\widehat b=2.790<\widehat a=3.807$ implies a heavier lower tail. We also
obtain faster mean reversion, $\widehat\kappa=0.080$, with a half-life of
8.7 years. Non-Gaussian permanent heterogeneity therefore improves fit and
reduces the persistence attributed to the OU state. We next examine whether
it also accounts for earnings dynamics.

\subsection{Limitations of the Gaussian OU benchmark}
\label{subsec:OU-limitations}

We assess the joint restriction of Gaussian persistent dynamics and
Gaussian transitory shocks, retaining non-Gaussian permanent
heterogeneity. We examine two-year changes, conditional means and variances,
and reversals after large losses and gains. We compare the data with 300
panels that we simulate from specification~(5) on the observed age schedules,
using its freely estimated initial dispersion. The data provide 22,619
two-year pairs and 17,586 triples with two successive two-year intervals.

\paragraph{Heavy-tailed changes}
Differencing eliminates $\alpha_i$ regardless of its distribution,
leaving the persistent and transitory components, both Gaussian under
the benchmark. Let $\rho\coloneqq\e^{-\kappa}$ and $\eta_i(t)$ denote annual persistence and Gaussian
OU innovation over two years. The change satisfies
\begin{equation}
	d_i(t)\coloneqq u_i(t+2)-u_i(t)=(\rho^2-1)X_i(t)+\eta_i(t)+\epsilon_i(t+2)-\epsilon_i(t).
 \label{eq:two-year-earnings-change}
\end{equation}
At a fixed initial age, $d_i(t)$ is therefore Gaussian.

\begin{figure}[!htb]
 \centering
 \includegraphics[width=0.7\textwidth]{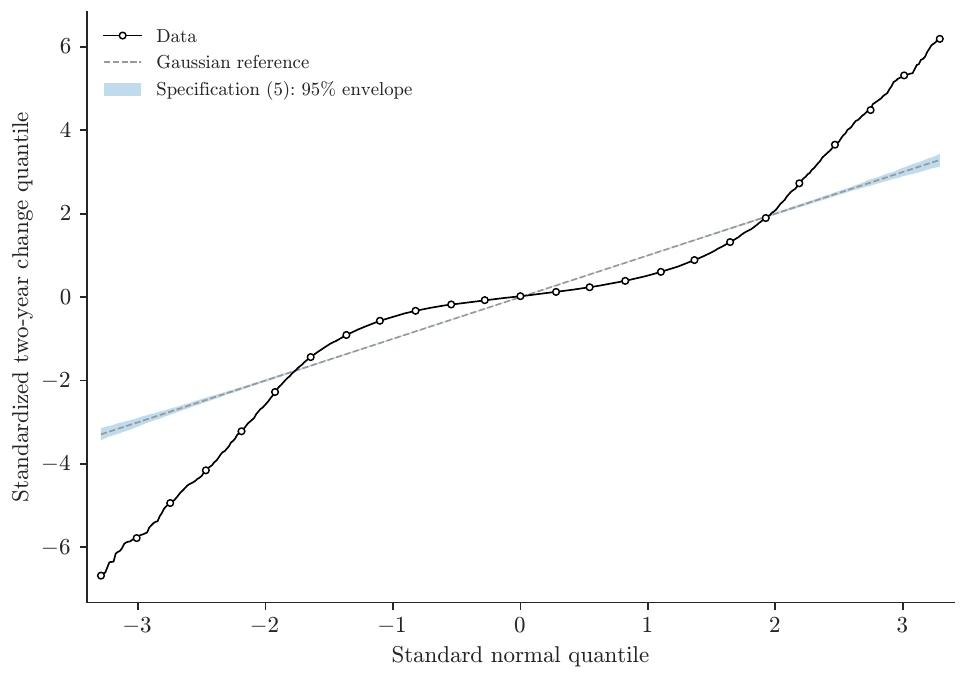}
 \caption{Q--Q plot of two-year changes in annual log-earnings residuals.}
 \caption*{\footnotesize\textit{Note:} We plot change quantiles against
 standard normal quantiles, demeaning and standardizing changes by their
 own standard deviation in the data and in each simulated panel.
 Probabilities range from $0.0005$ to $0.9995$; the dashed line is the
 Gaussian reference. For the benchmark comparisons in this section, we
 report medians and pointwise 95 percent ranges (2.5th--97.5th percentiles)
 across 300 panels that we simulate at the fitted parameters on the observed age
 schedules, excluding parameter uncertainty. In these simulation figures,
 thin black lines with circles denote the data; blue dashed lines (where
 shown) and bands give the simulation medians and ranges, respectively.
 We use specification (5) here.}
 \label{fig:benchmark-change-QQ}
\end{figure}

In Figure~\ref{fig:benchmark-change-QQ}, we
confront this prediction with the data. Both tails of the change distribution depart sharply from the Gaussian
reference. Excess kurtosis is $10.82$, compared with approximately zero in
the simulated panels
(Table~\ref{tab:benchmark-dynamics}). Pooling ages can mix Gaussian
distributions with different variances, which is why the figure also shows
the simulated envelope on the actual age schedules. That envelope remains
close to the Gaussian reference and far from the empirical tails. The
fitted benchmark therefore misses large changes despite allowing
non-Gaussian permanent heterogeneity.

\begin{figure}[!htb]
 \centering
 \includegraphics[width=\textwidth]{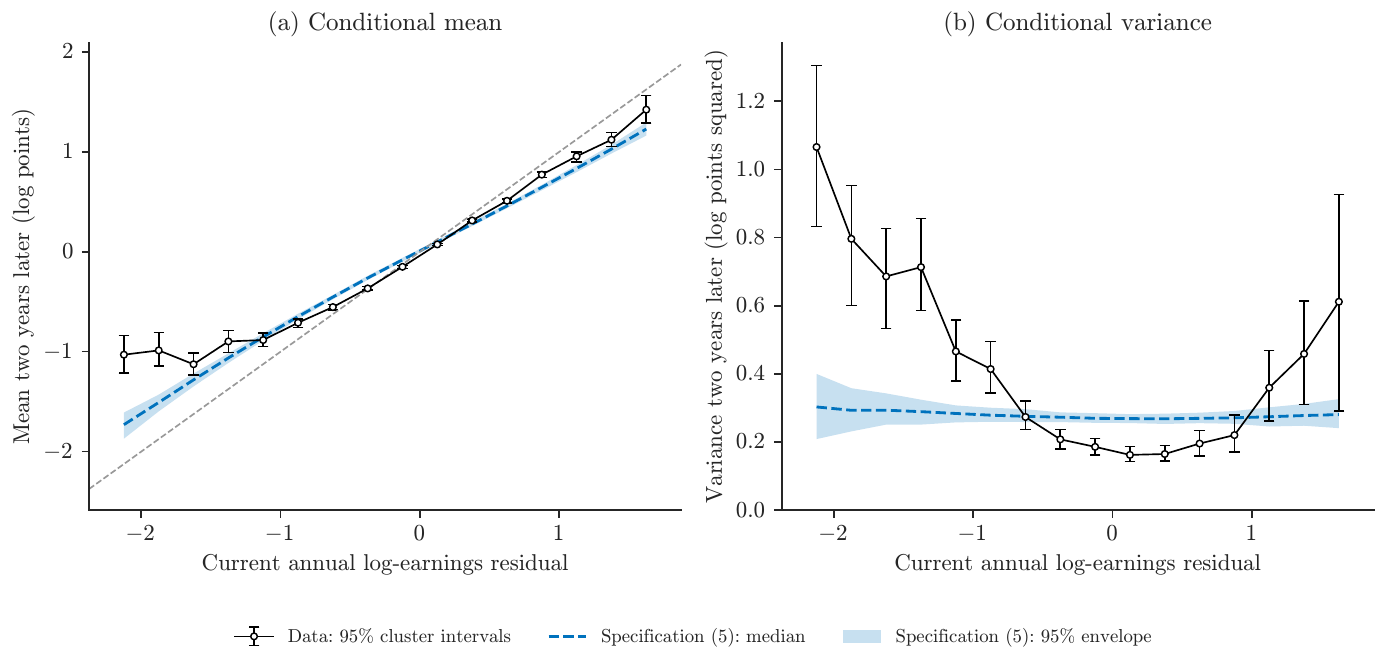}
 \caption{Conditional mean and variance of annual log-earnings residuals two years later.}
 \caption*{\footnotesize\textit{Note:} Within each bin of width 0.25, we
 regress the next residual on a constant and the current residual. The
 left panel plots the fitted mean at the bin midpoint; the right plots
 the residual variance, using divisor $n-2$. We only show bins with at least 100
 empirical pairs. Black error bars are pointwise 95 percent percentile
 intervals from 500 individual-cluster bootstrap samples. We apply the
 same bins and estimators to panels simulated from specification (5),
 using the conventions of Figure~\ref{fig:benchmark-change-QQ}. The dashed line in the
 left panel is the 45-degree line.}
 \label{fig:benchmark-conditional-moments}
\end{figure}

\paragraph{Conditional mean and dispersion}
Figure~\ref{fig:benchmark-conditional-moments} reports the mean
(left) and variance (right) of earnings residuals two years later,
conditional on the current residual. Within-bin linear regressions give
the mean at the bin midpoint and the variance around the fitted line.

The conditional mean rises with current earnings, but the fitted OU model
does not reproduce its shape. The benchmark predicts too much downward
adjustment at high earnings and too little recovery at very low earnings.
Conditional variance is U-shaped, with dispersion rising toward both
tails, especially on the lower side. The fitted benchmark gives a much
flatter profile, overstating central dispersion and missing the sharp
increase at low earnings.

These comparisons concern the observed residual $u_{it}$, rather than
the latent state $X_i(t)$. The OU state has a linear conditional mean and
a transition variance independent of its current value. With asymmetric,
non-Gaussian permanent heterogeneity, however, the composition of
$\alpha_i$, $X_i(t)$, and transitory noise varies with $u_{it}$. The
observed conditional mean need not be linear, nor its conditional
variance flat. We simulate the fitted model with all these components
to obtain the relevant comparison. Yet, its permanent heterogeneity
does not account for the observed conditional profiles.

\paragraph{Asymmetric reversals}
Conditional variance describes the size of future fluctuations, but not
whether earnings tend to reverse direction after a large move. Gaussian mean reversion and
Gaussian transitory noise can generate reversals, yet they treat positive and negative
moves symmetrically: reversing both signs leaves the joint distribution
of successive changes unchanged, even after pooling ages. The benchmark therefore
predicts the same reversal probability after gains and losses of equal
magnitude in log points. This provides a third diagnostic.

To define a large move, we ask whether adjusted earnings, $\exp(u_i(t))$,
fall to at most half their previous level or at least double. These events
correspond to $d_i(t)\le \log(1/2)=-\log 2$ and $d_i(t)\ge \log 2$, respectively.
These thresholds have equal magnitude and opposite signs in logs. We call
the next two-year change a reversal if it has the opposite sign; this does
not require earnings to return fully to their previous level.

Table~\ref{tab:benchmark-dynamics} shows that the data violate the
symmetry prediction. The next change reverses direction after $78.6\%$ of
large losses but only $59.6\%$ of large gains, compared with simulated
medians of about $74.0\%$ for both. The empirical loss--gain difference is
$19.0$ percentage points, with a 95 percent individual-cluster bootstrap
interval of $[14.5,23.7]$. The asymmetry also concerns the amount recovered:
earnings retrace at least half of the initial log change after $53.5\%$ of large
losses but only $25.1\%$ of large gains. The diagnostics leave open whether persistent dynamics, transitory
shocks, or both require a richer specification.

\begin{table}[!htb]
 \centering
 \caption{Annual earnings dynamics: data and the Gaussian OU benchmark.}
 \label{tab:benchmark-dynamics}
 \begin{tabular}{lrrr}
  \toprule
  Statistic & Data & Spec. (5) & 95\% simulation range \\
  \midrule
  \multicolumn{4}{l}{\textit{Two-year changes}} \\
  Standard deviation & $0.536$ & $0.555$ & $[0.549,0.561]$ \\
  Skewness & $-0.499$ & $0.001$ & $[-0.035,0.033]$ \\
  Excess kurtosis & $10.82$ & $0.00$ & $[-0.06,0.06]$ \\
  \addlinespace
  \multicolumn{4}{l}{\textit{Subsequent two-year reversal (percent)}} \\
  Opposite sign after a large loss & $78.63$ & $73.95$ & $[71.81,75.92]$ \\
  Opposite sign after a large gain & $59.62$ & $74.10$ & $[71.96,75.90]$ \\
  Loss--gain gap (percentage points) & $19.01$ & $0.00$ & $[-3.22,2.85]$ \\
  At least half retraced after a loss & $53.54$ & $39.39$ & $[37.22,41.40]$ \\
  At least half retraced after a gain & $25.12$ & $39.41$ & $[37.59,41.44]$ \\
  \bottomrule
 \end{tabular}
 \caption*{\footnotesize\textit{Note:} Change moments use 22,619 two-year
 pairs. Reversals use 17,586 triples, including 861 large losses and 832
 large gains, defined by $\abs{\text{initial changes}} \ge\log 2$.
 Opposite sign means a positive next change after a loss or a negative next
 change after a gain. At least half retraced means
 $-d_i(t+2)/d_i(t)\ge 1/2$. Rates condition on usable positive earnings at
 all three dates. We use specification (5) and the simulation
 conventions of Figure~\ref{fig:benchmark-change-QQ}.}
\end{table}

\section{Models and estimation}
\label{sec:models-estimation}

We compare four models of adjusted log earnings.

\subsection{Model specifications}
\label{subsec:model-specifications}

Throughout, $t$ measures years since age 30, and
$X_{it}\coloneqq X_i(t)$ at integer ages. We write annual persistence as
$\rho\coloneqq\e^{-\kappa}\in(0,1)$, with $\kappa>0$. The discrete-time
models evolve annually, including years without an observed earnings
measurement.

All specifications use the centered asymmetric Laplace permanent effect
in \eqref{eq:normal-laplace-effect}, imposing $\sigma_G=0$
and estimating the tail rates $a,b>0$. We initialize
$X_i(0)\sim\cN(0,\sigma_0^2)$ with a freely estimated $\sigma_0>0$ and
propagate this law to each individual's entry age. The transition laws
admit stationary distributions; the age-30 law need not be stationary.
We reestimate all parameters within each specification.

Our baseline observation equation is
\begin{equation}
 u_{ij}=\alpha_i+X_i(t_{ij})+\epsilon_{ij}. \label{eq:models-observation}
\end{equation}
One specification adds a shared loss component, as described below.
In every model, the independent transitory shock is either
$\epsilon_{ij}\sim\cN(0,\sigma_\epsilon^2)$ or
$\epsilon_{ij}\sim\cM(p_\epsilon,\delta_\epsilon,
\sigma_{\epsilon,1},\sigma_{\epsilon,2})$, where
\begin{equation}
 \cM(p,\delta,\sigma_1,\sigma_2)
 \coloneqq p\,\cN(-(1-p)\delta,\sigma_1^2)
 +(1-p)\,\cN(p\delta,\sigma_2^2)
 \label{eq:centered-normal-mixture}
\end{equation}
denotes a centered two-component Gaussian mixture, with $0<p<1$,
$\delta\ge 0$, and $\sigma_1,\sigma_2>0$. This four-parameter
mixture has mean zero and ordered component means; it adds three
parameters relative to Gaussian transitory noise. Individuals are
independent. Within each model, permanent effects, initial states, Brownian
motions, and the iid primitive shock sequences are mutually independent.

We begin with the Gaussian OU process for the persistent state. GKOS, GMR,
and PLV introduce non-Gaussian risk through mixture innovations, losses and
scarring, and state-dependent variance, respectively.

\paragraph{Ornstein--Uhlenbeck (OU) process}
\label{par:OU}

The OU state follows \eqref{eq:OU-dynamics}. At annual dates,
it is exactly a Gaussian AR(1) with persistence $\rho$ and innovation
variance $\sigma^2(1-\rho^2)/(2\kappa)$.

\paragraph{Guvenen--Karahan--Ozkan--Song (GKOS) process}
\label{par:GKOS}

We adapt the constant-mixture specification (Model~3) of
\citet{GuvenenKarahanOzkanSong2021} by introducing mean reversion. The
persistent state follows
\begin{equation}
	X_{it}=\rho X_{i,t-1}+\eta_{it}, \qquad \eta_{it}\sim\cM(p_\eta,\delta_\eta,\sigma_{\eta,1},\sigma_{\eta,2}).
 \label{eq:GKOS-dynamics}
\end{equation}
At the boundary $p_\eta\in\{0,1\}$, innovations are Gaussian; matching
the component variance to the OU innovation variance recovers the exact
annual OU state transition.
The mixture probability $p_\eta$ and its component distributions are constant
across ages and earnings levels. Unequal component means and variances
allow skewness and excess kurtosis in persistent innovations while
preserving linear conditional mean reversion.

\paragraph{Guvenen--McKay--Ryan (GMR) process}
\label{par:GMR}

We adapt the nonemployment and scarring specification (Model~2) of
\citet{GuvenenMcKayRyan2026}, replacing its random walk with mean reversion
and adding the common independent transitory shock. Retaining
their notation, let
\begin{equation}
 \zeta_{it}=
 \begin{cases}
  0,&\text{with probability }p_\zeta,\\
  \log(1-\ell_{it}),&\text{with probability }1-p_\zeta,
 \end{cases}
 \qquad c_{it}\coloneqq\zeta_{it}-\E[\zeta_{it}],
 \label{eq:GMR-loss}
\end{equation}
where $0<p_\zeta<1$ and $\ell_{it}$ has an exponential distribution with
rate $\lambda>0$, conditional on $0<\ell_{it}<1$. Centering accommodates
the zero-mean earnings residuals. The state and observation equations are
\begin{subequations}
\label{eq:GMR-model}
\begin{align}
 X_{it}&=\rho X_{i,t-1}+\eta_{it}+\psi c_{it},
 &\eta_{it}&\sim\cN(0,\sigma_\eta^2),
 \label{eq:GMR-state}\\
 u_{ij}&=\alpha_i+X_i(t_{ij})+(1-\psi)c_{i,t_{ij}}+\epsilon_{ij},
 \label{eq:GMR-observation}
\end{align}
\end{subequations}
with $\sigma_\eta>0$ and $0<\psi<1$. At the boundary $p_\zeta=1$,
the loss vanishes and $\psi$ drops out; matching $\sigma_\eta^2$ to the
OU innovation variance again gives the annual OU state transition.
For $t\ge1$, a loss innovation $c_{it}$ affects earnings by $c_{it}$
on impact and by $\psi\rho^k c_{it}$ after $k\ge1$ years, so scarring decays.
At age 30, the common Gaussian state is independent of $c_{i0}$;
state innovations begin at $t=1$. The non-Gaussian loss is present
under both specifications of $\epsilon$.

\paragraph{Piecewise linear variance (PLV) process}
\label{par:PLV}

The heavy-tailed changes and U-shaped conditional dispersion in
\S\ref{subsec:OU-limitations} motivate a parsimonious specification
with linear mean reversion and instantaneous variance that can increase
with the persistent state's distance from zero, at different rates on
either side.

For $x\in\R$, let $x^{+}\coloneqq\max\set{x,0}$ and
$x^{-}\coloneqq\max\set{-x,0}$ be the positive and negative parts of $x$.
We define the instantaneous variance function
\begin{equation}
 v(x)\coloneqq\sigma^2+c_{+}x^{+}+c_{-}x^{-},
 \label{eq:PLV-variance}
\end{equation}
where $\sigma>0$ and $c_{+},c_{-}\ge 0$. The persistent
component follows the diffusion
\begin{equation}
	\diff X_i(t)=-\kappa X_i(t)\diff t+\sqrt{v(X_i(t))}\diff B_i(t), \label{eq:PLV-diffusion}
\end{equation}
where $B_i$ is a standard Brownian motion.

Setting $c_{+}=c_{-}=0$ recovers the OU process. PLV belongs to the class of
diffusions with linear mean reversion studied by
\citet{BibbySkovgaardSorensen2005}. Its variance is affine on each
half-line. When both slopes are positive,
$v(x)$ is V-shaped, with its minimum at zero; $c_{-}>c_{+}$ makes variance
rise more steeply at negative persistent states.

Proposition~\ref{prop:PLV-stationary-distribution} gives exponential
stationary tails on each side with a positive slope. Proposition~\ref{prop:linear-drift-moments} shows that PLV preserves
the conditional mean of OU, $\E[X(t+h)\mid X(t)=x]=\e^{-\kappa h}x$, while
stationary latent changes have positive limiting excess kurtosis as
$h\downarrow0$ whenever $c_{+}+c_{-}>0$. Proposition~\ref{prop:diffusion-change-symmetry} shows that stationary latent and
observed changes remain symmetric even when the level distribution is
skewed. However, it does not imply equal reversal
probabilities after equally large gains and losses; unequal slopes or
asymmetric transitory shocks can break the additional sign symmetry.

\paragraph{Covariance equivalence}
\label{par:covariance-equivalence}

The four processes can share the stationary covariances targeted in
earnings calibrations \citep[e.g.,][]{StoreslettenTelmerYaron2004}.

\begin{prop}[Covariance equivalence]
\label{prop:covariance-equivalence}
Consider stationary versions of the GKOS, GMR, and PLV models, with the
observation equation \eqref{eq:models-observation} (including the shared
loss in GMR) and any transitory-shock distribution with finite variance.
For each such specification, there is a stationary OU specification with
the same persistence $\rho$ and the same permanent-effect variance whose
observed residuals have the same mean and the same autocovariance function
at all integer lags.
\end{prop}

See the derivations following
Proposition~\ref{prop:linear-drift-moments} in
Appendix~\ref{sec:PLV-properties} for the proof. The models can share
stationary covariances while differing in conditional earnings risk.

\subsection{Likelihood estimation}
\label{subsec:likelihood-estimation}

We estimate all eight specifications using the HMM likelihood in
\S\ref{sec:panel-likelihood}.
The likelihood conditions on each individual's observed ages and gaps,
under the observation-schedule independence assumption in
\S\ref{subsec:short-panel-model}. Let $\theta$ collect the parameters of
the specification we estimate.

We approximate the persistent state by a finite-state Markov chain while
leaving earnings continuous. We use conditional moment matching
\citep{FarmerToda2017QE} for OU, GKOS, and the Gaussian autoregressive
step in GMR, and a generator
approximation for PLV (Appendix~\ref{sec:numerical-implementation}).

Conditional on the permanent effect $\alpha_i=\alpha$, OU, GKOS, and PLV
have observation density $f_{\epsilon,\theta}(u_{ij}-\alpha-x)$, where
$f_{\epsilon,\theta}$ is the Gaussian or Gaussian-mixture density in
\S\ref{subsec:model-specifications}. For GMR, we use a joint transition and observation
kernel that integrates its shared loss and preserves the dependence
between current earnings and the persistent state. The HMM forward
updates in \eqref{eq:panel-filter}--\eqref{eq:panel-filter-log-likelihood}
give the conditional whole-history likelihood
$L_{i,N}(\theta\mid\alpha)$ on an $N$-point state grid.

We integrate the whole-history likelihood over the common centered
asymmetric-Laplace density $f_{\alpha,\theta}$. Let $K_\alpha$ denote the number
of quadrature nodes for the permanent effect, with $\alpha_{ik}(\theta)$
the $k$th node for individual $i$. The exact individual likelihood and
its numerical approximation are
\begin{subequations}
\begin{align}
 L_i(\theta)
 &=\int_{\R}L_i(\theta\mid\alpha)f_{\alpha,\theta}(\alpha)\diff\alpha, \\
 L_{i,N,K_\alpha}(\theta)
 &\coloneqq\sum_{k=1}^{K_\alpha} w_{ik}(\theta)
       L_{i,N}(\theta\mid\alpha_{ik}(\theta)),
 \label{eq:integrated-panel-likelihood}
\end{align}
\end{subequations}
where each effective weight $w_{ik}(\theta)$ includes the quadrature
weight and the permanent-effect density at that node. We use
individual-specific nodes to resolve each history's integrand while
retaining a common prior. Appendix~\ref{subsec:common-estimation-settings}
describes the quadrature, parameter transformations, and optimization.

We jointly estimate the dynamic parameters, initial dispersion,
transitory-shock distribution, and permanent-effect tail rates by
maximizing the sum of individual log likelihoods, as in
\eqref{eq:panel-discrete-estimator}. For model comparison, we treat
individuals as independent sampling units and the first-stage adjustment
in \eqref{eq:PSID-residualization} as fixed.
Table~\ref{tab:estimate-accuracy} reports the available grid, range, and
quadrature refinements at the empirical estimates;
Appendix~\ref{subsec:OU-validation} documents the validation and scope of
these checks.

\subsection{Model comparison}
\label{subsec:model-comparison}

Table~\ref{tab:model-comparison} compares the four models under Gaussian
(left) and Gaussian-mixture (right) independent transitory shocks.
GKOS retains its persistent innovation mixture and GMR its shared loss
in both comparisons.

\begin{table}[!htb]
 \centering
 \caption{Selected estimates and log likelihoods.}
 \label{tab:model-comparison}
 \newcommand{\mpar}[2]{\makebox[2em][l]{$#1$}$#2$}%
 \begin{tabular}{lrrrrrrrr}
  \toprule
  & \multicolumn{4}{c}{Gaussian $\epsilon$} & \multicolumn{4}{c}{Gaussian-mixture $\epsilon$} \\
  \cmidrule(lr){2-5}\cmidrule(lr){6-9}
  & OU & GKOS & GMR & PLV & OU & GKOS & GMR & PLV \\
  \midrule
  \multicolumn{9}{l}{\textit{Common parameters}} \\
  $\widehat\rho$ & $0.923$ & $0.715$ & $0.963$ & $0.838$ & $0.979$ & $0.973$ & $0.969$ & $0.962$ \\
  $\widehat\sigma_0$ & $0.213$ & $0.467$ & $0.256$ & $0.502$ & $0.220$ & $0.256$ & $0.263$ & $0.347$ \\
  $\widehat a$ & $3.807$ & $2.492$ & $2.702$ & $2.584$ & $3.237$ & $3.596$ & $3.130$ & $2.981$ \\
  $\widehat b$ & $2.790$ & $2.779$ & $4.442$ & $2.698$ & $3.315$ & $3.880$ & $4.349$ & $3.325$ \\
  $\widehat\sigma_\epsilon$ & $0.314$ & $0.025^{\mathrm b}$ & $0.025^{\mathrm b}$ & $0.033$ & & & & \\
  $\widehat p_\epsilon$ & & & & & $0.184$ & $0.127$ & $0.153$ & $0.108$ \\
  $\widehat\delta_\epsilon$ & & & & & $0.470$ & $0.558$ & $0.367$ & $0.537$ \\
  $\widehat\sigma_{\epsilon,1}$ & & & & & $0.908$ & $0.934$ & $0.850$ & $0.981$ \\
  $\widehat\sigma_{\epsilon,2}$ & & & & & $0.025^{\mathrm b}$ & $0.069$ & $0.029$ & $0.025$ \\
  \addlinespace
  \multicolumn{9}{l}{\textit{Model-specific parameters}} \\
  $\widehat\sigma$ & $0.247$ & & & $0.135$ & $0.129$ & & & $0.080$ \\
  $\widehat c_+$ & & & & $0.137$ & & & & $0.054$ \\
  $\widehat c_-$ & & & & $0.835$ & & & & $0.155$ \\
  $\widehat p_\eta$ & & $0.126$ & & & & $0.119$ & & \\
  $\widehat\delta_\eta$ & & $0.323$ & & & & $0.075$ & & \\
  $\widehat\sigma_{\eta,1}$ & & $0.997$ & & & & $0.427$ & & \\
  $\widehat\sigma_{\eta,2}$ & & $0.128$ & & & & $0.077$ & & \\
  $\widehat\sigma_\eta$ & & & $0.143$ & & & & $0.112$ & \\
  $\widehat p_\zeta$ & & & $0.799$ & & & & $0.944$ & \\
  $\widehat\lambda$ & & & $0.804$ & & & & $1.912$ & \\
  $\widehat\psi$ & & & $0.142$ & & & & $0.519$ & \\
  \midrule
  $d_\theta$ & 6 & 9 & 9 & 8 & 9 & 12 & 12 & 11 \\
  LL & $-22{,}595$ & $-14{,}808$ & $-14{,}610$ & $-14{,}290$ & $-13{,}694$ & $-12{,}850$ & $-13{,}038$ & $-12{,}765$ \\
  AIC & $45{,}203$ & $29{,}634$ & $29{,}237$ & $28{,}597$ & $27{,}406$ & $25{,}724$ & $26{,}100$ & $25{,}551$ \\
  \bottomrule
 \end{tabular}
 \caption*{\footnotesize\textit{Note:} Annual persistence is $\rho=\e^{-\kappa}$.
 OU, GKOS, and the Gaussian step of GMR use the \citeauthor{FarmerToda2017QE}
 chain with $N=321$ nodes on $[-4,4]$; PLV uses the generator with $N=641$
 nodes on $[-8,8]$ and $r=4$. We optimize with $K_\alpha=64$ and report log
 likelihoods (LL) with $K_\alpha=128$; Table~\ref{tab:estimate-accuracy} checks
 their accuracy. $d_\theta$ is the number of free parameters, and AIC is
 $2d_\theta-2\,\mathrm{LL}$. Superscript $\mathrm b$ marks estimates at their
 lower bound, the grid spacing. Lowering OU's $\sigma_{\epsilon,2}$ with
 the other parameters fixed raises LL by less than 2 in the reported
 sensitivity check (Appendix~\ref{subsec:OU-validation}). Blank entries are
 inapplicable.}
\end{table}

With Gaussian transitory shocks, all three alternatives improve on OU
by more than 7,700 log likelihood points. PLV has the highest
likelihood and lowest AIC. Allowing Gaussian-mixture independent transitory shocks improves every
model by at least 1,500 log likelihood points. OU with mixture shocks
even fits better than PLV with Gaussian shocks. We therefore focus the
comparison of persistent dynamics on the mixture specifications.

With mixture transitory shocks, PLV has the highest likelihood and lowest
AIC, followed by GKOS, GMR, and OU. Its log likelihood exceeds those of
GKOS and GMR by about 85 and 270 points, respectively. All three
alternatives improve on OU by more than 650 points. PLV uses one fewer
parameter than GKOS or GMR. Table~\ref{tab:vuong} reports the Vuong tests of
\S\ref{subsec:model-comparison-theory}, using individual log likelihood
differences for the non-nested pairs. With mixture shocks, PLV and GKOS
are both significantly closer to the data than GMR, but the test only
weakly favors PLV over GKOS ($z=1.56$, one-sided $p=0.060$). PLV's
advantage of 85 log points therefore gives a numerical ranking without
clearly separating the two leading specifications.

\begin{table}[!htb]
 \centering
 \caption{Vuong tests between non-nested models.}
 \label{tab:vuong}
 \begin{tabular}{lrrrr}
  \toprule
  & \multicolumn{2}{c}{Gaussian $\epsilon$} & \multicolumn{2}{c}{Gaussian-mixture $\epsilon$} \\
  \cmidrule(lr){2-3}\cmidrule(lr){4-5}
  Comparison & $\ell^A-\ell^B$ & $z$ & $\ell^A-\ell^B$ & $z$ \\
  \midrule
  PLV vs GKOS & $517.4$ & $5.55$ & $85.4$ & $1.56$ \\
  PLV vs GMR & $319.2$ & $2.53$ & $273.3$ & $4.43$ \\
  GKOS vs GMR & $-198.2$ & $-1.48$ & $187.8$ & $4.28$ \\
  \bottomrule
 \end{tabular}
 \caption*{\footnotesize\textit{Note:} For each model pair $A$ vs $B$,
 $\ell^A-\ell^B$ is the difference in log likelihoods at the estimates in
 Table~\ref{tab:model-comparison}, and $z=\sqrt I\bar d/\widehat\sigma_d$
 is the statistic of \S\ref{subsec:model-comparison-theory}, with
 $I=4{,}683$. Positive values favor model $A$. Vuong's correction for the number
 of parameters changes no statistic by more than 0.1. We omit OU, which
 every other model nests.}
\end{table}

The transitory specification also changes the estimated decomposition of
earnings risk. With Gaussian independent transitory shocks, all three
alternatives estimate much less dispersion in $\epsilon$ than OU. When
we allow a transitory mixture and jointly re-estimate the PLV parameters,
its overall transitory standard deviation rises from $0.033$ to $0.363$,
while the persistent state's estimated half-life lengthens from $3.9$ to
$18.0$ years. The fitted mixture combines a tightly concentrated
component with a small component capturing large shocks. The estimated
PLV variance increases on both sides of zero, more steeply at negative
persistent states.

\subsection{Fit diagnostics}
\label{subsec:model-fit-diagnostics}

We compare the data with 100 panels simulated from each specification at
the estimates in Table~\ref{tab:model-comparison}. The panels retain the observed ages
and gaps, start from the estimated Gaussian law at age 30, and evolve
annually through unobserved years. For OU, GKOS, and GMR, we simulate the
exact processes at the estimated parameters. For PLV, we use 100 steps
per year, each an exact Ornstein--Uhlenbeck step with the variance held
at its value at the start of the step. Dynamic statistics follow
\S\ref{subsec:OU-limitations}.

\begin{figure}[!htb]
 \centering
 \includegraphics[width=\textwidth]{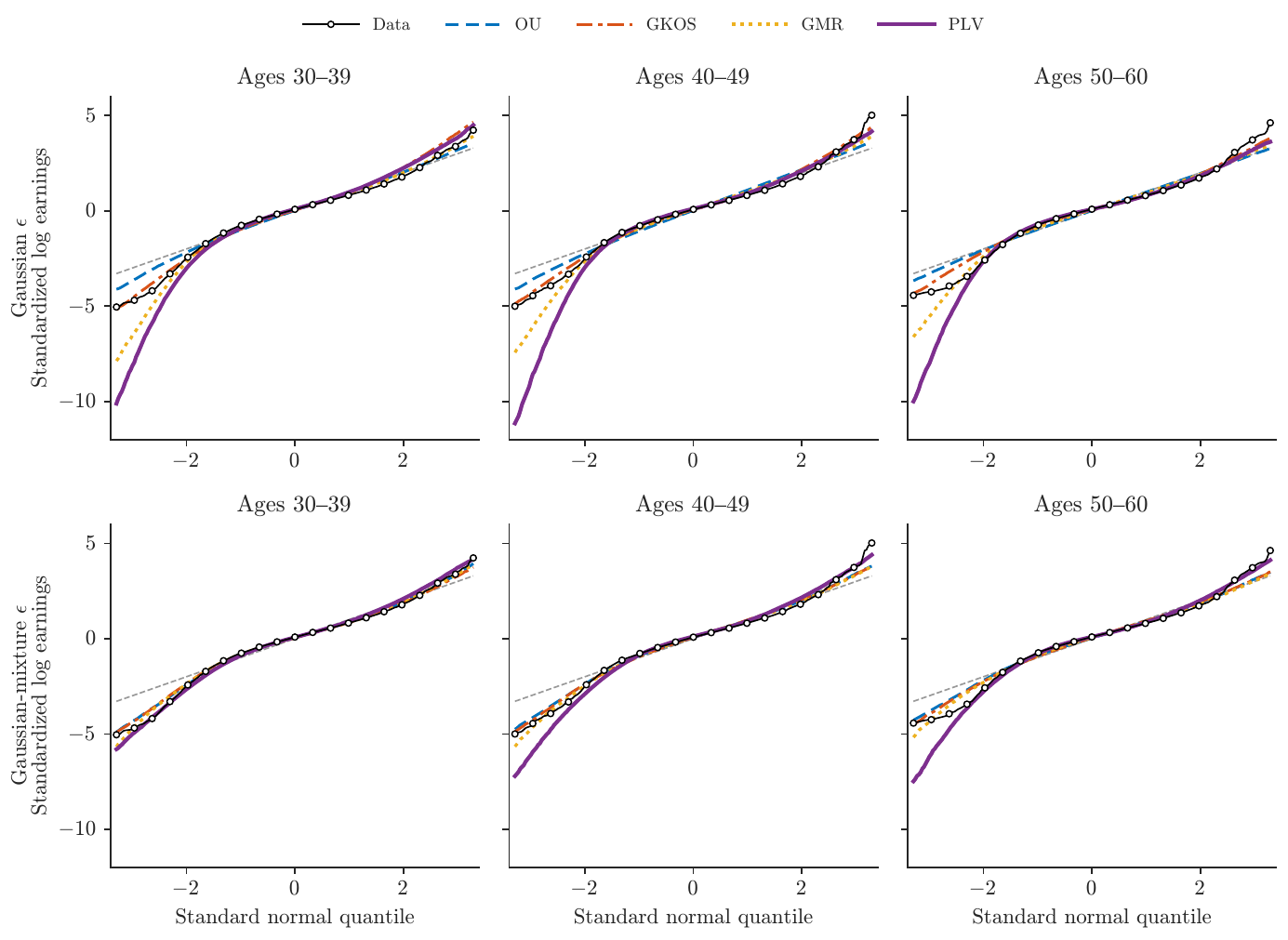}
 \caption{Standardized log-earnings quantiles by model and age.}
 \caption*{\footnotesize\textit{Note:} Columns are age groups; rows use
 Gaussian and Gaussian-mixture independent transitory shocks. Model curves
 are pointwise medians across 100 simulated panels. We standardize empirical
 and model quantiles using each age group's empirical mean and standard
 deviation. Probabilities range from $0.0005$ to $0.9995$. The dashed gray
 line is the Gaussian reference; all panels use common axis limits.}
 \label{fig:model-level-QQ}
\end{figure}

\paragraph{Earnings levels}
Figure~\ref{fig:model-level-QQ} shows that mixture transitory shocks
improve the fit to earnings levels, although tail discrepancies remain.
The central quantiles are broadly similar across models. With Gaussian
transitory shocks, GKOS comes closest to the most extreme lower
quantiles, while GMR and PLV produce too heavy a lower tail. Mixture
shocks reduce these discrepancies, but PLV's lower tail remains too
heavy after age 40. All models understate the most extreme upper
quantiles after age 40, with mixture PLV coming closest.

\paragraph{Heavy-tailed changes}

\begin{figure}[!htb]
 \centering
 \includegraphics[width=\textwidth]{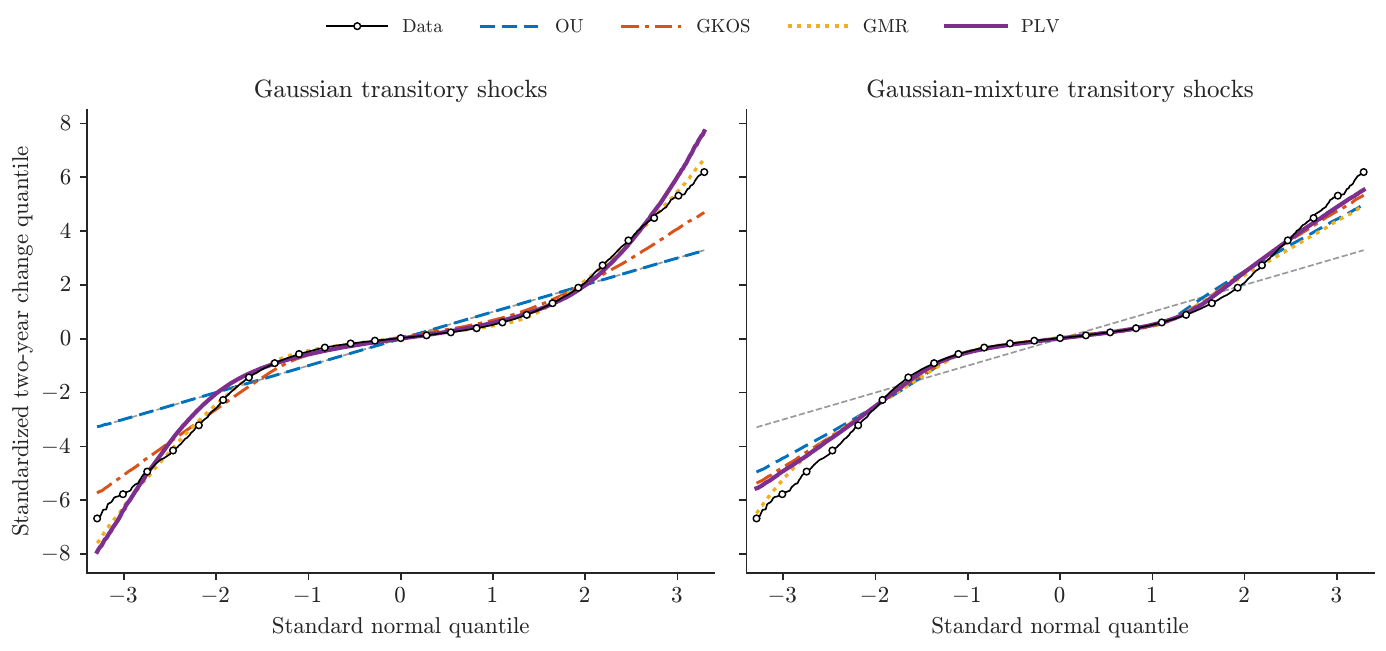}
 \caption{Two-year log-earnings changes under the four models.}
 \caption*{\footnotesize\textit{Note:} The left and right panels use
 Gaussian and Gaussian-mixture independent transitory shocks, respectively.
 We demean and standardize changes by their own standard deviation in
 the data and each simulated panel. Quantile probabilities and simulation
 medians follow Figure~\ref{fig:model-level-QQ}. The dashed
 gray line is the Gaussian reference.}
 \label{fig:model-change-QQ}
\end{figure}

Permanent effects cancel in earnings changes, so
Figure~\ref{fig:model-change-QQ} isolates the persistent and transitory
components. With Gaussian transitory shocks, OU and GKOS produce tails
that are too thin, while GMR and PLV produce tails that are too heavy.
Mixture transitory shocks generate heavy-tailed changes even under OU,
but all four mixture specifications understate the most extreme
standardized changes.

\begin{figure}[!htb]
 \centering
 \includegraphics[width=\textwidth]{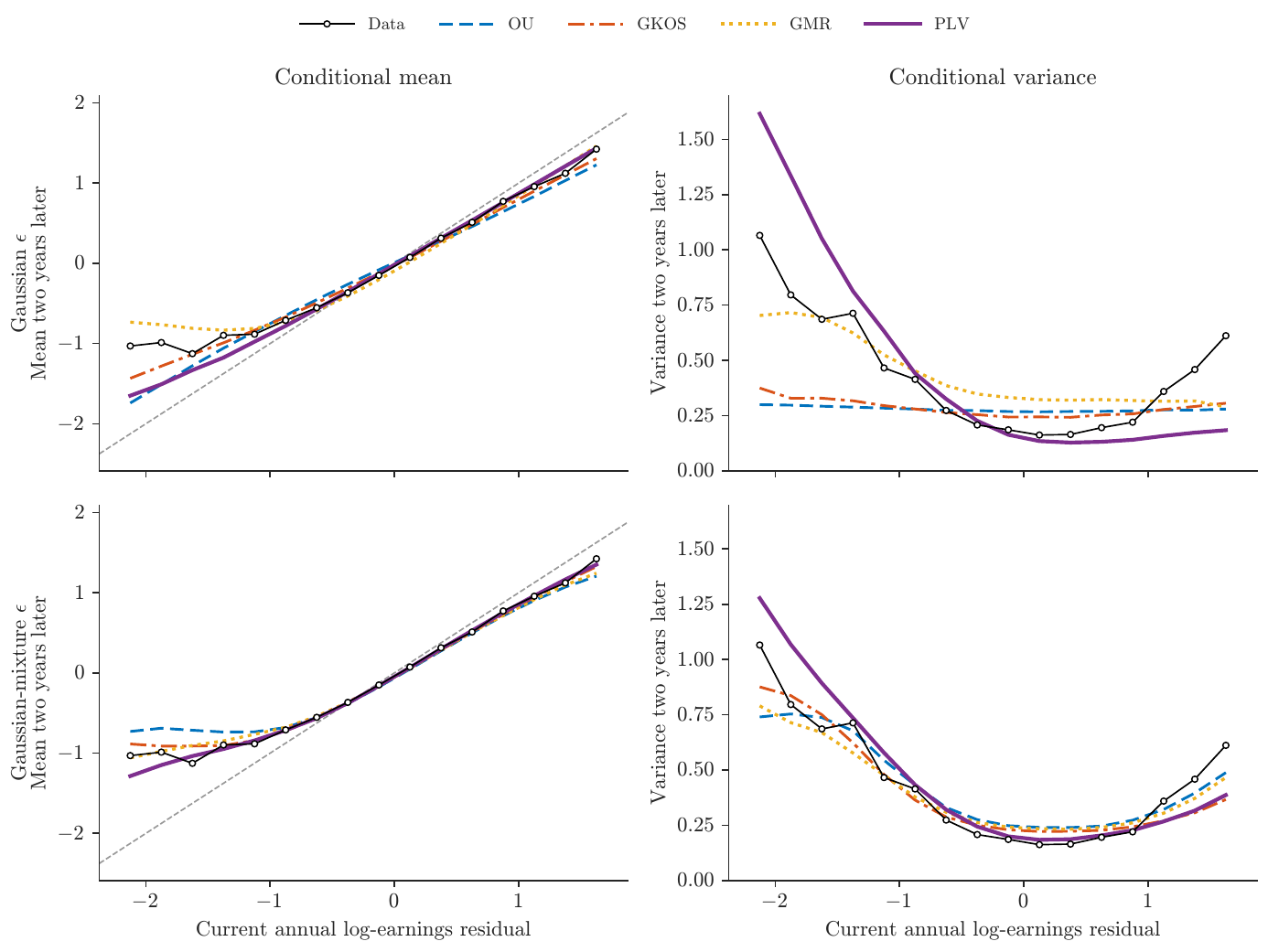}
 \caption{Conditional earnings moments under the four models.}
 \caption*{\footnotesize\textit{Note:} Rows use Gaussian and Gaussian-mixture
 independent transitory shocks; columns, bins, estimators, and empirical
 confidence intervals follow Figure~\ref{fig:benchmark-conditional-moments}.
 Model curves are pointwise medians across 100 simulated panels. Dashed
 lines in the mean panels are 45-degree lines.}
 \label{fig:model-conditional-moments}
\end{figure}

\paragraph{Conditional mean and dispersion}
Figure~\ref{fig:model-conditional-moments} shows that the conditional
variance profile also depends on the transitory specification. With
Gaussian transitory shocks, GMR and PLV generate a pronounced rise at low
earnings. PLV overstates this rise at the lower extreme, and GMR
overstates dispersion at central earnings. No Gaussian specification
generates the rise at high earnings.

With mixture shocks, all four models, including OU, produce a U-shaped
variance profile. Thus a U-shaped profile alone does not identify
state-dependent persistent volatility. Mixture PLV fits the center most
closely but overstates variance at the lower extreme, and all four
models understate variance at the upper end. The conditional means track
the data through the central and upper bins. In the lowest bins, mixture
OU predicts too much recovery and mixture PLV too little.

\paragraph{Asymmetric reversals}
Table~\ref{tab:model-change-diagnostics} shows more frequent reversals
after large losses than gains under the non-Gaussian specifications,
including OU with mixture transitory shocks. Reversal asymmetry therefore
also arises without state-dependent persistent volatility. Among the
mixture specifications, PLV comes closest to the opposite-sign reversal
rates and their gap. All mixture specifications, including PLV,
overpredict half-retracement after losses.

\begin{table}[!htb]
 \centering
 \caption{Two-year earnings changes and subsequent reversals.}
 \label{tab:model-change-diagnostics}
 \begin{tabular}{lrrrrrrrrr}
  \toprule
  & & \multicolumn{4}{c}{Gaussian $\epsilon$} & \multicolumn{4}{c}{Gaussian-mixture $\epsilon$} \\
  \cmidrule(lr){3-6}\cmidrule(lr){7-10}
  & Data & OU & GKOS & GMR & PLV & OU & GKOS & GMR & PLV \\
  \midrule
  \multicolumn{10}{l}{\textit{Two-year changes}} \\
  SD & $0.536$ & $0.555$ & $0.548$ & $0.713$ & $0.569$ & $0.634$ & $0.594$ & $0.597$ & $0.589$ \\
  Skewness & $-0.499$ & $0.001$ & $-0.756$ & $-0.656$ & $-0.237$ & $-0.005$ & $-0.038$ & $-0.555$ & $-0.048$ \\
  Excess kurtosis & $10.82$ & $0.00$ & $5.34$ & $14.92$ & $20.49$ & $5.05$ & $6.36$ & $7.67$ & $7.08$ \\
  \addlinespace
  \multicolumn{10}{l}{\textit{Opposite sign (\%)}} \\
  After a loss & $78.63$ & $73.91$ & $82.56$ & $94.14$ & $67.55$ & $86.57$ & $84.09$ & $86.16$ & $81.11$ \\
  After a gain & $59.62$ & $73.90$ & $54.05$ & $49.60$ & $46.09$ & $62.88$ & $60.52$ & $60.37$ & $59.03$ \\
  Gap (pp) & $19.01$ & $0.01$ & $28.47$ & $44.60$ & $21.47$ & $23.52$ & $23.55$ & $25.75$ & $21.88$ \\
  \addlinespace
  \multicolumn{10}{l}{\textit{At least half retraced (\%)}} \\
  After a loss & $53.54$ & $39.15$ & $40.94$ & $83.58$ & $46.14$ & $72.37$ & $62.34$ & $62.89$ & $64.24$ \\
  After a gain & $25.12$ & $39.49$ & $22.43$ & $9.03$ & $15.55$ & $25.81$ & $21.47$ & $27.31$ & $22.93$ \\
  \bottomrule
 \end{tabular}
 \caption*{\footnotesize\textit{Note:} Model entries are medians across
 100 simulated panels at the estimates in Table~\ref{tab:model-comparison}. Samples and
 definitions follow Table~\ref{tab:benchmark-dynamics}. Gap is the median within-panel loss--gain difference in
 opposite-sign reversal rates, in percentage points.}
\end{table}

\section{Economic consequences of model choice}
\label{sec:saving-application}

We now ask how the choice of earnings process and the method used to
discipline that process change saving and welfare. We use a deliberately
simple saving problem to measure how alternative earnings inputs change
economic predictions while holding preferences, prices, and mean
resources fixed. We first compare the fitted processes and counterfactual
processes with a common log-earnings covariance function. We then
calibrate each model to moments of earnings changes and levels and
compare the economic implications with those of the likelihood estimates. The main
results use Gaussian-mixture transitory shocks. Allowing mixtures
improves the log likelihood by more than 1,500 points in every model
relative to Gaussian transitory shocks (\S\ref{subsec:model-comparison}).

\subsection{Saving under alternative earnings dynamics}
Households maximize $\E\sum_{t=0}^{\infty}\beta^t u(c_t)$ subject to
\begin{equation*}
 a_{t+1}=Ra_t+Y_t-c_t\ge0,\qquad c_t>0,
\end{equation*}
where $u(c)\coloneqq c^{1-\gamma}/(1-\gamma)$. We set
$\beta=0.96$, $\gamma=2$, and $R=1.02$. Households observe their
persistent state and current income. To isolate risk over time, we set
permanent effects to zero and use stationary earnings states. We treat the transitory shock $\epsilon$ as earnings risk in every
model.\footnote{Part of the transitory shock in survey data is likely
reporting error \citep{BoundBrownDuncanRodgers1994}, which affects measured
but not actual earnings. Treating the entire shock as earnings risk
therefore overstates transitory income risk.} We normalize mean raw
earnings to $\E y_t=1$ and set
total income to $Y_t=y_t+b$, with $b=0.1$. The transfer is thus
$10\%$ of mean raw earnings.\footnote{For comparison, the 2022 federal
poverty guideline for a one-person household in the contiguous states
and District of Columbia was \$13,590 \citep{HHS2022PovertyGuidelines},
about $15\%$ of mean annual earnings of \$92,832 in our 2022 PSID sample.}
This common income floor also ensures finite optimal value despite the
heavy lower earnings tails with PLV.

We measure precautionary wealth by stationary mean assets. For welfare,
we average lifetime value from zero assets over each model's stationary
earnings law. Let $\cW_m$ denote this expected value under earnings
process $m$. Its constant consumption equivalent (CE), the constant
consumption level that delivers $\cW_m$, is
\begin{equation}
 \mathrm{CE}_m\coloneqq
 [(1-\gamma)(1-\beta)\cW_m]^{1/(1-\gamma)}.
 \label{eq:saving-welfare}
\end{equation}

The fitted processes use the likelihood estimates in
Table~\ref{tab:model-comparison}. To examine what covariance targets
leave undetermined, we retain the fitted OU process and adjust the other
three models to match its stationary annual log-earnings covariance
function:
\begin{equation}
 \operatorname{Cov}(\log y_t,\log y_{t+h})
 =V\rho^h+T\mathbf{1}\set{h=0},\qquad h=0,1,\dotsc,
 \label{eq:saving-covariance-target}
\end{equation}
where $\rho$, $V$, and $T$ are the fitted OU persistence, persistent
variance, and transitory variance. Appendix~\ref{subsec:saving-discretization}
describes how we adjust persistence and rescale the persistent and
transitory components, including which standardized distributions we
preserve. Table~\ref{tab:saving-application} reports mean assets and
CE under the fitted and covariance-matched processes.

\begin{table}[!htb]
 \centering
 \caption{Precautionary wealth and welfare under alternative earnings dynamics.}
 \label{tab:saving-application}
 \begin{tabular}{lrrrr}
   \toprule
   & OU & GKOS & GMR & PLV \\
   \midrule
   Mean assets, fitted & $4.805$ & $7.374$ & $5.592$ & $10.508$ \\
   Mean assets, covariance-matched & $4.805$ & $4.636$ & $4.313$ & $4.193$ \\
   \addlinespace
   CE, fitted & $0.793$ & $0.746$ & $0.802$ & $0.739$ \\
   CE, covariance-matched & $0.793$ & $0.796$ & $0.805$ & $0.832$ \\
   \bottomrule
  \end{tabular}
 \caption*{\footnotesize\textit{Note:} Gaussian-mixture transitory shocks,
 estimates in Table~\ref{tab:model-comparison}. Mean raw earnings
 equal one, and the transfer is $b=0.1$, so the no-risk CE
 benchmark is $1.1$. Matched processes share the
 fitted OU covariance function \eqref{eq:saving-covariance-target}. CE is
 defined in \eqref{eq:saving-welfare}.}
\end{table}

Relative to OU, fitted GKOS and GMR raise mean assets by 53 and 16
percent, and fitted PLV more than doubles them. GKOS and PLV lower CE by 6--7 percent,
whereas GMR raises it by 1 percent. The fitted comparison reflects
differences in second moments and distributional shape: fitted PLV has a
larger persistent variance than OU (0.68 versus 0.40) and a heavier upper
tail in earnings levels (Appendix~\ref{sec:PLV-properties}).

Matching mean raw earnings and the entire stationary log-earnings
covariance function narrows the range of mean assets and CE. Yet all
alternatives imply less wealth and higher CE than OU. Matched GMR and
PLV lower mean assets by 10 and 13 percent, and matched PLV raises CE by 5
percent. PLV moves from the
highest wealth among the fitted processes to the lowest among the
covariance-matched processes.

\begin{figure}[!htb]
 \centering
 \includegraphics[width=\textwidth]{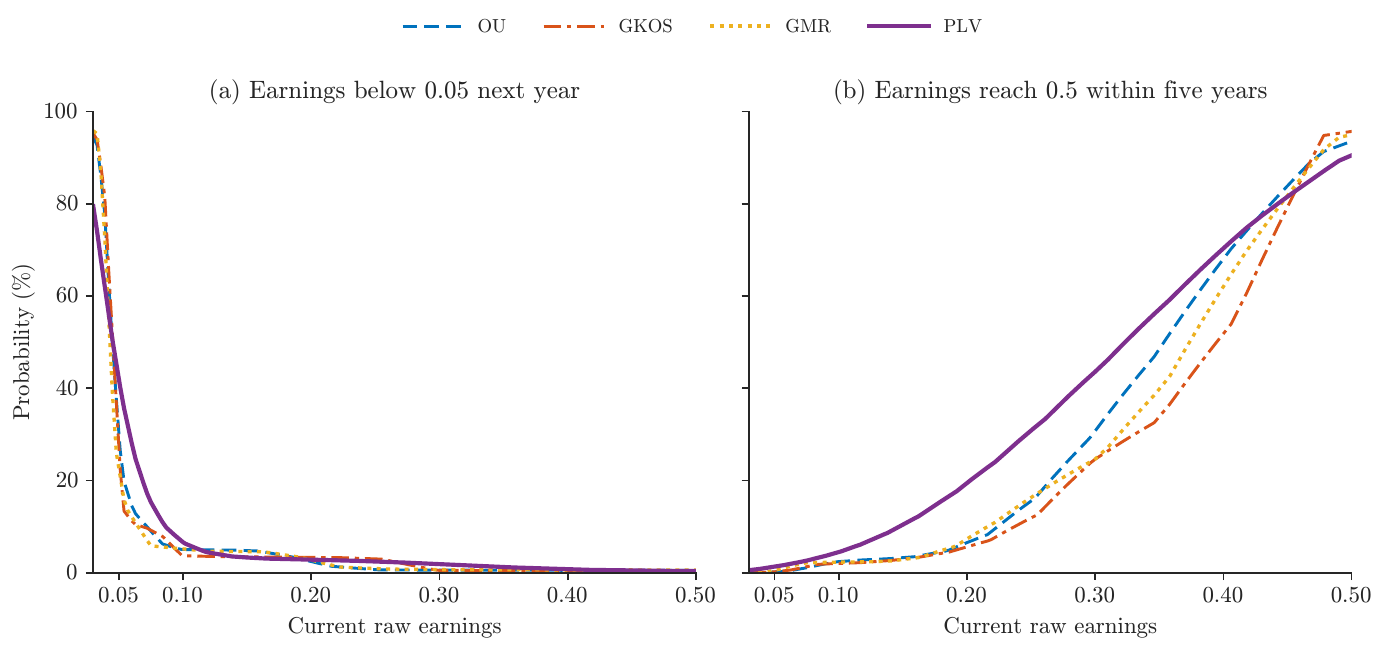}
 \caption{Conditional earnings risk under covariance-matched dynamics.}
 \caption*{\footnotesize\textit{Note:} Gaussian-mixture transitory shocks.
 All processes share the fitted OU log-earnings covariance function and
 mean raw earnings of one. The left panel shows the probability of raw
 earnings below $0.05$ next year. The right panel shows the probability of
 raw earnings of at least $0.5$ at one of the next five annual observations.
 We set current centered transitory shocks, including the shared GMR loss,
 to zero and integrate future shocks jointly with the persistent transition.}
 \label{fig:saving-recovery}
\end{figure}

Figure~\ref{fig:saving-recovery} shows what the common covariances leave
undetermined about downside risk and recovery prospects. From current earnings of $0.05$, the probability of earnings below
$0.05$ next year is 40 percent under PLV and 27 percent under OU. From
current earnings of $0.2$, the probability of earnings of at least $0.5$
within five years is 20 percent under PLV and 7 percent under OU. The
full earnings transition law and optimal saving policies determine mean
assets and CE. The fitted processes and
saving-policy profiles appear in Figure~\ref{fig:saving-recovery-policy}
in Appendix~\ref{subsec:saving-discretization}.

The transitory specification also affects the fitted comparison. After
re-estimating each model with Gaussian transitory shocks, fitted GKOS,
GMR, and PLV imply 44, 16, and 35 percent
\emph{less} wealth than OU (Table~\ref{tab:saving-gaussian} in
Appendix~\ref{subsec:saving-discretization}).

\subsection{Calibrating to moments of earnings changes and levels}
\label{subsec:moment-selection}

The covariance-matched results show what second moments leave
undetermined. We next ask whether targeting higher moments, tail
statistics, conditional moments, and moments of levels resolves these
economic differences. Recent calibrations use such targets
\citep{GuvenenKarahanOzkanSong2021,GuvenenMcKayRyan2026}. We follow their
designs as closely as our biennial panel allows and calibrate each model,
with Gaussian-mixture transitory shocks, to 70 moments of the earnings
residuals in \eqref{eq:PSID-residualization}.
Table~\ref{tab:calibration-moments} groups the targets into eight sets.
Sets~1--2 contain the standard deviation, skewness, and excess kurtosis of
two- and six-year changes by age and recent earnings (RE), as in
\citet{GuvenenKarahanOzkanSong2021}. We use three age groups (30--39,
40--49, and 50--60) and measure recent earnings by the tercile of the
residual within its age group. Sets~3--4 contain mean subsequent changes
after extreme two-year changes, which approximate the short- and
long-horizon impulse responses in \citet{GuvenenKarahanOzkanSong2021}.
Sets~5--6 contain the 10th and 90th percentiles and the tail
probabilities of changes, as in \citet{GuvenenMcKayRyan2026}. Sets~7--8
contain moments of earnings levels by age: the variance and
autocovariances of log earnings, which both papers target in the form of
the variance over the life cycle, and the 10th and 90th percentiles.

Because the targets include levels, we calibrate the initial dispersion
$\sigma_0$ and the permanent-effect rates $(a,b)$ together with the
dynamic and transitory parameters. We minimize a weighted sum of squared
relative discrepancies, giving equal weight to each set and equal weight
within sets. We fixed the targets and weights before computing any
results. Appendix~\ref{subsec:calibration-exercise} describes the moment
calculations, parameter restrictions, and search, and lists all 70
targets. Appendix~\ref{subsec:calibration-changes} reports an earlier
calibration to 18 moments of earnings changes only, which leads to the
same conclusions.

\begin{table}[!htb]
 \centering
 \caption{Calibration targets, fit, and likelihoods.}
 \label{tab:calibration-moments}
 \begin{tabular}{cp{0.40\textwidth}rrrrr}
  \toprule
  Set & Targets & Number & OU & GKOS & GMR & PLV \\
  \midrule
  1 & Two-year changes by age $\times$ RE: SD, skewness, kurtosis & 27 & $0.284$ & $0.247$ & $0.259$ & $0.262$ \\
  2 & Six-year changes by RE: SD, skewness, kurtosis & 9 & $0.326$ & $0.207$ & $0.198$ & $0.182$ \\
  3 & Mean change 2, 4 years after extreme changes & 4 & $0.230$ & $0.167$ & $0.144$ & $0.127$ \\
  4 & Mean change 6, 10 years after extreme changes & 4 & $0.157$ & $0.123$ & $0.090$ & $0.094$ \\
  5 & P10, P90 of 2-, 4-, 6-year changes & 6 & $0.059$ & $0.066$ & $0.057$ & $0.171$ \\
  6 & Shares of two-year changes beyond $\pm1.2$ & 2 & $0.156$ & $0.069$ & $0.094$ & $0.152$ \\
  7 & Variance, autocovariances of levels by age & 12 & $0.156$ & $0.084$ & $0.094$ & $0.194$ \\
  8 & P10, P90 of levels by age & 6 & $0.090$ & $0.044$ & $0.047$ & $0.067$ \\
  \midrule
  \multicolumn{3}{l}{Objective, likelihood estimate} & $0.187$ & $0.114$ & $0.148$ & $0.149$ \\
  \multicolumn{3}{l}{Objective, calibrated} & $0.0406$ & $0.0206$ & $0.0197$ & $0.0276$ \\
  \addlinespace
  \multicolumn{3}{l}{LL, likelihood estimate} & $-13{,}694$ & $-12{,}850$ & $-13{,}038$ & $-12{,}765$ \\
  \multicolumn{3}{l}{LL, calibrated} & $-15{,}104$ & $-14{,}209$ & $-13{,}877$ & $-13{,}956$ \\
  \bottomrule
 \end{tabular}
 \caption*{\footnotesize\textit{Note:} Gaussian-mixture transitory shocks.
 For each set, entries are root mean squared arc-percent deviations
 $2(m_k-d_k)/(\abs{m_k}+\abs{d_k})$ of the calibrated model moments $m_k$
 from the data moments $d_k$. Model moments use 200 simulated panels with
 draws independent of the search. The objective is the weighted sum of
 squared arc-percent deviations; a miss of one bootstrap standard error
 in every target gives about 0.0092. LL is the log likelihood, computed
 as in Table~\ref{tab:model-comparison}.}
\end{table}

Table~\ref{tab:calibration-moments} shows that calibration lowers the
moment objective in every model, by 78 to 87 percent relative to the
likelihood estimates. GMR and GKOS have the lowest calibrated objective
values. All four calibrations closely match the percentiles of changes
(set~5) and of levels (set~8). GKOS and GMR also come closer than OU
and PLV to the tail probabilities of changes and to the variance and
autocovariances of levels (sets~6--7), although OU, GKOS, and GMR
understate the variance of log earnings at ages 50--60 (0.48 to 0.55,
compared with 0.68). The largest discrepancies occur in the shape of
two-year changes within age and recent-earnings cells (set~1). For
example, changes after low recent earnings are positively skewed in the
data, but every calibration overstates this skewness at ages 50--60
(0.61 to 0.90, compared with 0.28). Every calibration also overstates
the mean recovery over the next two years after changes in the bottom
decile (0.49 to 0.54, compared with 0.40). Thus no calibration matches all
targets within sampling error: the lowest objectives are about twice the
objective of a miss of one standard error in every target.

\begin{samepage}
The close fit to many targets does not extend to earnings histories.
The bottom rows of Table~\ref{tab:calibration-moments} show that
calibration lowers the log likelihood by 839 to 1,410 points relative to
the likelihood estimates. All four calibrations have log likelihoods
below every likelihood-estimated specification with mixture shocks in
Table~\ref{tab:model-comparison}, including OU, whose persistent dynamics
are Gaussian.

\end{samepage}

We next ask how calibration changes saving and welfare, and whether
calibrations with similar target fit imply similar economic outcomes.
Table~\ref{tab:calibration} compares economic
outcomes at the fitted and calibrated parameters. We also report
\emph{near-optimal} candidates, whose objectives are less than 0.0092
above the reported objective. The tolerance approximates the objective
for a miss of one bootstrap standard error in every target.

\begin{table}[!htb]
 \centering
 \caption{Saving and welfare under likelihood estimation and calibration.}
 \label{tab:calibration}
 \begin{tabular}{lrrrr}
  \toprule
  & OU & GKOS & GMR & PLV \\
  \midrule
  Mean assets, fitted & $4.80$ & $7.37$ & $5.59$ & $10.51$ \\
  Mean assets, calibrated & $1.63$ & $4.69$ & $4.98$ & $3.76$ \\
  Mean assets, near-optimal & 1.4--3.6 & 2.2--8.2 & 2.9--7.2 & 3.1--9.3 \\
  \addlinespace
  CE, fitted & $0.793$ & $0.746$ & $0.802$ & $0.739$ \\
  CE, calibrated & $1.029$ & $0.893$ & $0.854$ & $0.677$ \\
  CE, near-optimal & 0.92--1.04 & 0.70--1.00 & 0.66--0.97 & 0.67--0.77 \\
  \bottomrule
 \end{tabular}
 \caption*{\footnotesize\textit{Note:} Gaussian-mixture transitory shocks.
 Mean assets and CE follow Table~\ref{tab:saving-application}.  Near-optimal ranges
 cover the reported calibration and evaluated candidates with objectives
 less than 0.0092 above it
 (Tables~\ref{tab:calibration-alternatives-ou}--\ref{tab:calibration-alternatives-plv}).}
\end{table}

Table~\ref{tab:calibration} shows that calibration changes the economic
answer. Every calibration implies less wealth than the corresponding
likelihood estimate: mean assets fall from 4.8 to 1.6 for OU, from 7.4
to 4.7 for GKOS, from 5.6 to 5.0 for GMR, and from 10.5 to 3.8 for PLV.
CE rises for OU, GKOS, and GMR and falls for PLV.

The targets also leave room for different economic answers within each
model. Among the near-optimal candidates, mean assets range from 1.4 to
3.6 for OU, from 2.2 to 8.2 for GKOS, from 2.9 to 7.2 for GMR, and from
3.1 to 9.3 for PLV
(Tables~\ref{tab:calibration-alternatives-ou}--\ref{tab:calibration-alternatives-plv}).
Thus calibrations that fit the targets about equally well imply mean
assets that differ by a factor of 2.5 to 3.7 within each model. These
ranges describe evaluated candidates. They are neither confidence
intervals nor exhaustive bounds over calibrations satisfying the
tolerance.

Across the benchmark likelihood estimates, covariance-matched processes,
and reported calibrations with Gaussian-mixture transitory shocks, the
four models imply mean assets between 1.6 and 10.5
and CE between 0.68 and 1.03, although
every process is disciplined by the same data and preferences, prices,
mean earnings, and transfers are common. The ranking of models also
depends on the method: PLV implies the most wealth among the fitted
processes and the least among the covariance-matched processes, whereas
GMR implies the most among the calibrated processes and OU the least.

\section{Conclusion}
\label{sec:conclusion}

Quantitative macroeconomics usually disciplines models by calibration to
selected moments. Calibration is transparent, but it uses only part of
the information in the data. We showed that the HMM approach makes
likelihood estimation feasible for nonlinear, non-Gaussian latent
processes with permanent heterogeneity and irregular observation dates,
and that the likelihood leads to different economic answers.

Our application yields three results. First, the likelihood ranks four
earnings processes that can share the same covariance function: PLV and
GKOS fit the data better than GMR and OU. Second, close target fit does
not extend to complete earnings histories. Calibrations to 70 moments of
earnings changes and levels, in the style of \citet{GuvenenKarahanOzkanSong2021,GuvenenMcKayRyan2026}, fit the
histories worse than every likelihood-estimated specification with
mixture transitory shocks, including OU, whose persistent dynamics are
Gaussian. Third, the method of choosing parameters changes the economic
answer. In the same saving model, every calibration implies less wealth
than the corresponding likelihood estimate, and calibrations of one
process that fit the targets about equally well imply mean assets that
differ by a factor of 2.5 to 3.7. Even the direction depends on the
targets: when we target changes only, calibrated GKOS and GMR imply more
than twice the wealth of their likelihood estimates.

The likelihood is not a substitute for economic judgment. Every model we
consider is misspecified, and the likelihood-preferred specifications
still miss features of the data, such as the extreme kurtosis of
earnings changes. However, the likelihood provides a common scale on
which to compare misspecified models. We therefore recommend that
quantitative studies compare candidate models by their likelihood before
using them for counterfactual and policy analysis, and use targeted
moments to diagnose where the selected model fails.

\appendix

\section{Properties of the PLV process}
\label{sec:PLV-properties}

We first derive the PLV invariant law, then establish moment and symmetry
results for broader diffusion classes and their implications for
identification. We suppress the individual index and use
\eqref{eq:models-observation} unless stated otherwise. Stationary statements
refer to initialization at the invariant law.

\subsection{Stationary distribution and tails}

Following \citet{BibbySkovgaardSorensen2005}, linear mean reversion
can coexist with a non-Gaussian invariant law.

\begin{prop}[Stationary distribution]
\label{prop:PLV-stationary-distribution}
Suppose $\kappa,\sigma>0$ and $c_{+},c_{-}\ge 0$. For $c>0$ and $y\ge 0$, define
\begin{equation}
 r_c\coloneqq\frac{2\kappa\sigma^2}{c^2},
\qquad \lambda_c\coloneqq\frac{2\kappa}{c},
\qquad g_c(y)\coloneqq
 \left(1+\frac{cy}{\sigma^2}\right)^{r_c-1}
 \e^{-\lambda_c y},
 \label{eq:PLV-tail-parameters}
\end{equation}
and set $g_0(y)\coloneqq\e^{-\kappa y^2/\sigma^2}$. Then the diffusion
\eqref{eq:PLV-diffusion} has a unique nonexplosive strong solution and
is positive recurrent with the unique invariant density $\pi$
\begin{equation}
 \pi(x)=C
 \begin{cases}
  g_{c_{+}}(x),&x\ge 0,\\
  g_{c_{-}}(-x),&x<0,
 \end{cases}
 \qquad
 C^{-1}\coloneqq \int_0^\infty
       [g_{c_{+}}(y)+g_{c_{-}}(y)]\diff y.
 \label{eq:PLV-stationary-density}
\end{equation}
All polynomial moments of $\pi$ are finite, and its mean and unique mode are zero.
\end{prop}

\begin{proof}
Let $L\coloneqq c_{+}+c_{-}$. The triangle inequality and $\abs{x^{\pm}-y^{\pm}}\le \abs{x-y}$ imply
\begin{align*}
 \abs{v(x)-v(y)}&=\abs{c_+(x^+-y^+)+c_-(x^--y^-)}\\
 &\le c_{+}\abs{x^{+}-y^{+}}+c_{-}\abs{x^{-}-y^{-}}\le L\abs{x-y}.
\end{align*}
Since $v(x)\ge\sigma^2>0$, the derivative of $r\mapsto\sqrt r$ is bounded
by $1/(2\sigma)$ on $[\sigma^2,\infty)$. The mean value theorem therefore
implies
\begin{equation*}
 \abs{\sqrt{v(x)}-\sqrt{v(y)}}
 \le\frac{1}{2\sigma}\abs{v(x)-v(y)}
 \le\frac{L}{2\sigma}\abs{x-y}.
\end{equation*}
Hence both the drift and diffusion coefficients are globally Lipschitz.
They also have at most linear growth, so existence and
uniqueness follow from \citet[Ch.~15, Theorem~16.5]{KarlinTaylor1981}.
The stationary-density formula in
\citet[Ch.~15, equation~(5.34)]{KarlinTaylor1981} gives
\begin{equation}
 \pi(x)\ \propto\ \frac{1}{v(x)}
 \exp\left(-2\kappa\int_0^x\frac{y}{v(y)}\diff y\right).
 \label{eq:PLV-stationary-density-formula}
\end{equation}
For $c,z\ge 0$, we have
\begin{equation}
 \int_0^z\frac{y}{\sigma^2+cy}\diff y
 =\begin{cases*}
\frac{z^2}{2\sigma^2} & if $c=0$,\\
\frac{z}{c}-\frac{\sigma^2}{c^2}
     \log\left(1+\frac{cz}{\sigma^2}\right) & if $c>0$.
\end{cases*}
\label{eq:PLV-scale-integral}
\end{equation}
Evaluating \eqref{eq:PLV-stationary-density-formula} on each half-line
therefore gives \eqref{eq:PLV-stationary-density}. The two expressions
have the same multiplicative constant because $v$ is continuous at zero.

The density is integrable. We define the scale density by
\begin{equation}
 s(x)\coloneqq\exp\left(2\kappa\int_0^x\frac{y}{v(y)}\diff y\right) \ge1. \label{eq:PLV-scale-density}
\end{equation}
The scale function $S(x)\coloneqq\int_0^x s(y)\diff y$ therefore diverges
to $-\infty$ and $+\infty$ at the respective boundaries. Together with
the finite speed measure, whose density is proportional to $\pi$,
this establishes positive recurrence and
uniqueness of the invariant distribution by the scale and speed criteria
of \citet[Ch.~15, Sections~6--7]{KarlinTaylor1981}.
Its exponential or Gaussian tails imply that all polynomial
moments exist. Integrating $(v\pi)'=-2\kappa x\pi$ over $\R$, with
$v\pi\to0$ at both infinities, gives $\E_\pi[X]=0$.
Finally, for $c>0$,
\begin{equation*}
 \frac{\diff}{\diff z}\log g_c(z)
 =-\frac{c+2\kappa z}{\sigma^2+cz}<0,\qquad z>0.
\end{equation*}
The Gaussian expression is also strictly decreasing for $z>0$.
Thus the unique mode is zero.
\end{proof}

With positive slopes, \eqref{eq:PLV-stationary-density} gives
\begin{equation}
\pi(x)\sim C_{+}x^{r_{c_{+}}-1}\e^{-\lambda_{c_{+}}x},\qquad
\pi(-x)\sim C_{-}x^{r_{c_{-}}-1}\e^{-\lambda_{c_{-}}x}
 \label{eq:PLV-stationary-tails}
\end{equation}
as $x\to\infty$ for positive constants $C_{+},C_{-}$. A larger slope gives
a heavier tail, holding $\kappa$ fixed; $\sigma^2$ affects the center and
power factors but not the exponential decay rates. A zero slope gives a
Gaussian tail on that side; when both vanish, the invariant law is
$\cN(0,\sigma^2/(2\kappa))$. Integrating these tails shows that $\exp(X)$ has upper and lower
Pareto-type tail exponents $\lambda_{c_{+}}$ and $\lambda_{c_{-}}$, with
slowly varying logarithmic factors. This connects the diffusion
to the earnings patterns in \S\ref{subsec:permanent-heterogeneity}
and \citet{Toda2012JEBO}.

Finite polynomial moments of $X$ do not guarantee a finite mean of
$\exp(X)$. For $c_{+}>0$, \eqref{eq:PLV-stationary-tails} gives
$\E_\pi[\exp(X)]<\infty$ if and only if
$\lambda_{c_{+}}=2\kappa/c_{+}>1$; for $c_{+}=0$, this mean is finite.

Adding the independent permanent effect in
\eqref{eq:normal-laplace-effect} selects the slower exponential
rate on each side, so $\exp(u)$ has tail exponents
$\min\{a,\lambda_{c_{+}}\}$ and $\min\{b,\lambda_{c_{-}}\}$, with
$\lambda_0\coloneqq\infty$ for a Gaussian side. Equal rates can change
the logarithmic factors but not the exponents; Gaussian and finite
Gaussian-mixture components leave the exponents unchanged. Tail exponents
alone therefore cannot distinguish permanent heterogeneity from
persistent shocks, whereas earnings changes eliminate the permanent effect.

\subsection{Moments and identification}
\label{subsec:diffusion-moments}

The conditional identities below hold for general variance functions $v$
with linear mean-reverting drift $-\kappa x$, without stationarity.
We state the covariance and short-horizon kurtosis results under
stationarity and impose explicit assumptions on invariant moments. These
moment assumptions do not follow from Lipschitz continuity alone.

\begin{prop}[Persistence and earnings risk]
\label{prop:linear-drift-moments}
Consider \eqref{eq:PLV-diffusion} with $\kappa>0$ and any
$v:\R\to(0,\infty)$ whose square root is globally Lipschitz, and suppose
$\E[X(0)^2]<\infty$.
For $h>0$, the conditional mean and variance satisfy
\begin{subequations}
\begin{align}
 \E[X(t+h)\mid X(t)=x]&=\e^{-\kappa h} x,
 \label{eq:linear-drift-conditional-mean} \\
\var(X(t+h)\mid X(t)=x)&=\int_0^h\e^{-2\kappa(h-s)}
       \E[v(X(t+s))\mid X(t)=x]\diff s.
 \label{eq:linear-drift-conditional-variance}
\end{align}
\end{subequations}
For each fixed $x$, this conditional variance is $hv(x)+o(h)$ as
$h\downarrow0$. If we initialize $X$ at an invariant law $\pi$ with finite
second moment, write $V_\pi\coloneqq\var_\pi(X)$. Observed residuals under
\eqref{eq:models-observation} at distinct dates satisfy
\begin{equation}
 \operatorname{Cov}(u(t),u(t+h))
 =\var(\alpha_i)+\e^{-\kappa h} V_\pi.
 \label{eq:linear-drift-stationary-covariance}
\end{equation}
Moreover, if $\E_\pi[X^4]<\infty$, stationary latent changes satisfy
\begin{equation}
 \lim_{h\downarrow0}
 \frac{\E_\pi[(X(t+h)-X(t))^4]}
      {\E_\pi[(X(t+h)-X(t))^2]^2}-3
 =\frac{3\var_\pi(v(X))}{\E_\pi[v(X)]^2}.
 \label{eq:diffusion-short-horizon-kurtosis}
\end{equation}
The limit is strictly positive if $v(X)$ is nonconstant under $\pi$, and
zero otherwise.
\end{prop}

\begin{proof}
Applying It\^o's formula
\citep[Ch.~15, Theorem~16.4]{KarlinTaylor1981} to $\e^{\kappa t}X(t)$
gives
\begin{equation}
 X(t+h)=\e^{-\kappa h}X(t)+
 \int_t^{t+h}\e^{-\kappa(t+h-s)}\sqrt{v(X(s))}\diff B(s).
 \label{eq:linear-drift-integrating-factor}
\end{equation}
The Lipschitz and initial-moment assumptions ensure finite second moments.
The stochastic integral has conditional mean zero by
\citet[Ch.~15, Theorem~16.1]{KarlinTaylor1981}, giving
\eqref{eq:linear-drift-conditional-mean}. The It\^o isometry in the same theorem gives
\eqref{eq:linear-drift-conditional-variance}; continuity and the moment bounds
imply $\E[v(X(t+s))\mid X(t)=x]\to v(x)$ as $s\downarrow0$, giving
the stated short-horizon expansion. At stationarity,
\eqref{eq:linear-drift-conditional-mean} implies $\E_\pi[X]=0$, so
\begin{equation*}
 \operatorname{Cov}(X(t),X(t+h))
 =\E_\pi[X(t)\E[X(t+h)\mid X(t)]]
 =\e^{-\kappa h} V_\pi.
\end{equation*}
Independence of the permanent effect and the transitory shocks then gives
\eqref{eq:linear-drift-stationary-covariance}.

For the kurtosis result, work under stationarity, use the norm
$\norm{Y}_4\coloneqq(\E_\pi\abs{Y}^4)^{1/4}$, and write
\begin{equation*}
 \frac{X(h)-X(0)}{\sqrt h}
 =\sqrt{v(X(0))}\,Z_h+\frac{R_h}{\sqrt h},
 \qquad Z_h\coloneqq\frac{B(h)-B(0)}{\sqrt h}.
\end{equation*}
Since $Z_h\sim\cN(0,1)$ is independent of $X(0)$, the leading term is a
Gaussian variance mixture. To justify the approximation, define
\begin{equation*}
 R_h=-\kappa\int_0^h X(s)\diff s+\int_0^h H_s\diff B(s),
 \qquad H_s\coloneqq\sqrt{v(X(s))}-\sqrt{v(X(0))}.
\end{equation*}
The standard increment bound
\citep[Theorem~4.5.4]{KloedenPlaten1992} gives
$\E_\pi\abs{X(s)-X(0)}^4=O(s^2)$. Lipschitz continuity of $\sqrt v$
then gives $\E_\pi\abs{H_s}^4=O(s^2)$. The
Burkholder--Davis--Gundy inequality
\citep[Ch.~3, Theorem~3.28]{KaratzasShreve1991} and Cauchy--Schwarz imply
\begin{equation*}
 \E_\pi\abs*{\int_0^h H_s\diff B(s)}^4
 \le C\E_\pi\left(\int_0^h H_s^2\diff s\right)^2
 \le Ch\int_0^h\E_\pi\abs{H_s}^4\diff s=O(h^4).
\end{equation*}
The drift integral also has $L^4$ norm $O(h)$, so
$\norm{R_h/\sqrt h}_4=O(\sqrt h)\to0$. H\"older's inequality therefore
implies that the normalized increments' second and fourth moments
approach those of the variance mixture:
\begin{align*}
 \lim_{h\downarrow0}\frac{\E_\pi[(X(h)-X(0))^2]}{h}
 &=\E_\pi[v(X)],\\
 \lim_{h\downarrow0}\frac{\E_\pi[(X(h)-X(0))^4]}{h^2}
 &=3\E_\pi[v(X)^2].
\end{align*}
The stationary change has mean zero. Taking the ratio and subtracting
three gives \eqref{eq:diffusion-short-horizon-kurtosis}, since
$\E_\pi[v(X)^2]-\E_\pi[v(X)]^2=\var_\pi(v(X))$.
The limit is positive exactly when $v(X)$ is nonconstant under $\pi$.
\end{proof}

For PLV, Proposition~\ref{prop:PLV-stationary-distribution} establishes
the regularity conditions and finite stationary moments required above.
Since its invariant density is strictly positive on $\R$, the limiting
excess kurtosis is positive whenever at least one variance slope is positive.

Equation~\eqref{eq:linear-drift-stationary-covariance} implies that every
stationary diffusion in this class has the same earnings covariance
function as an OU process with the same $\kappa$, $V_\pi$, permanent-effect
variance, and transitory-shock variance
\citep[Theorem~2.1]{BibbySkovgaardSorensen2005}. At annual dates, the GKOS
AR(1) has the same form, with stationary state variance
$V_X=\var(\eta_t)/(1-\rho^2)$.

For the stationary GMR process, let $S_c\coloneqq\var(c_t)$ and
$\tau_\epsilon^2\coloneqq\var(\epsilon)$, allowing either transitory-shock
specification. The state equation gives
\begin{equation*}
 V_X=\frac{\sigma_\eta^2+\psi^2S_c}{1-\rho^2},
 \qquad \operatorname{Cov}(X_t,c_t)=\psi S_c.
\end{equation*}
Future losses are independent of $u_t$, and the AR(1) recursion gives
$\operatorname{Cov}(u_t,X_{t+h})=\rho^h\operatorname{Cov}(u_t,X_t)$.
Thus, for integer $h\ge1$,
\begin{align*}
 \var(u_t)
 &=\var(\alpha_i)+V_X+(1-\psi^2)S_c+\tau_\epsilon^2,\\
 \operatorname{Cov}(u_t,u_{t+h})
 &=\var(\alpha_i)+\rho^h[V_X+\psi(1-\psi)S_c].
\end{align*}
An OU model matches both expressions with stationary state variance
$V_X+\psi(1-\psi)S_c$ and transitory variance
$\tau_\epsilon^2+(1-\psi)S_c$, keeping the same persistence and
permanent-effect variance. Hence all four stationary specifications can
share all earnings second moments. Distinguishing them requires
conditional dispersion, higher moments, or other distributional
information. Age-specific covariances under the common nonstationary
initialization can still differ.

The conditional mean formula also implies that a mean-zero initial law
remains mean zero, with mean-reversion half-life $\log 2/\kappa$.
The conditional moments concern $X$, whereas the empirical profiles
condition on $u$, mixing over permanent effects, persistent states, and
transitory noise. For finite $h$, the variance formula also averages future
state-dependent risk. Observed changes approach the difference of two
transitory shocks as $h\downarrow0$, so
\eqref{eq:diffusion-short-horizon-kurtosis} characterizes latent changes.

\subsection{Time reversibility and change symmetry}

The symmetry restrictions below follow from time reversibility of
stationary one-dimensional diffusions on an interval.

\begin{prop}[Symmetric changes and reversals]
\label{prop:diffusion-change-symmetry}
Let $X$ be a regular, conservative, positive recurrent, time-homogeneous
diffusion on an interval $\cX\subseteq\R$, and suppose it starts from its invariant law
$\pi$. Then $X$ is time reversible, and $X(t+h)-X(t)$ has a distribution
symmetric about zero for every $h>0$. For observations
\eqref{eq:models-observation} with an independent permanent effect and iid
transitory shocks independent of $X$, observed changes are also symmetric
about zero. Successive observed changes over equal intervals satisfy
\begin{equation}
 (D_1,D_2)\overset{d}{=}(-D_2,-D_1),
 \qquad
 \begin{cases}
 D_1\coloneqq u(t+h)-u(t),\\
 D_2\coloneqq u(t+2h)-u(t+h).
 \end{cases}
 \label{eq:diffusion-time-reversal}
\end{equation}
\end{prop}

\begin{proof}
A regular one-dimensional diffusion on an interval is symmetric with
respect to its speed measure \citep[Theorem~8.1]{Kent1978}. Under the stated
recurrence and conservativeness conditions, the normalized speed measure
is the invariant probability law $\pi$, and the stationary process is time
reversible \citep[Section~9]{Kent1978}. Its transition kernel $P_h$ therefore
satisfies detailed balance:
\begin{equation*}
 \pi(\diff x)P_h(x,\diff y)=\pi(\diff y)P_h(y,\diff x).
\end{equation*}
Thus $(X(t),X(t+h))$ is exchangeable, so its difference is symmetric about
zero. The permanent effect cancels from $u(t+h)-u(t)$, while the difference
of the iid transitory shocks is symmetric and independent of the latent
change. This proves the same conclusion for observed changes.

Reversibility of the stationary triple gives
\begin{equation*}
 (X(t),X(t+h),X(t+2h))
 \overset d=(X(t+2h),X(t+h),X(t)).
\end{equation*}
The independent permanent effect and iid transitory shocks preserve
reversibility of the observed triple. Taking successive differences gives
\eqref{eq:diffusion-time-reversal}.
\end{proof}

For PLV, Proposition~\ref{prop:PLV-stationary-distribution} verifies
the conditions for stationary reversibility. With unequal slopes,
Gaussian age-30 initialization can yield asymmetric finite-age changes.
If $c_{+}=c_{-}$, reflecting
$X$ and $B$ preserves the dynamics. With symmetric transitory shocks and
a symmetric initial law, the changes also satisfy
$(D_1,D_2)\overset d=(-D_1,-D_2)$. This applies both under the invariant
law and under Gaussian age-30 initialization.

With unequal slopes or asymmetric transitory shocks, this additional
sign symmetry need not hold. Time reversal exchanges both the order and
signs of successive changes; it does not equate reversal probabilities
after equally large initial losses and gains. Unequal slopes alone do
not determine the direction or size of this gap.

There is a further identification restriction. Let $\widetilde X$ have
the slopes $(c_{-},c_{+})$ in place of $(c_{+},c_{-})$ and the reflected
initial law. Its variance function is $\widetilde v(x)=v(-x)$; reflecting
$X$ and $B$ in \eqref{eq:PLV-diffusion} and using uniqueness gives
\begin{equation*}
 (\widetilde X(t))_{t\ge0}\overset d=(-X(t))_{t\ge0}.
\end{equation*}
Consequently, swapping slopes preserves every latent covariance and,
with the same independent permanent and transitory components, every
observed covariance. This holds under the common Gaussian age-30
initialization, whose law is symmetric, as well as under the respective
invariant laws.

Under those invariant laws, Proposition~\ref{prop:diffusion-change-symmetry}
also implies that $\widetilde X(t+h)-\widetilde X(t)$ and
$X(t+h)-X(t)$ have the same distribution for every $h>0$. The permanent
effect cancels from earnings changes, and the independent transitory
difference is symmetric for any iid transitory law, including an
asymmetric Gaussian mixture. Thus stationary marginal earnings-change
distributions cannot distinguish the two slope orientations, even if
we observe the entire distribution. Information about this orientation
must come from other features, such as earnings levels, conditional
profiles, or joint histories of changes.

\printbibliography

\newpage

\begin{refsection}

\begin{center}
    {\Large \textbf{Online Appendix (not for publication)}}
\end{center}

\section{Asymptotic theory of HMM estimation}
\label{sec:asymptotic-theory}

This appendix develops the asymptotic theory that underlies the inference
procedure in \S\ref{subsec:panel-inference}. We maintain the
setting and notation of \S\ref{sec:panel-likelihood}. The number of
individuals $I$ grows, while each history remains bounded. We first fix
the grid and then let it refine with $I$.

\subsection{Fixed-grid inference}
\label{subsec:panel-fixed-grid}

We first fix the grid $\cX_N$ and let $I\to\infty$. Recall the
discretized population objective $\cQ_N$, its maximizer $\theta_N$, and
the individual score $s_{i,N}$ and Hessian $H_{i,N}$ in
\eqref{eq:panel-discrete-score-hessian} from
\S\ref{subsec:panel-inference}.

At fixed $N$, the discretized quasi-maximum likelihood estimator is an
$M$-estimator. We impose standard regularity conditions.

\begin{asmp}[Fixed-grid regularity]
\label{asmp:panel-fixed-grid}
Consider the following conditions:
\begin{enumerate}
 \item\label{item:panel-fixed-grid-identification}
 (Identification)
 The function $\cQ_N$ is finite and continuous on the compact
 parameter space $\Theta$, and its unique maximizer
 is $\theta_N$.

 \item\label{item:panel-fixed-grid-ulln}
 (Uniform convergence of criterion)
 The sample criterion satisfies the uniform law of large numbers
 \begin{equation*}
  \plim_{I\to\infty}
  \sup_{\theta\in\Theta}
  \abs*{
  \frac{1}{I}\sum_{i=1}^I\ell_{i,N}(\theta)
  -\cQ_N(\theta)
  }=0.
 \end{equation*}

 \item\label{item:panel-fixed-grid-smoothness}
 (Interior and smoothness)
 The parameter $\theta_N$ belongs to the interior of $\Theta$. On an
 open neighborhood $U\subset\interior\Theta$ of $\theta_N$,
 $\ell_{i,N}(\theta)$ is twice continuously differentiable in $\theta$
 almost surely.

 \item\label{item:panel-fixed-grid-score}
 (Score regularity)
 Differentiation under the expectation is valid at $\theta_N$, and
 $\E_0\norm{s_{i,N}(\theta_N)}^2<\infty$.

 \item\label{item:panel-fixed-grid-hessian}
 (Uniform convergence of Hessian)
 The expected Hessian $\E_0[H_{i,N}(\theta)]$ is continuous at
 $\theta_N$, and the sample Hessian satisfies the uniform law of large
 numbers
 \begin{equation*}
  \plim_{I\to\infty}
  \sup_{\theta\in U}
  \norm*{
  \frac{1}{I}\sum_{i=1}^I H_{i,N}(\theta)
  -\E_0[H_{i,N}(\theta)]
  }=0.
 \end{equation*}

 \item\label{item:panel-fixed-grid-nonsingularity}
 (Nonsingularity)
 The matrix $A_N\coloneqq-\E_0[H_{i,N}(\theta_N)]$ is nonsingular.
\end{enumerate}
\end{asmp}

These conditions yield the standard consistency and asymptotic normality
results for the discretized $M$-estimator at a fixed grid.
Assumption~\ref{asmp:panel-fixed-grid} is deliberately high-level but
standard for smooth finite-dimensional $M$-estimation.  Conditions
\ref{item:panel-fixed-grid-identification}--%
\ref{item:panel-fixed-grid-ulln} deliver consistency, while the remaining
conditions permit a quadratic expansion and a central limit theorem.  At a
fixed grid, they typically follow from smooth transition probabilities and
observation densities, suitable integrable envelopes, and local
identification. In the earnings model below, we use a Gaussian observation
density and integrate over permanent heterogeneity before forming the
individual log likelihood.

\begin{prop}[Fixed-grid consistency]
\label{prop:panel-fixed-grid-consistency}
Under Assumption~\ref{asmp:panel-fixed-grid}%
\ref{item:panel-fixed-grid-identification}--%
\ref{item:panel-fixed-grid-ulln}, we have $\plim_{I\to\infty}\widehat\theta_{I,N}=\theta_N$.
\end{prop}

\begin{proof}
This is an application of the argmax consistency theorem \citep[Theorem~5.7]{vanderVaart1998}.
\end{proof}

\begin{prop}[Fixed-grid asymptotic normality]
\label{prop:panel-fixed-grid-normality}
Under Assumption~\ref{asmp:panel-fixed-grid}%
\ref{item:panel-fixed-grid-identification}--%
\ref{item:panel-fixed-grid-nonsingularity}, letting $B_N\coloneqq
 \E_0[
 s_{i,N}(\theta_N)
 s_{i,N}(\theta_N)^\top]$, as $I\to\infty$ we have
\begin{equation}
 \sqrt I(\widehat\theta_{I,N}-\theta_N)
 \dto\cN(0,A_N^{-1}B_NA_N^{-1}).
 \label{eq:panel-fixed-grid-normality}
\end{equation}
\end{prop}

\begin{proof}
This follows from the standard argument for smooth $M$-estimators,
analogous to \citet[Theorem~5.23]{vanderVaart1998}.
\end{proof}

Even when the continuous-state model is correct, its finite-grid likelihood
is generally misspecified. The sandwich covariance
in \eqref{eq:panel-fixed-grid-normality} therefore does not ordinarily
reduce to $A_N^{-1}$.

\subsection{Continuous-state inference}
\label{subsec:panel-continuous-state}

We now allow the grid size $N$ to increase with the cross-sectional
sample size $I$.  Let $N=N_I\to\infty$ as $I\to\infty$.  Consistency requires a qualitative
approximation of the continuous-state likelihood, whereas inference
centered at the continuous-state parameter requires a quantitative bound
on the discretization error.

\begin{asmp}[Refining-grid regularity]
\label{asmp:panel-refining-grid}
Consider the following conditions:
\begin{enumerate}
 \item\label{item:panel-refining-grid-identification}
 (Identification)
 The function $\cQ$ is finite and continuous on the compact parameter space
 $\Theta$, and its unique maximizer is $\theta_0$.

 \item\label{item:panel-refining-grid-positivity}
 (Positivity)
 For every $\theta\in\Theta$ and sufficiently large $N$, we have
 $L_i(\theta)>0$ and $L_{i,N}(\theta)>0$ almost surely.

 \item\label{item:panel-refining-grid-envelope}
 (Integrable envelope)
 For some integrable $M_i\ge0$,
 $\abs{\ell_{i,N}(\theta)}\le M_i$ for every sufficiently large $N$ and $\theta\in\Theta$.

 \item\label{item:panel-refining-grid-lipschitz}
 (Uniform Lipschitz continuity)
 There is a nonnegative random variable $C_i$ with $\E_0C_i<\infty$
 such that, for every sufficiently large $N$ and
 $\theta,\theta'\in\Theta$,
 \begin{equation*}
  \abs{\ell_{i,N}(\theta)-\ell_{i,N}(\theta')}
  \le C_i\norm{\theta-\theta'}.
 \end{equation*}

 \item\label{item:panel-refining-grid-approximation}
 (Likelihood approximation)
 As $N\to\infty$, we have
 \begin{equation}
  \E_0\left[\sup_{\theta\in\Theta}
  \abs{\ell_{i,N}(\theta)-\ell_i(\theta)}\right]\to0.
  \label{eq:panel-log-likelihood-approximation}
 \end{equation}
\end{enumerate}
\end{asmp}

These conditions make the discretized sample criterion converge uniformly
to the continuous-state population objective as both $I$ and $N$ increase.

\begin{prop}[Continuous-state consistency]
\label{prop:panel-refining-grid}
Under Assumption~\ref{asmp:panel-refining-grid}, for every deterministic
sequence $N_I\to\infty$, we have $\plim_{I\to\infty}\widehat\theta_{I,N_I}=\theta_0$.
\end{prop}

\begin{proof}
Recall the sample objective $\widehat \cQ_{I,N}$ in
\eqref{eq:panel-discrete-estimator}.
We first establish the triangular-array uniform law
\begin{equation*}
 \sup_{\theta\in\Theta}
 \abs*{\widehat \cQ_{I,N_I}(\theta)-\cQ_{N_I}(\theta)}
 =o_{\mathrm p}(1).
\end{equation*}
Fix $\delta>0$.  Compactness of $\Theta$ gives a finite $\delta$-net
$\set{\theta_1,\dots,\theta_K}$.  For each $\theta\in\Theta$, choose
$k(\theta)$ such that $\norm{\theta-\theta_{k(\theta)}}\le\delta$.
Assumption~\ref{asmp:panel-refining-grid}%
\ref{item:panel-refining-grid-lipschitz} implies
\begin{multline}
 \sup_{\theta\in\Theta}
 \abs*{\widehat \cQ_{I,N_I}(\theta)-\cQ_{N_I}(\theta)}\\
\le
 \max_{1\le k\le K}
 \abs*{\widehat \cQ_{I,N_I}(\theta_k)-\cQ_{N_I}(\theta_k)}
 +\delta\left(\frac{1}{I}\sum_{i=1}^I C_i+\E_0C_i\right).
 \label{eq:panel-proof-triangular-ulln-bound}
\end{multline}
We next show that every term in the finite maximum converges to zero in
probability.  Fix $k$ and write
$Z_{i,I}\coloneqq\ell_{i,N_I}(\theta_k)$.  By
Assumption~\ref{asmp:panel-refining-grid}%
\ref{item:panel-refining-grid-envelope}, $\abs{Z_{i,I}}\le M_i$ with
$\E_0M_i<\infty$.  For $b>0$, define
$Z_{i,I}^{(b)}\coloneqq
Z_{i,I}\mathbf 1_{\set*{M_i\le b}}$.  Independence across
individuals and Chebyshev's inequality give
\begin{equation*}
 \E_0\abs*{\frac{1}{I}\sum_{i=1}^I
 \left(Z_{i,I}^{(b)}-\E_0Z_{i,I}^{(b)}\right)}^2
 \le\frac{b^2}{I}.
\end{equation*}
For the discarded part, the triangle inequality yields
\begin{equation*}
	\E_0\abs*{\frac{1}{I}\sum_{i=1}^I
 \left[Z_{i,I}-Z_{i,I}^{(b)}
 -\E_0\left(Z_{i,I}-Z_{i,I}^{(b)}\right)\right]}\le 2\E_0[M_i\mathbf 1_{\set*{M_i>b}}].
\end{equation*}
The right-hand side tends to zero as $b\to\infty$.  Taking $I\to\infty$
for fixed $b$ and then $b\to\infty$ proves $\widehat \cQ_{I,N_I}(\theta_k)-\cQ_{N_I}(\theta_k)
 =o_{\mathrm p}(1)$. 
Because the net is finite, the maximum in
\eqref{eq:panel-proof-triangular-ulln-bound} is also
$o_{\mathrm p}(1)$. The weak law also gives
$I^{-1}\sum_{i=1}^I C_i\pto\E_0C_i$. We can first choose $\delta$ arbitrarily small and then let
$I\to\infty$ in \eqref{eq:panel-proof-triangular-ulln-bound}, which proves
the claimed uniform law.

Assumption~\ref{asmp:panel-refining-grid}%
\ref{item:panel-refining-grid-approximation} also implies
\begin{align*}
 \sup_{\theta\in\Theta}\abs{\cQ_N(\theta)-\cQ(\theta)}
 &\le \E_0\left[\sup_{\theta\in\Theta}
 \abs{\ell_{i,N}(\theta)-\ell_i(\theta)}\right]
 \to0.
\end{align*}
Consequently,
\begin{equation}
 \sup_{\theta\in\Theta}
 \abs*{\widehat \cQ_{I,N_I}(\theta)-\cQ(\theta)}
 =o_{\mathrm p}(1).
 \label{eq:panel-proof-uniform-exact-objective}
\end{equation}

It remains to apply identification.  For any $\epsilon>0$, define
\begin{equation*}
 \Theta_\epsilon
 \coloneqq\set*{\theta\in\Theta:
 \norm{\theta-\theta_0}\ge\epsilon}.
\end{equation*}
If $\Theta_\epsilon$ is nonempty, continuity of $\cQ$, compactness of
$\Theta_\epsilon$, and uniqueness of its maximizer imply
\begin{equation*}
 \eta_\epsilon
 \coloneqq \cQ(\theta_0)-\sup_{\theta\in\Theta_\epsilon}\cQ(\theta)>0.
\end{equation*}
On the event that the supremum in
\eqref{eq:panel-proof-uniform-exact-objective} is smaller than
$\eta_\epsilon/3$,
\begin{equation*}
 \widehat \cQ_{I,N_I}(\theta_0)
 >\sup_{\theta\in\Theta_\epsilon}
 \widehat \cQ_{I,N_I}(\theta).
\end{equation*}
Hence no maximizer of the sample objective belongs to $\Theta_\epsilon$.
The probability of this event tends to one by
\eqref{eq:panel-proof-uniform-exact-objective}, so
$\Pr(\norm{\widehat\theta_{I,N_I}-\theta_0}\ge\epsilon)\to0$ for every
$\epsilon>0$.
\end{proof}

Consistency imposes no relative rate between $I$ and $N_I$. Suppose
the grid becomes fine on every compact subset of the state space and its
endpoints approach the boundaries.  On an evenly spaced grid, standard
quadrature formulas such as the composite trapezoidal and Simpson rules
approximate transition integrals under familiar smoothness and tail
conditions \citep{DavisRabinowitz1984}.  If the resulting discrete initial
distributions and transition kernels converge weakly, uniformly in
$\theta$, then bounded path length, continuity, uniform integrability, and
suitable control of likelihoods near zero allow the convergence to pass
through the successive transition integrations and the logarithm. Obtaining
\eqref{eq:panel-log-likelihood-approximation} requires these additional
uniformity conditions because weak convergence alone does not control
logarithms or parameter variation. This rate-free conclusion
parallels \citet[Theorem~5.1]{Farmer2021}.

Consistency alone does not make the grid error negligible relative to the
sampling error.  To address this issue, define the
individual score and Hessian of the exact continuous-state log likelihood by
\begin{equation*}
 s_i(\theta)\coloneqq\nabla\ell_i(\theta),
 \qquad
 H_i(\theta)\coloneqq\nabla^2\ell_i(\theta).
\end{equation*}
In Equation~\eqref{eq:panel-discrete-score-hessian}, we define the corresponding
discretized derivatives $s_{i,N}$ and $H_{i,N}$.

\begin{asmp}[Continuous-state normality]
\label{asmp:panel-continuous-normality}
Consider the following conditions:
\begin{enumerate}
 \item\label{item:panel-continuous-smoothness}
 (Interior and smoothness)
 The parameter $\theta_0$ belongs to the interior of $\Theta$. On an open
 neighborhood $U\subset\interior\Theta$ of $\theta_0$,
 $\ell_i(\theta)$ and $\ell_{i,N}(\theta)$ are twice continuously
 differentiable in $\theta$ almost surely for every sufficiently large $N$.

 \item\label{item:panel-continuous-score}
 (Score regularity)
We have $\E_0s_i(\theta_0)=0$ and $\E_0\norm{s_i(\theta_0)}^2<\infty$.

 \item\label{item:panel-continuous-hessian}
 (Hessian regularity)
 We have $\E_0\sup_{\theta\in U}\norm{H_i(\theta)}<\infty$, and
 \begin{equation*}
  \plim_{I\to\infty}\sup_{\theta\in U}
  \norm*{\frac{1}{I}\sum_{i=1}^I H_i(\theta)
  -\E_0H_i(\theta)}=0.
 \end{equation*}

 \item\label{item:panel-continuous-nonsingularity}
 (Nonsingularity)
 The matrix $A\coloneqq-\E_0H_i(\theta_0)$ is nonsingular.

 \item\label{item:panel-continuous-score-approximation}
 (Score approximation)
 As $N\to\infty$,
 \begin{equation}
  \E_0\norm{s_{i,N}(\theta_0)-s_i(\theta_0)}^2\to0.
  \label{eq:panel-score-l2-error}
 \end{equation}

 \item\label{item:panel-continuous-hessian-approximation}
 (Hessian approximation)
 As $N\to\infty$,
 \begin{equation}
  \E_0\left[\sup_{\theta\in U}
  \norm{H_{i,N}(\theta)-H_i(\theta)}\right]\to0.
  \label{eq:panel-hessian-approximation}
 \end{equation}

 \item\label{item:panel-continuous-score-bias}
 (Score-bias rate)
 Along the sequence $N_I$,
 \begin{equation}
  \sqrt I\norm{\E_0s_{i,N_I}(\theta_0)}\to0.
  \label{eq:panel-score-bias-rate}
 \end{equation}
\end{enumerate}
\end{asmp}

These additional conditions make the discretization error negligible at the
$\sqrt I$ scale and permit inference for the continuous-state target.

\begin{prop}[Continuous-state asymptotic normality]
\label{prop:panel-continuous-normality}
Under Assumptions~\ref{asmp:panel-refining-grid} and
\ref{asmp:panel-continuous-normality}, letting $B\coloneqq\E_0[s_i(\theta_0)s_i(\theta_0)^\top]$, as $I\to\infty$ we have
\begin{equation}
 \sqrt I(\widehat\theta_{I,N_I}-\theta_0)
 \dto\cN(0,A^{-1}BA^{-1}).
 \label{eq:panel-continuous-normality}
\end{equation}
\end{prop}

\begin{proof}
For brevity, write $\widehat\theta_I\coloneqq
\widehat\theta_{I,N_I}$ and define the approximate sample score and Hessian
by
\begin{equation*}
 \widehat S_I(\theta)
 \coloneqq\frac{1}{I}\sum_{i=1}^I s_{i,N_I}(\theta),
 \qquad
 \widehat H_I(\theta)
 \coloneqq\frac{1}{I}\sum_{i=1}^I H_{i,N_I}(\theta).
\end{equation*}
We first derive the limiting distribution of the score at $\theta_0$.
Let $d_{i,I}\coloneqq s_{i,N_I}(\theta_0)-s_i(\theta_0)$. 
Since $\E_0s_i(\theta_0)=0$, we may write
\begin{equation}
 \sqrt I\widehat S_I(\theta_0)
 =\frac{1}{\sqrt I}\sum_{i=1}^I s_i(\theta_0)
 +\frac{1}{\sqrt I}\sum_{i=1}^I
 \left(d_{i,I}-\E_0d_{i,I}\right)
 +\sqrt I\E_0s_{i,N_I}(\theta_0).
 \label{eq:panel-proof-score-decomposition}
\end{equation}
The first term converges in distribution to $\cN(0,B)$ by the
multivariate central limit theorem.  Independence across individuals and
Assumption~\ref{asmp:panel-continuous-normality}%
\ref{item:panel-continuous-score-approximation} imply
\begin{equation*}
 \E_0\norm*{\frac{1}{\sqrt I}\sum_{i=1}^I
 \left(d_{i,I}-\E_0d_{i,I}\right)}^2=\operatorname{tr}\var_0(d_{i,I})\le\E_0\norm{d_{i,I}}^2\to0.
\end{equation*}
The second term in \eqref{eq:panel-proof-score-decomposition} is therefore
$o_{\mathrm p}(1)$, while its third
term is $o(1)$ by
Assumption~\ref{asmp:panel-continuous-normality}%
\ref{item:panel-continuous-score-bias}.  It follows that
\begin{equation}
 \sqrt I\widehat S_I(\theta_0)
 \dto\cN(0,B).
 \label{eq:panel-proof-score-limit}
\end{equation}

We next establish convergence of the approximate sample Hessian.  Let
\begin{equation*}
 Z_{i,N}\coloneqq\sup_{\theta\in U}
 \norm{H_{i,N}(\theta)-H_i(\theta)}.
\end{equation*}
By Assumption~\ref{asmp:panel-continuous-normality}%
\ref{item:panel-continuous-hessian-approximation} and Markov's inequality,
for every $\epsilon>0$,
\begin{equation*}
 \Pr\left(\frac{1}{I}\sum_{i=1}^I Z_{i,N_I}>\epsilon\right)
 \le\frac{\E_0Z_{i,N_I}}{\epsilon}\to0.
\end{equation*}
Therefore,
\begin{equation*}
 \sup_{\theta\in U}
 \norm*{\widehat H_I(\theta)
 -\frac{1}{I}\sum_{i=1}^I H_i(\theta)}=o_{\mathrm p}(1).
\end{equation*}
Combining this result with
Assumption~\ref{asmp:panel-continuous-normality}%
\ref{item:panel-continuous-hessian} gives
\begin{equation}
 \sup_{\theta\in U}
 \norm{\widehat H_I(\theta)-\E_0H_i(\theta)}=o_{\mathrm p}(1).
 \label{eq:panel-proof-hessian-uniform}
\end{equation}
The map $\theta\mapsto\E_0H_i(\theta)$ is continuous at $\theta_0$:
the Hessian is continuous almost surely by
Assumption~\ref{asmp:panel-continuous-normality}%
\ref{item:panel-continuous-smoothness}, and the integrable bound in
Assumption~\ref{asmp:panel-continuous-normality}%
\ref{item:panel-continuous-hessian} permits dominated convergence.

By Proposition~\ref{prop:panel-refining-grid},
$\widehat\theta_I$ converges in probability to $\theta_0$. Choose $r>0$
so that $U$ contains the closed ball of radius $r$ centered at $\theta_0$.
With probability approaching one, $\widehat\theta_I$ belongs to this ball.
The interior of $\Theta$ contains the convex ball, so
the estimator is interior and the line segment joining $\theta_0$ to
$\widehat\theta_I$ lies in $U$.  On this event, the first-order condition
and the fundamental theorem of calculus give
\begin{equation}
 0=\widehat S_I(\widehat\theta_I)
 =\widehat S_I(\theta_0)
 +\bar H_I(\widehat\theta_I-\theta_0),
 \label{eq:panel-proof-score-expansion}
\end{equation}
where
\begin{equation*}
 \bar H_I\coloneqq\int_0^1
 \widehat H_I\left(\theta_0
 +u(\widehat\theta_I-\theta_0)\right)\diff u.
\end{equation*}
Uniform convergence in \eqref{eq:panel-proof-hessian-uniform}, continuity of
$\E_0H_i(\theta)$ at $\theta_0$, and consistency of
$\widehat\theta_I$ imply
\begin{equation}
 \bar H_I=-A+o_{\mathrm p}(1).
 \label{eq:panel-proof-average-hessian}
\end{equation}
Since $A$ is nonsingular, $\bar H_I$ is nonsingular with probability
approaching one, and $-\bar H_I^{-1}=A^{-1}+o_{\mathrm p}(1)$.
Multiplying \eqref{eq:panel-proof-score-expansion} by $\sqrt I$ and using
\eqref{eq:panel-proof-score-limit} and
\eqref{eq:panel-proof-average-hessian} yields
\begin{equation*}
 \sqrt I(\widehat\theta_{I,N_I}-\theta_0)
 =-\bar H_I^{-1}\sqrt I\widehat S_I(\theta_0)
 \dto\cN(0,A^{-1}BA^{-1}). \qedhere
\end{equation*}
\end{proof}

When the continuous-state model is correct and the
information identity holds, $B=A$.  The estimator then has asymptotic
variance $A^{-1}$ and attains the parametric information bound for
$\theta_0$ in the continuous-state model.  Under misspecification, the
sandwich form in \eqref{eq:panel-continuous-normality} applies instead.

A sufficient quantitative condition for
\eqref{eq:panel-score-bias-rate} is a bound
$\norm{\E_0s_{i,N}(\theta_0)}\le a_N$ with
$\sqrt Ia_{N_I}\to0$.  Under the same local curvature conditions,
this score bias also bounds $\norm{\theta_N-\theta_0}$ to first order.
With bounded histories, independence lets the centered score approximation
errors average out under \eqref{eq:panel-score-l2-error}, while their mean must
be $o(I^{-1/2})$. This gives the short-panel counterpart to
\citet[Theorems~5.1 and 5.3]{Farmer2021}, whose analysis controls
approximation errors along a growing time series.

Additional smoothness relates the required rate to the grid.
The fixed parameter dimension $d_\theta$ determines the dimensions of the
score and Hessian but does not enter the rate condition.  The state dimension
$d_x$ does enter because it determines how the mesh width varies with the
total number of grid points.  Consider a rectangular tensor-product grid
whose coordinate ranges and resolutions are comparable.  Let $R_N$ denote
the common order of its side lengths, let $h_N$ be the largest mesh width,
and let $\tau_N$ bound the error from truncating the state space.  Since $N$
is the total number of grid points, a balanced grid satisfies
$h_N\asymp R_NN^{-1/d_x}$.  Suppose, uniformly on a neighborhood of
$\theta_0$, that for some constant $C<\infty$, we have
\begin{multline*}
 \norm{\E_0s_{i,N}(\theta_0)}
 +\left(\E_0\norm{s_{i,N}(\theta_0)-s_i(\theta_0)}^2\right)^{1/2}
 +\E_0\left[\sup_{\theta\in U}
 \norm{H_{i,N}(\theta)-H_i(\theta)}\right]
 \\
\le C(R_N^{d_x}h_N^r+\tau_N).
\end{multline*}
The tensor-product composite trapezoidal rule has $r=2$ when the relevant
integrands have bounded second derivatives, whereas the tensor-product
composite Simpson rule has $r=4$ when they have bounded fourth derivatives
and each coordinate has an even number of subintervals
\citep{DavisRabinowitz1984}.  It is therefore sufficient for centered
normal inference that
\begin{equation}
 \sqrt I\bigl(R_{N_I}^{d_x}h_{N_I}^r+\tau_{N_I}\bigr)\to0.
 \label{eq:panel-grid-rate}
\end{equation}
For a balanced grid on a compact state space, $R_N=O(1)$ and $\tau_N=0$,
while $h_N=O(N^{-1/d_x})$. Condition
\eqref{eq:panel-grid-rate} then reduces to $\sqrt I N_I^{-r/d_x}\to 0$, or equivalently $N_II^{-d_x/(2r)}\to\infty$. Thus it is sufficient that $N_I/I^{d_x/4}\to\infty$ for the trapezoidal
rule or $N_I/I^{d_x/8}\to\infty$ for Simpson's rule. These are illustrative
sufficient rates: to apply them to a particular HMM approximation, we
must verify the stated uniform bound for its likelihood derivatives
across all transition and integration approximations. Simpson's quadrature
order alone does not determine the filter's overall accuracy. On an
expanding grid, both the volume bound
$R_N^{d_x}$ and the tail error $\tau_N$ enter
\eqref{eq:panel-grid-rate}.

These rates describe a balanced full tensor grid, whose node count
grows exponentially in $d_x$ at a fixed resolution per coordinate.  When this curse of dimensionality is consequential, sparse-grid or
dimension-adaptive quadrature can reduce the number of nodes under suitable
mixed-smoothness conditions \citep{GerstnerGriebel1998}. The
consistency and normality propositions remain applicable. To apply either
proposition with such a rule, we would replace the tensor-grid bound in
\eqref{eq:panel-grid-rate} with the error bound for the chosen grid and
quadrature rule. In the earnings application, $N$ discretizes the scalar
persistent state and $K_\alpha$ nodes integrate the permanent effect; approximation
requirements apply jointly to both.

\section{OU likelihood with normal--Laplace effects}
\label{sec:OU-normal-laplace-likelihood}

We derive the exact likelihood for specification~(5) in
Table~\ref{tab:OU-benchmarks}.

\subsection{Covariance and centering}
Let individual $i$ have $n_i\coloneqq J_i+1$ observations and let $\mathbf u_i\coloneqq(u_{i0},\dots,u_{iJ_i})^\top$ be the residuals.
We write $\iota_{n_i}$ for the $n_i$-vector of ones and recall
$\omega=\sigma/\sqrt{2\kappa}$ from \S\ref{subsec:OU-benchmark}.
We suppress parameter dependence in the covariance and auxiliary quantities.
The covariance function of the OU component follows from the solution
\begin{equation*}
 X_i(t)=\e^{-\kappa t}X_i(0)
 +\sigma\int_0^t\e^{-\kappa(t-r)}\diff B_i(r).
\end{equation*}
In particular, the autocovariance is
\begin{equation}
 \cov\left(X_i(s),X_i(t)\right)
 =\omega^2\e^{-\kappa\abs{t-s}}
 +(\sigma_0^2-\omega^2)\e^{-\kappa(t+s)}.
 \label{eq:OU-covariance}
\end{equation}
Let $C_i$ denote the $n_i\times n_i$ matrix that we form by evaluating
\eqref{eq:OU-covariance} at the observed dates. After we integrate
out the Gaussian random-intercept component $G_i$, the covariance matrix of
all Gaussian terms in the observation vector is
\begin{equation*}
 \Sigma_i
 \coloneqq C_i+\sigma_G^2\iota_{n_i}\iota_{n_i}^\top
 +\sigma_\epsilon^2 I_{n_i}.
\end{equation*}

Now define the uncentered asymmetric-Laplace variable
$W_i\coloneqq V_{a,i}-V_{b,i}$. Its mean and density are
\begin{equation*}
 m\coloneqq\E W_i=\frac1a-\frac1b,
 \qquad
 f_W(w)=\frac{ab}{a+b}
 \begin{cases}
  \e^{-aw},&w\ge0,\\
  \e^{bw},&w<0.
 \end{cases}
\end{equation*}
The random effect in \eqref{eq:normal-laplace-effect} is therefore
$\alpha_i=G_i+W_i-m$. Conditional on $W_i=w$, the observation vector is
Gaussian: $\mathbf u_i\mid W_i=w
 \sim\cN((w-m)\iota_{n_i},\Sigma_i)$.

\subsection{Integrated likelihood}
Writing $\phi_n( \cdot ;\Sigma)$ for the density of a
zero-mean $n$-variate Gaussian variable with covariance $\Sigma$, the
individual likelihood contribution is
\begin{multline}
 L_i(\theta)
 =\frac{ab}{a+b}\left\{
 \int_0^\infty
 \phi_{n_i}\left(\mathbf u_i-(w-m)\iota_{n_i};\Sigma_i\right)\e^{-aw}\diff w
 \right.\\
+\left.\int_{-\infty}^0
 \phi_{n_i}\left(\mathbf u_i-(w-m)\iota_{n_i};\Sigma_i\right)\e^{bw}\diff w
 \right\}.
 \label{eq:OU-likelihood-integral}
\end{multline}

Both integrals in \eqref{eq:OU-likelihood-integral} are Gaussian tail
integrals. To evaluate them, define
\begin{equation*}
 \mathbf z_i\coloneqq\mathbf u_i+m\iota_{n_i},
 \quad
 q_i\coloneqq\iota_{n_i}^\top\Sigma_i^{-1}\iota_{n_i},
 \quad
 \widehat w_i\coloneqq
 \frac{\iota_{n_i}^\top\Sigma_i^{-1}\mathbf z_i}{q_i}, \quad
 S_i\coloneqq\mathbf z_i^\top\Sigma_i^{-1}\mathbf z_i-q_i\widehat w_i^2.
\end{equation*}
Completing the square in the conditional Gaussian density gives
\begin{equation*}
 \phi_{n_i}(\mathbf z_i-w\iota_{n_i};\Sigma_i)
 =(2\pi)^{-n_i/2}\abs{\Sigma_i}^{-1/2}
 \e^{-S_i/2}
 \e^{-\frac{q_i}{2}(w-\widehat w_i)^2}.
\end{equation*}
For the positive branch, another completion of the square yields
\begin{equation*}
A_{i,+}\coloneqq \int_0^\infty
 \e^{-\frac{q_i}{2}(w-\widehat w_i)^2-aw}\diff w=\sqrt{\frac{2\pi}{q_i}}
 \e^{-a\widehat w_i+\frac{a^2}{2q_i}}
 \Phi\left(\sqrt{q_i}\widehat w_i-\frac{a}{\sqrt{q_i}}\right),
\end{equation*}
where $\Phi$ is the standard Gaussian distribution function. Similarly, the
negative branch is
\begin{equation*}
 A_{i,-}\coloneqq \int_{-\infty}^0
 \e^{-\frac{q_i}{2}(w-\widehat w_i)^2+bw}\diff w=\sqrt{\frac{2\pi}{q_i}}
 \e^{b\widehat w_i+\frac{b^2}{2q_i}}
 \Phi\left(-\sqrt{q_i}\widehat w_i-\frac{b}{\sqrt{q_i}}\right).
\end{equation*}
The exact individual likelihood is then
\begin{equation}
 L_i(\theta)
 =\frac{ab}{a+b}(2\pi)^{-n_i/2}
 \abs{\Sigma_i}^{-1/2}\e^{-S_i/2}(A_{i,+}+A_{i,-}).
 \label{eq:OU-exact-likelihood}
\end{equation}
Independence across individuals gives the log likelihood
$\ell(\theta)\coloneqq\sum_{i=1}^I\log L_i(\theta)$.

\subsection{Stable evaluation}
For numerical evaluation, we obtain $q_i$, $\widehat w_i$, $S_i$, and
$\log\abs{\Sigma_i}$ from a Cholesky factorization rather than form
$\Sigma_i^{-1}$ explicitly. We evaluate the Gaussian tail terms on the log
scale, using the scaled complementary error function for large negative
arguments, and combine $A_{i,+}$ and $A_{i,-}$ with a log-sum-exp operation.
These steps avoid cancellation when either tail rate is large.

We can alternatively apply the Kalman innovations decomposition to
$\mathbf z_i$ and $\iota_{n_i}$ to obtain the same quadratic terms and
determinant. Both evaluations give the same likelihood, and in both we
evaluate the asymmetric-Laplace integral analytically. As $a,b\to\infty$, $W_i-m$
converges to zero and
\eqref{eq:OU-exact-likelihood} converges to the Gaussian random-intercept
likelihood for specification~(4), providing a direct nesting check on the
implementation.

\section{Numerical implementation and validation}
\label{sec:numerical-implementation}

We construct the likelihoods used in \S\ref{subsec:likelihood-estimation}
from known transition densities for OU, GKOS, and the Gaussian step of GMR
(Appendix~\ref{subsec:transition-density-discretization}). For PLV, whose
transition density has no closed form, we approximate the diffusion
generator (Appendix~\ref{subsec:diffusion-generator}).
We also apply the generator construction to OU, whose exact likelihood
provides a benchmark for numerical accuracy. We describe the common
grid settings, permanent-effect integration, and optimization in
Appendix~\ref{subsec:common-estimation-settings}, then validate the
numerical methods in Appendix~\ref{subsec:OU-validation}.

Let $N$ denote the number of filtering nodes and $K_\alpha$ the number
of permanent-effect quadrature nodes. For the generator construction, we
also use $M$ generator nodes, suppressing dependence on
$M$ in $P_{\theta,N}$, $\nu_{\theta,N,i}$, and $L_{i,N,K_\alpha}$.
The unconditional state transition over an integer gap $\Delta$ is
\begin{equation}
 P_{\theta,N}(\Delta)\coloneqq P_{\theta,N}(1)^\Delta.
 \label{eq:gap-transition}
\end{equation}

\subsection{Discretization from transition densities}
\label{subsec:transition-density-discretization}

For OU, GKOS, and the Gaussian step of GMR, the annual transition density
is available in closed form. We discretize it by the method of
\citet{FarmerToda2017QE} on an evenly spaced grid $x_1<\cdots<x_N$.\footnote{Version 0.1.0 of the \citet{FarmerToda2017QE} MATLAB and
Python implementation imposed a $10^{-8}$ lower bound on reference weights
for numerical stability. Because it affected the log likelihood, we
replaced it with log-space scaling and a relative \texttt{realmin}
safeguard in version 0.2.0, which we use here; see
\url{https://github.com/alexisakira/discretization}.}

The continuous-time OU process has an exact Gaussian AR(1) representation
at annual dates. We therefore use the same AR(1) construction for OU and
GKOS, with Gaussian and Gaussian-mixture innovations, respectively. For
each starting node $x_n$, we set
$q_{nk}\propto f_{\eta,\theta}(x_k-\rho x_n)$ and normalize across $k$,
where $f_{\eta,\theta}$ is the persistent innovation density. We choose
transition probabilities by minimizing relative entropy subject to conditional-moment constraints:
\begin{align*}
 \min_{p_1,\dots,p_N\ge0}\quad
 &\sum_{k=1}^N p_k\log\frac{p_k}{q_{nk}}\\
 \text{subject to}\quad
 &\sum_{k=1}^N p_k=1,\qquad
   \sum_{k=1}^N p_kx_k=\rho x_n, \qquad \sum_{k=1}^N p_k(x_k-\rho x_n)^2=\var_\theta(\eta_t).
\end{align*}
The resulting row forms $P_{\theta,N}(1)$. The reference weights retain
the shape of the innovation density beyond these two moments. For
stationary AR(1) processes, \citet[Proposition~3.4 and Corollary~3.5]{FarmerToda2017QE}
give sufficient conditions on the grid under which we can match both
conditional moments at every node, even with non-Gaussian innovations.
Our grids satisfy these conditions at the reported estimates for OU,
GKOS, and the Gaussian step of GMR.

For GMR, write $Z_t\coloneqq\rho X_{t-1}+\eta_t$. For $t\ge1$,
\eqref{eq:GMR-model} becomes
$X_t=Z_t+\psi c_t$ and $u_t=\alpha+Z_t+c_t+\epsilon_t$.
We discretize the Gaussian transition from $X_{t-1}$ to $Z_t$ by the same
moment method, obtaining a matrix $A_\theta$. Let $s_k(z,c)$ be the linear
interpolation weight on node $x_k$ for $z+\psi c$, assigning values
outside the grid to the nearest endpoint. If $F_{c,\theta}$ denotes the
centered loss law, the marginal state transition and joint transition
and observation kernel are
\begin{align*}
 [P_{\theta,N}(1)]_{nk}
 &=\sum_{m=1}^N[A_\theta]_{nm}
       \int s_k(x_m,c)F_{c,\theta}(\diff c),\\
 [H_{\theta,N}(u,\alpha)]_{nk}
 &=\sum_{m=1}^N[A_\theta]_{nm}
       \int s_k(x_m,c)f_{\epsilon,\theta}(u-\alpha-x_m-c)
       F_{c,\theta}(\diff c).
\end{align*}
Thus we update a row of filtered probabilities $r^\top$ over an observed
gap $\Delta\ge1$ to
$r^\top P_{\theta,N}(1)^{\Delta-1}H_{\theta,N}(u,\alpha)$,
then normalize it by its sum to obtain the conditional likelihood
contribution. This integrates the same current loss in both equations.
We include the no-loss point mass explicitly and apply Gauss--Legendre
quadrature to the continuous loss component. For speed, we tabulate the continuous
kernel in earnings residuals and evaluate it by cubic
interpolation. Gaussian-mixture transitory noise enters by summing the
two component kernels.

We initialize OU and GKOS with the Gaussian age-30 law on the grid and
propagate it to entry age using the annual transition matrix.
For GMR, the first update at $t_{i0}\ge1$ is
$p_{\theta,N,0}^\top P_{\theta,N}(1)^{t_{i0}-1}
H_{\theta,N}(u_{i0},\alpha)$, where $p_{\theta,N,0}$ contains the Gaussian
age-30 cell probabilities. At $t_{i0}=0$, $X_0$ is independent of $c_0$,
so we instead multiply those probabilities by
\begin{equation*}
 g_{\theta,0}(u\mid x,\alpha)
 =\int f_{\epsilon,\theta}(u-\alpha-x-(1-\psi)c)
       F_{c,\theta}(\diff c).
\end{equation*}

For validation, we also construct an OU generator approximation
(Appendix~\ref{subsec:diffusion-generator}) and compare it with the
transition-density construction in Appendix~\ref{subsec:OU-validation}.

\subsection{Discretization through the diffusion generator}\label{subsec:diffusion-generator}

PLV has no closed-form transition density, so we construct its annual
transition matrix by approximating the diffusion generator. Setting both
variance slopes $c_\pm$ to zero gives the OU case. We use this case to validate
the generator approach against an exact likelihood in
Appendix~\ref{subsec:OU-validation}; our empirical OU estimates use the
transition-density construction of
Appendix~\ref{subsec:transition-density-discretization}.
Both diffusion generators have the form
\begin{equation}
 \cL_\theta f(x)
 =-\kappa x f'(x)+\frac12v_\theta(x)f''(x),
 \qquad
 v_\theta(x)\coloneqq\sigma^2+c_{+}x^++c_{-}x^-.
 \label{eq:PLV-generator}
\end{equation}
We approximate the operator by a finite-state
generator $Q_\theta$, whose off-diagonal entries are nonnegative and
whose rows sum to zero. Its matrix exponential $\exp(hQ_\theta)$
propagates the chain over $h$ years without an Euler time step.

\paragraph{Scale and speed densities}
Following the scale and speed representation of
\citet[Ch.~15, Section~6]{KarlinTaylor1981}, we use
\eqref{eq:PLV-scale-integral} to write the scale density \eqref{eq:PLV-scale-density} explicitly as
\begin{equation}
 s_\theta(x)\coloneqq
 \begin{cases*}
  \e^{\lambda_{c_+}x}
  (1+c_+x/\sigma^2)^{-r_{c_+}},
    & if $x\ge0$ and $c_+>0$,\\
  \e^{-\lambda_{c_-}x}
  (1-c_-x/\sigma^2)^{-r_{c_-}},
    & if $x<0$ and $c_->0$, \\
  \e^{\kappa x^2/\sigma^2},
    & otherwise,
 \end{cases*}
 \label{eq:PLV-scale-speed}
\end{equation}
where we use $r_c$ and $\lambda_c$ from
\eqref{eq:PLV-tail-parameters}. The speed density $m_\theta(x)\coloneqq2/[v_\theta(x)s_\theta(x)]$ is
proportional to the invariant density in
Proposition~\ref{prop:PLV-stationary-distribution}.

\paragraph{Finite-state generator}
We adapt the scale-and-speed Markov chain construction of
\citet{LiYing2021} by assigning speed mass to midpoint cells.
We truncate the state space to the finite interval $[z_1,z_M]$ and take
ordered generator nodes $z_1<\cdots<z_M$ that include zero. We define
\begin{equation*}
 C_j\coloneqq
 \begin{cases*}
  [z_1,(z_1+z_2)/2], & if $j=1$,\\
  [(z_{j-1}+z_j)/2,(z_j+z_{j+1})/2], & if $2\le j\le M-1$,\\
  [(z_{M-1}+z_M)/2,z_M], & if $j=M$.
 \end{cases*}
\end{equation*}
Thus the endpoint cells stop at the finite truncation boundaries. Define
\begin{align*}
 \mu_j&\coloneqq\int_{C_j}m_\theta(x)\diff x,
 &&j=1,\dots,M,\\
 r_j&\coloneqq 1/\int_{z_j}^{z_{j+1}}s_\theta(x)\diff x,
 &&j=1,\dots,M-1,
\end{align*}
which we evaluate by Gauss--Legendre quadrature, splitting cells at the
variance kink at zero.
The nonzero off-diagonal entries of the generator are
\begin{equation}
 (Q_\theta)_{j,j+1}\coloneqq\frac{r_j}{\mu_j},
 \qquad
 (Q_\theta)_{j+1,j}\coloneqq\frac{r_j}{\mu_{j+1}},
 \quad j=1,\dots,M-1.
 \label{eq:diffusion-neighbor-rates}
\end{equation}
We set each diagonal entry equal to minus its row's total outgoing rate.
We impose reflecting boundaries by omitting outward transitions at the
two endpoints. The rates satisfy detailed balance with weights $\mu_j$:
$\mu_j(Q_\theta)_{j,j+1}=r_j=
\mu_{j+1}(Q_\theta)_{j+1,j}$. Thus the finite chain preserves the
normalized stationary cell masses on that interval.

\paragraph{Projection to the filtering grid}
The one-year transition matrix on the generator grid is
$T_\theta\coloneqq\exp(Q_\theta)$. We use $N$ filtering nodes nested within
the $M$ generator nodes, writing $x_n\coloneqq x_{n,N}=z_{j(n)}$.
With $r$ generator intervals per filtering interval, $M=r(N-1)+1$.

Let $p_{\theta,1}(y\mid x)$ be the one-year transition density with
reflecting boundaries at $z_1,z_M$, and let $q_k$ be positive spatial
quadrature weights. To approximate
$\int f(y)p_{\theta,1}(y\mid x_n)\diff y$ by
$\sum_k q_kp_{\theta,1}(x_k\mid x_n)f(x_k)$, we first convert the
fine-grid cell probabilities into density evaluations:
\begin{equation*}
 [T_\theta]_{j(n),j(k)}
 \approx\int_{C_{j(k)}}p_{\theta,1}(y\mid x_n)\diff y
 \approx\frac{p_{\theta,1}(x_k\mid x_n)}{m_\theta(x_k)}\mu_{j(k)}.
\end{equation*}
The filtering transition is therefore
\begin{align}
 d_n&\coloneqq\frac{q_n m_\theta(x_n)}{\mu_{j(n)}}, &
 [P_{\theta,N}(1)]_{nk}
 &\coloneqq\frac{[T_\theta]_{j(n),j(k)}d_k}
        {\sum_{\ell=1}^N[T_\theta]_{j(n),j(\ell)}d_\ell}.
 \label{eq:diffusion-projection}
\end{align}
The factor $d_k$ converts a fine-grid probability into a filtering-grid
quadrature contribution; row normalization makes the matrix stochastic.
On a fine uniform generator grid, $d_k\approx q_k/\Delta z$ at interior
nodes. We compute $T_\theta$ using all generator states, then retain the
selected entries in \eqref{eq:diffusion-projection}. This is a
quadrature reduction rather than aggregation of fine cells into coarse
cells, so the reduced chain need not preserve the exact stationary cell
probabilities.

\paragraph{Quadrature weights and resolution}
On a uniform filtering grid with odd $N$ and spacing $\Delta x$, composite
Simpson weights are
\begin{align*}
 (q_1,\dots,q_N)
 &=\frac{\Delta x}{3}(1,4,2,4,\dots,2,4,1).
\end{align*}
For a transformed grid, we apply this rule in the uniform coordinate and
include the transformation's Jacobian. These are integration weights,
not probabilities of equal-width cells. Increasing $M$ improves the
generator approximation; increasing $N$ resolves the filtering integrals.
Row normalization cannot compensate for insufficient filtering resolution.
For integer observation gaps, we iterate the reduced annual matrix as in
\eqref{eq:gap-transition}.

\paragraph{Initialization}
We first integrate the Gaussian initial law over the fine-grid cells
$C_j$ and normalize on the truncated interval, obtaining the row vector
$p_{\theta,M,0}$. At entry time $t_{i0}>0$, we propagate this distribution
on the fine grid and apply the same probability-to-quadrature conversion:
\begin{align}
 [\nu_{\theta,N,i}]_n
 &\coloneqq\frac{[p_{\theta,M,0}\exp(t_{i0}Q_\theta)]_{j(n)}d_n}
        {\sum_{\ell=1}^N
          [p_{\theta,M,0}\exp(t_{i0}Q_\theta)]_{j(\ell)}d_\ell}.
 \label{eq:diffusion-initialization}
\end{align}
The numerator approximates the entry-age density at $x_n$ multiplied by
$q_n$; we normalize these contributions to obtain state probabilities.
We collect the entries in the column vector $\nu_{\theta,N,i}$,
consistent with \eqref{eq:panel-discrete-likelihood}.
At $t_{i0}=0$, we use the Gaussian density directly on the filtering grid,
with probabilities proportional to
$q_n\varphi(x_n/\sigma_0)/\sigma_0$. This initialization retains the
freely estimated Gaussian initial law rather than replacing it with
stationary probabilities.

\paragraph{Generator grid choices}
For PLV, we use $N=641$ evenly spaced filtering nodes on $[-8,8]$ to
accommodate the greater dispersion at negative states.

For the PLV generator, we use $M=r(N-1)+1$ with $r=4$, so $M=2{,}561$ and the generator
spacing is $1/160$. The cost of the matrix exponential grows like $M^3$,
so we do not refine further during estimation. Instead, we recompute the likelihood with
$r=8$ at the reported estimates (Table~\ref{tab:estimate-accuracy}).

\subsection{Common numerical settings and optimization}
\label{subsec:common-estimation-settings}

We feed both transition constructions into the HMM likelihood. We integrate
over permanent heterogeneity and optimize the resulting criterion in the
same way across models.

During estimation, we hold the filtering grids fixed. For OU, GKOS, and
GMR, we use $N=321$ evenly spaced nodes on $[-4,4]$, so the spacing is
$1/40$, the same as for PLV (Appendix~\ref{subsec:diffusion-generator}). We bound each standard deviation of a transitory shock or of a GKOS
innovation component below by the filtering spacing of $1/40=0.025$, because the filter
must resolve the transitory density (Appendix~\ref{subsec:OU-validation}).
At the reported estimates, we recompute the likelihood on a grid twice as
fine and on a range twice as wide (Table~\ref{tab:estimate-accuracy}).

We use individual-specific quadrature to cover a broad range of permanent
effects while resolving the region that contributes most to each
integral. We obtain reference quantities from specification~(5) of
Table~\ref{tab:OU-benchmarks} at its maximum likelihood estimate
$\widehat\theta^{(5)}$. This specification has a closed-form likelihood
and posterior (Appendix~\ref{sec:OU-normal-laplace-likelihood}).

For the window, let $\widehat\alpha_i^{\mathrm{GLS}}$ be the generalized
least squares estimate under the specification-(5) covariance and let
$\tau_i$ be its standard error under that reference covariance. We use
this reference scale to set the integration window:
\begin{equation*}
 \begin{aligned}
 A_i^-&=1.5\min\set{\widehat\alpha_i^{\mathrm{GLS}}-8\tau_i,-4.5},\\
 A_i^+&=1.5\max\set{\widehat\alpha_i^{\mathrm{GLS}}+8\tau_i,4.5}.
 \end{aligned}
\end{equation*}
Separately, we compute the reference posterior moments
\begin{equation*}
 \mu_i=\E_{\widehat\theta^{(5)}}[\alpha_i\mid\mathbf u_i],
 \qquad
 s_i^2=\var_{\widehat\theta^{(5)}}(\alpha_i\mid\mathbf u_i).
\end{equation*}
We use these moments to place nodes near the reference posterior's mass
and split at the current permanent-effect density's mode, where its slope
changes. The three interior cutpoints are
\begin{equation*}
 \mu_i-2\max\{s_i,0.08\},\qquad
 \frac1b-\frac1a,\qquad
 \mu_i+2\max\{s_i,0.08\}.
\end{equation*}
We clip them to $[A_i^-,A_i^+]$ and sort them. We hold the windows and
reference moments fixed across models and throughout estimation; only the
prior-mode cutpoint moves with $a,b$. The integrand uses each model's
current likelihood and permanent-effect density.

We apply $L$-point Gauss--Legendre quadrature on each of the four
subintervals, giving $K_\alpha=4L$ nodes; zero-width intervals receive zero
weight. We optimize with $L=16$ ($K_\alpha=64$) and report likelihoods with
$L=32$ ($K_\alpha=128$). We assess these quadrature choices in the OU
benchmark and at the empirical estimates in Appendix~\ref{subsec:OU-validation}.

Multiplying the quadrature weights by the common permanent-effect
density gives $w_{ik}(\theta)$ in
\eqref{eq:integrated-panel-likelihood}. We combine
whole-history likelihoods by log-sum-exp. We compute scores by an exact
reverse pass of the filter. This pass gives the derivatives of the log
likelihood with respect to every transition probability, initial
probability, grid node, transitory-shock parameter, and permanent-effect
node. The maps from the parameters to the transition matrices, initial
laws, and quadrature nodes and weights are cheap to evaluate, and we
differentiate them by central differences. The PLV matrix exponential is
the exception: we differentiate it by the adjoint of its Fr\'echet
derivative, which costs about three matrix exponentials in total instead of two
per parameter. For GMR, the tabulated loss kernels carry analytic
derivatives. At cutpoint ties we follow the
selected ordering branch and omit zero-width intervals.

We optimize in transformed parameters: logarithms of positive
parameters, logits of probabilities, and $\rho=\e^{-\e^{\eta}}$ for
discrete-time persistence. We leave each mixture separation $\delta$ in
\eqref{eq:centered-normal-mixture} unrestricted in sign. Swapping the
component labels, $(p,\delta,\sigma_1,\sigma_2)\mapsto
(1-p,-\delta,\sigma_2,\sigma_1)$, leaves the distribution unchanged, so
$\delta=0$, the symmetric scale mixture, is an interior point. We relabel
the estimates so that $\delta\ge0$ when we report them.

We maximize with bounded L-BFGS-B, holding the grid fixed. Apart from the
lower bounds on standard deviations described above, the bounds are wide
and do not bind at the estimates. We stop when the projected gradient
falls below 0.02 in the transformed parameters, or when five consecutive
iterations together raise the log likelihood by less than 0.01. We
use the second rule when progress becomes negligible. The quasi-Newton
curvature serves only to guide optimization.

We start OU with Gaussian transitory shocks at the exact estimate of
specification~(5) in Table~\ref{tab:OU-benchmarks}. We initialize GKOS and
GMR with Gaussian transitory shocks near this estimate, perturbing the
additional parameters away from the OU restrictions. For their versions
with mixture transitory shocks, we combine each model's
Gaussian-specification estimates with the fitted OU transitory mixture.
We initialize both PLV specifications from estimates on an earlier sample.
We resume interrupted runs from saved parameter values.

\subsection{Validation and numerical accuracy}
\label{subsec:OU-validation}

We validate both numerical transition constructions on OU, for which we
can evaluate the likelihood exactly. This provides a known benchmark for
the generator approach that we use for PLV. We include the \citet{Rouwenhorst1995} method for discretizing the Gaussian AR(1) process as an
additional comparison.
Specification~(5) of Table~\ref{tab:OU-benchmarks} has a Gaussian OU
state, Gaussian transitory noise, and the centered asymmetric-Laplace
permanent effect. Appendix~\ref{sec:OU-normal-laplace-likelihood} derives
its likelihood in closed form. We use the same 28,422 observations for
4,683 men as in \S\ref{subsec:data-adjustment}. Write $\ell$ for the exact
log likelihood, $\widehat\theta$ for its maximizer, and $\ell_{N,K_\alpha}$
for the HMM log likelihood. At $\widehat\theta$, the
estimates are $\widehat\kappa=0.0797$, $\widehat\sigma=0.2467$,
$\widehat\sigma_0=0.2129$, $\widehat\sigma_\epsilon=0.3145$,
$\widehat a=3.807$, and $\widehat b=2.790$, so the stationary standard
deviation of the state is $\omega=\sigma/\sqrt{2\kappa}=0.618$.

\paragraph{Design}
We first evaluate the numerical likelihood and its score at
$\widehat\theta$, holding the model and parameters fixed while we vary the
numerical approximation. We re-estimate the parameters separately in
Table~\ref{tab:OU-refits}, then check numerical refinements at the empirical
estimates in Table~\ref{tab:estimate-accuracy}. Because
the exact score vanishes at $\widehat\theta$, the numerical score
$s_{N}\coloneqq\nabla\ell_{N,K_\alpha}(\widehat\theta)$ measures how far the
numerical estimator moves. Let $H\coloneqq\nabla^2\ell(\widehat\theta)$ be
the exact Hessian. To first order, the numerical estimator satisfies
$\widehat\theta_N-\widehat\theta\approx(-H)^{-1}s_N$, and the exact log
likelihood loses $\frac12s_N^\top(-H)^{-1}s_N$. We report the signed error
$\ell_{N,K_\alpha}(\widehat\theta)-\ell(\widehat\theta)$ and the largest
implied numerical bias in units of reference standard errors, using the
inverse observed information $(-H)^{-1}$. In the tables, SE denotes this
reference scale from the observed information for the OU benchmark. We use it solely to
express numerical parameter displacements. The signed error matters for
model comparison, which uses likelihood levels. The implied bias matters
for estimation, which uses likelihood slopes. We work in the transformed
parameters of Appendix~\ref{subsec:common-estimation-settings} (logs of
positive parameters); the ratios of numerical bias to these reference
scales do not depend on this choice to first order. The re-estimation exercise also checks this
first-order approximation.

\paragraph{Permanent-effect quadrature}
To isolate the quadrature error, we combine the exact Gaussian likelihood
conditional on $\alpha$ with the permanent-effect quadrature of
Appendix~\ref{subsec:common-estimation-settings}.
Table~\ref{tab:OU-quadrature-accuracy} reports the result.
$K_\alpha=32$ loses 1.4 log points and moves $\widehat a$ by 0.2 standard
errors. $K_\alpha=64$ is accurate to $2\times10^{-4}$, and $K_\alpha=128$ is
exact to rounding. These results support the quadrature settings in
Appendix~\ref{subsec:common-estimation-settings} for the OU benchmark.
The subsequent OU benchmark comparisons
use $K_\alpha=128$, so the quadrature error is negligible and we can
isolate the state approximation error.

\begin{table}[!htb]
 \centering
 \caption{Permanent-effect quadrature with an exact state.}
 \label{tab:OU-quadrature-accuracy}
 \begin{tabular}{rrrr}
  \toprule
  & \multicolumn{2}{c}{At $\widehat\theta$} & At $\widetilde\theta$ \\
  \cmidrule(lr){2-3}\cmidrule(lr){4-4}
  $K_\alpha$ & Signed LL error & Max $\abs{\text{bias}}$/SE & Signed LL error \\
  \midrule
  32  & $-1.3868$ & 0.204 & $-2.0500$ \\
  64  & 0.0002 & 0.000 & $-0.0063$ \\
  128 & 0.0000 & 0.000 & 0.0000 \\
  256 & 0.0000 & 0.000 & 0.0000 \\
  \bottomrule
 \end{tabular}
 \caption*{\footnotesize\textit{Note:} We integrate out the state exactly with
 the Kalman filter and use quadrature only for the integral over $\alpha$.
 $\widetilde\theta$ is the parameter value with small transitory noise
 defined below Table~\ref{tab:OU-refits}. Entries of 0.0000 are below
 $10^{-8}$ in absolute value.}
\end{table}

\paragraph{Transition approximations at the exact estimate}
We compare the transition-density and generator constructions, with
Rouwenhorst as a third benchmark. All three approximate the annual OU
transition, which is the Gaussian AR(1) $X_{t+1}=\rho X_t+\eta_{t+1}$ with
$\rho=\e^{-\kappa}$ and $\var(\eta_{t+1})=\omega^2(1-\rho^2)$. The method
of \citet{Rouwenhorst1995} places $N$ evenly spaced nodes on
$[-\sqrt{N-1}\,\omega,\sqrt{N-1}\,\omega]$. We use this same grid for the
other two chains, so that at each $N$ the three chains differ only in
their transition probabilities. The second chain is
\citet{FarmerToda2017QE}, which matches the conditional mean and variance
by relative entropy with Gaussian reference weights
(Appendix~\ref{subsec:transition-density-discretization}). The third chain
is the diffusion generator of Appendix~\ref{subsec:diffusion-generator} on a uniform generator grid with
$M=4(N-1)+1$ nodes, projected onto the filtering grid with composite
Simpson weights. We take $N=2^k+1$ for $k=5,\dots,9$. All three chains
reach an observation gap $\Delta$ through
\eqref{eq:gap-transition}. Timings are for one likelihood evaluation with
our Python implementation on a 16-core laptop computer.

Table~\ref{tab:OU-chain-accuracy} shows that the three chains behave very
differently. The \citeauthor{FarmerToda2017QE} chain is accurate to
0.008 log points at $N=33$ and exact to rounding from $N=65$. The
Rouwenhorst chain overstates the log likelihood by 293 at $N=33$ and
still by 24 at $N=513$; its error roughly halves when $N$ doubles. At
$N=129$, a common choice, its score implies a bias of 5.1 standard errors
in $\kappa$. The generator lies in between: its error falls from 55 to 1.7
and is below 0.4 standard errors from $N=129$.

\begin{table}[!htb]
 \centering
 \caption{OU state approximations at the exact estimates ($K_\alpha=128$).}
 \label{tab:OU-chain-accuracy}
 \begin{tabular}{rrrrrrrrr}
  \toprule
  & & \multicolumn{2}{c}{Rouwenhorst} & \multicolumn{2}{c}{Farmer--Toda}
  & \multicolumn{2}{c}{Generator} & \\
  \cmidrule(lr){3-4}\cmidrule(lr){5-6}\cmidrule(lr){7-8}
  $N$ & $\Delta x$ & LL error & Bias/SE & LL error & Bias/SE & LL error & Bias/SE & Seconds \\
  \midrule
  33  & 0.218 & 293.23 & 16.42 & 0.0079 & 0.003 & 55.26 & 9.38 & 6 \\
  65  & 0.154 & 164.96 & 9.30 & 0.0000 & 0.000 & 13.63 & 0.85 & 12 \\
  129 & 0.109 & 89.18 & 5.10 & 0.0000 & 0.000 & 6.70 & 0.32 & 30 \\
  257 & 0.077 & 46.77 & 2.71 & 0.0000 & 0.000 & 3.36 & 0.16 & 70 \\
  513 & 0.055 & 24.03 & 1.41 & 0.0000 & 0.000 & 1.68 & 0.08 & 240 \\
  \bottomrule
 \end{tabular}
 \caption*{\footnotesize\textit{Note:} All chains use the Rouwenhorst grid
 of $N$ evenly spaced nodes on $\pm\sqrt{N-1}\,\omega$, with spacing
 $\Delta x$. LL error is $\ell_{N,K_\alpha}(\widehat\theta)-\ell(\widehat\theta)$,
 where $\ell(\widehat\theta)=-22{,}595.35$. Bias/SE is the largest implied
 first-order bias across the six parameters, in standard errors; for the
 Rouwenhorst chain and the generator, the largest bias is in $\kappa$.
 The generator uses $M=4(N-1)+1$ nodes. Seconds are wall times for one
 likelihood evaluation and are similar across the three chains. Entries of
 0.0000 are below $10^{-9}$ in absolute value.}
\end{table}

The differences come from how each chain approximates the conditional
distribution. \citeauthor{FarmerToda2017QE} start from the Gaussian density
at the nodes and adjust it only to match the first two conditional
moments. For a smooth integrand, the resulting sums behave like the
trapezoidal rule, which converges very fast once the spacing is below the
conditional standard deviation (0.24 here). The Rouwenhorst chain counts
how many of $N-1$ two-state chains are in the high state. Starting from
node $i$, the next node is the sum of two independent binomials,
$\mathrm{Bin}(i,p)+\mathrm{Bin}(N-1-i,1-p)$ with $p=(1+\rho)/2$. This
distribution has full support and the exact conditional mean and variance,
but it is skewed away from the center of the grid and has thinner tails
than the Gaussian. It approaches the Gaussian only through the central
limit theorem, which explains the slow convergence. The generator is a
second-order approximation of the diffusion, so its error falls by a factor
of four when the generator spacing halves.

Two further features matter for applications. First, the nonzero errors
reported in Table~\ref{tab:OU-chain-accuracy} are positive at this OU
benchmark. A cruder discretization
therefore gives a model a spurious likelihood advantage here, of 89 log
points for the Rouwenhorst chain at $N=129$. The errors can have either
sign at other parameters, as Table~\ref{tab:OU-hard-point} shows. Second,
the signed error and the implied
bias measure different things. At $N=513$, the Rouwenhorst likelihood is
24 log points too high, but its implied loss from biased estimates is only
1.9; most of its error is a nearly constant shift. A constant shift is
harmless for estimating one model but not for comparing models.

\paragraph{Generator resolution}
Table~\ref{tab:OU-generator-resolution} separates the filtering spacing
$\Delta x$ from the generator spacing $\Delta z$. At $N=129$, halving
$\Delta z$ divides the error by about four. The rows with $(N,M)=(129,1{,}025)$
and $(513,2{,}049)$ have the same $\Delta z$ and give the same error to four
digits, although their filtering grids differ by a factor of four. At these
parameters, $N=129$ already resolves the filter, and the remaining error
comes from the generator. For this benchmark, we obtain accurate
likelihoods with moderate $N$ and a finer generator grid (large $M$). The
middle rows use the estimation grid of
Appendix~\ref{subsec:common-estimation-settings}, $N=321$ evenly spaced
nodes on $[-4,4]$. With $r=4$ and $r=8$ generator intervals per filtering
interval, the generator misses the likelihood by 0.35 and 0.09 log points,
with implied biases below 0.01 standard errors. The $r=4$ row uses the
same generator spacing of $1/160$ as the PLV estimation grid, $N=641$ nodes
on $[-8,8]$ with $r=4$. This OU comparison assesses the construction at
that spacing; we check PLV separately in Table~\ref{tab:estimate-accuracy}.
The last row uses the cubic grid of our earlier estimates, which
concentrates nodes near zero. Its error at $\widehat\theta$ is 0.78, but its
spacing of about 0.04 near $\abs{x}=1$ makes it inaccurate when the
transitory noise is small (Table~\ref{tab:OU-hard-point}).

\begin{table}[!htb]
 \centering
 \caption{Generator refinement at the exact estimates ($K_\alpha=128$).}
 \label{tab:OU-generator-resolution}
 \begin{tabular}{rrrrrr}
  \toprule
  $N$ & $M$ & $\Delta x$ & $\Delta z$ & LL error & Bias/SE \\
  \midrule
  129 & 257   & 0.109 & 0.0546 & 26.29 & 1.24 \\
  129 & 513   & 0.109 & 0.0273 & 6.70 & 0.32 \\
  129 & 1,025 & 0.109 & 0.0137 & 1.68 & 0.08 \\
  129 & 2,049 & 0.109 & 0.0068 & 0.42 & 0.02 \\
  513 & 2,049 & 0.055 & 0.0137 & 1.68 & 0.08 \\
  \midrule
  321 & 1,281 & 0.025 & 0.0063 & 0.35 & 0.01 \\
  321 & 2,561 & 0.025 & 0.0031 & 0.09 & 0.00 \\
  \midrule
  321 & 1,281 & \multicolumn{2}{c}{cubic grid} & 0.78 & 0.02 \\
  \bottomrule
 \end{tabular}
 \caption*{\footnotesize\textit{Note:} The first five rows use uniform
 grids on the Rouwenhorst range $\pm\sqrt{N-1}\,\omega$. The next two rows
 use uniform grids on $[-4,4]$. The last row uses the cubic grid
 $x(s)=0.6s+19.4s^3$ on $[-20,20]$, which we construct from equally spaced
 coordinates $s\in[-1,1]$. See Table~\ref{tab:OU-chain-accuracy} for
 definitions.}
\end{table}

\paragraph{Re-estimation with numerical likelihoods}
We next re-estimate the OU model by maximizing $\ell_{N,K_\alpha}$ and
evaluate the exact likelihood at each numerical estimate
(Table~\ref{tab:OU-refits}). For the \citeauthor{FarmerToda2017QE} chain,
the numerical estimate coincides with $\widehat\theta$. For the generator with $N=129$, the actual
loss of 0.13 is close to the predicted 0.12, and each bias is within 0.02
standard errors of its first-order prediction. For the cubic grid, the
actual and predicted losses are both 0.001, and the biases agree to within
0.001 standard errors. Thus the first-order approximation is reliable when
the numerical likelihood is accurate. For the Rouwenhorst chain with $N=129$, the
first-order approximation predicts an exact loss of 25 log points; the
actual loss is 954. The numerical estimate has $\kappa=0.0093$ instead of
0.0797, so the half-life of a persistent shock rises from 9 to 75 years,
and $\sigma_\epsilon$ falls from 0.31 to 0.24. Thus the Rouwenhorst
likelihood has a spurious maximum far from $\widehat\theta$. Lower $\kappa$
widens the Rouwenhorst grid and coarsens its spacing, and lower
$\sigma_\epsilon$ makes the filter more sensitive to that spacing, so the
optimizer moves toward parameters where the approximation overstates the
likelihood the most.

\begin{table}[!htb]
 \centering
 \caption{Re-estimation with the numerical likelihood ($K_\alpha=128$).}
 \label{tab:OU-refits}
 \begin{tabular}{lrrrrrrrr}
  \toprule
  & \multicolumn{2}{c}{Exact LL loss} & \multicolumn{6}{c}{Bias/SE} \\
  \cmidrule(lr){2-3}\cmidrule(lr){4-9}
  Chain & Actual & Predicted & $\kappa$ & $\sigma$ & $\sigma_0$
  & $\sigma_\epsilon$ & $a$ & $b$ \\
  \midrule
  Rouwenhorst, $N=129$ & 953.70 & 24.97 & $-26.43$ & $-3.12$ & $-1.60$ & $-16.59$ & 1.24 & 0.61 \\
  Farmer--Toda, $N=129$ & 0.00 & 0.00 & 0.00 & 0.00 & 0.00 & 0.00 & 0.00 & 0.00 \\
  Generator, $N=129$ & 0.13 & 0.12 & $-0.34$ & $-0.12$ & 0.00 & $-0.06$ & 0.12 & 0.08 \\
  Generator, cubic grid & 0.00 & 0.00 & $-0.02$ & 0.00 & 0.01 & $-0.02$ & 0.01 & 0.01 \\
  \bottomrule
 \end{tabular}
 \caption*{\footnotesize\textit{Note:} $\widehat\theta_N$ maximizes
 $\ell_{N,K_\alpha}$. Exact LL loss is
 $\ell(\widehat\theta)-\ell(\widehat\theta_N)$, and the predicted loss is
 $\frac12s_N^\top(-H)^{-1}s_N$. Bias/SE is
 $\widehat\theta_N-\widehat\theta$ in standard errors. The generator with
 $N=129$ uses $M=513$; the cubic grid is that of
 Table~\ref{tab:OU-generator-resolution}.}
\end{table}

In this benchmark, the Rouwenhorst grid changes with the parameters. The example shows
that accuracy at the exact estimate alone can miss spurious maxima
elsewhere. For the empirical fits, we instead hold the grids fixed and
tie the parameter bounds to them: the lower bound on each transitory
standard deviation is at least the grid spacing, for the reason we
explain next. These choices guard against the failure illustrated here.
The refinements below assess sensitivity at the fitted parameters.

\paragraph{Small transitory noise}
Several models in \S\ref{subsec:likelihood-estimation} reach transitory
standard deviations of 0.02--0.04. We therefore repeat the
comparison at $\widetilde\theta$, which equals $\widehat\theta$ except
$\sigma_\epsilon=0.03$. Because $\widetilde\theta$ is not a maximizer, we
report only signed errors (Table~\ref{tab:OU-hard-point}). With $\Delta
x$ several times $\sigma_\epsilon$, all chains miss the likelihood by
hundreds to thousands of log points, with erratic signs. The
\citeauthor{FarmerToda2017QE} chain becomes accurate once $\Delta
x\approx2.6\sigma_\epsilon$ (error $-0.07$ at $N=257$). The Rouwenhorst chain
remains off by about 1,000 even at $N=513$, and the generator by 87.

\begin{table}[!htb]
 \centering
 \caption{Small transitory noise: signed log-likelihood errors at
 $\widetilde\theta$ ($\sigma_\epsilon=0.03$, $K_\alpha=128$).}
 \label{tab:OU-hard-point}
 \begin{tabular}{rrrrr}
  \toprule
  $N$ & $\Delta x/\sigma_\epsilon$ & Rouwenhorst & Farmer--Toda & Generator \\
  \midrule
  33  & 7.3 & $-7{,}516.1$ & $-12{,}794.6$ & $-11{,}801.1$ \\
  65  & 5.1 & 1,113.1 & $-2{,}763.3$ & $-2{,}356.5$ \\
  129 & 3.6 & 2,480.3 & $-173.5$ & $-1.9$ \\
  257 & 2.6 & 1,700.4 & $-0.07$ & 118.2 \\
  513 & 1.8 & 1,022.2 & 0.02 & 87.1 \\
  \midrule
  321 & \multicolumn{3}{l}{Generator, cubic grid ($M=1{,}281$)} & 78.2 \\
  \bottomrule
 \end{tabular}
 \caption*{\footnotesize\textit{Note:} $\ell(\widetilde\theta)=-30{,}675.88$.
 See Table~\ref{tab:OU-chain-accuracy} for the grids.}
\end{table}

The filter evaluates the transitory density only at the state nodes. The
conditional likelihood of an observation $u$,
$\sum_n\Pr(X=x_n)f_\epsilon(u-\alpha-x_n)$, is a trapezoidal approximation of
$\int f_X(x)f_\epsilon(u-\alpha-x)\diff x$, and the narrowest factor of
this integrand has width $\sigma_\epsilon$. After the update, the filtered
state is also concentrated within about $\sigma_\epsilon$ of $u-\alpha$. The
state grid must therefore resolve $\sigma_\epsilon$, whatever the
transition method. Integrating over $\alpha$ shifts the residual
continuously across the grid and smooths part of this error, which is why
the \citeauthor{FarmerToda2017QE} chain is accurate somewhat before $\Delta
x$ reaches $\sigma_\epsilon$. When the observations nearly pin down the
state, errors in the shape of the transition matter more, because the
likelihood then behaves like that of an observed Markov chain. This is why
the Rouwenhorst chain and the generator remain far off. The cubic grid is
fine near zero but has spacing of about 0.04 near $\abs{x}=1$, which explains
its error of 78.

\paragraph{Numerical choices for estimation}
These experiments motivate the settings with fixed grids in
Appendix~\ref{subsec:common-estimation-settings}. We use the
\citeauthor{FarmerToda2017QE} construction where the transition density
is available. For the empirical estimates, we use the generator only for
PLV, with generator spacing $1/160$; the OU generator serves as a
validation exercise. The following checks assess numerical sensitivity
at the fitted parameters.

\paragraph{Checks at the empirical estimates}
The OU benchmark tests the numerical constructions against a known
likelihood. It does not establish the same error for PLV or the other
non-Gaussian models, so we check numerical refinements separately at
all eight empirical estimates (Table~\ref{tab:estimate-accuracy}).

The OU benchmark establishes quadrature accuracy at the reference
parameters (Table~\ref{tab:OU-quadrature-accuracy}). When the transitory
noise is small, however, the history likelihood varies with $\alpha$ on
the scale of the grid spacing, and the quadrature needs more nodes.
Table~\ref{tab:estimate-accuracy} reports refinements with up to
$K_\alpha=512$. Among the completed quadrature checks, raising
$K_\alpha$ from 128 to 512 changes the log likelihood by at most 1.7.
With $K_\alpha=64$, the PLV-Gaussian likelihood is 9.6 lower. These
changes in likelihoods at fixed parameters are small relative to the differences
between models.

The lower bound on standard deviations binds for the small transitory
component of OU with mixture shocks and for $\sigma_\epsilon$ in GKOS and GMR
with Gaussian shocks. To measure its cost, we hold the other OU-mixture
parameters at the estimate and lower $\sigma_{\epsilon,2}$ on grids with
$N=641$ and $1{,}281$ nodes on $[-4,4]$, whose spacings allow it. The log
likelihood rises by 1.22 at $\sigma_{\epsilon,2}=0.02$, 1.69 at $0.015$,
and 1.76 at $0.0125$, and not further at $0.01$. The gain stays below two
log points when we vary this parameter and hold the others fixed. On the
estimation grid, the same computation stays
within 0.02 of the finer grids down to $\sigma_{\epsilon,2}=0.015$, about
0.6 times the spacing, consistent with Table~\ref{tab:OU-hard-point}.

\begin{table}[!htb]
 \centering
 \caption{Accuracy checks at the estimates.}
 \label{tab:estimate-accuracy}
 \begin{tabular}{lrrrrrrrr}
  \toprule
  & \multicolumn{2}{c}{OU} & \multicolumn{2}{c}{GKOS} & \multicolumn{2}{c}{GMR} & \multicolumn{2}{c}{PLV} \\
  \cmidrule(lr){2-3}\cmidrule(lr){4-5}\cmidrule(lr){6-7}\cmidrule(lr){8-9}
  Check & G & M & G & M & G & M & G & M \\
  \midrule
  $K_\alpha=64$ & $0.00$ & $0.00$ & $-0.24$ & $0.01$ & $0.00$ & $-0.03$ & $-9.61$ & $-2.33$ \\
  $K_\alpha=256$ & $0.00$ & $0.00$ & $0.51$ & $0.00$ & $0.00$ & $0.00$ & $-1.57$ & $0.08$ \\
  $K_\alpha=512$ & $0.00$ & $0.00$ & $0.46$ & $0.00$ & $0.00$ & $0.00$ & $-1.67$ & $0.10$ \\
  Grid twice as fine & $0.00$ & $0.00$ & $0.02$ & $0.01$ & $-0.67$ & $0.40$ & $0.87$ & $0.30$ \\
  Range twice as wide & $0.00$ & $0.00$ & $-1.74$ & $-0.75$ & $0.00$ & $-1.18$ & $-2.21$ & $-0.02$ \\
  Generator twice as fine & --- & --- & --- & --- & --- & --- & $-0.22$ & $-0.23$ \\
  \bottomrule
 \end{tabular}
 \caption*{\footnotesize\textit{Note:} Entries are changes in the log
 likelihood relative to the reported value in
 Table~\ref{tab:model-comparison} ($K_\alpha=128$ on the estimation
 grid). G and M denote Gaussian and Gaussian-mixture transitory shocks.
 The finer grid halves the spacing on the same range; the wider range
 doubles the range at the same spacing; the finer generator doubles $r$.}
\end{table}

\subsection{Implementation of the saving experiment}
\label{subsec:saving-discretization}
\label{subsec:saving-gaussian}

This appendix gives the covariance transformations, earnings transitions,
and household solution used in \S\ref{sec:saving-application}, followed
by additional conditional results. The numerical checks below concern
the fitted and covariance-matched processes under Gaussian and
Gaussian-mixture transitory shocks.

\paragraph{Covariance matching}
The targets $(\rho,V,T)$ in \eqref{eq:saving-covariance-target} come from
the fitted OU process. We use hats for the fitted parameters and stationary
moments. For GKOS, we set persistence to $\rho$ and multiply the component
means and standard deviations of its persistent innovation mixture by $s_\eta=\sqrt{(1-\rho^2)V/\widehat{\var}(\eta)}$. We scale its independent transitory shock by
$\sqrt{T/\widehat{\var}(\epsilon)}$. These transformations
preserve both standardized mixture distributions.

For PLV, let $s_X=\sqrt{V/\widehat V_X}$ and
$a_\kappa=(-\log\rho)/\widehat\kappa$. The transformed process
$X^*(t)=s_X\widehat X(a_\kappa t)$ has parameters $\kappa^*=-\log\rho$, $(\sigma^*)^2=a_\kappa s_X^2\widehat\sigma^2$, $c_\pm^*=a_\kappa s_X\widehat c_\pm$. It preserves the standardized stationary distribution while matching
persistence and stationary variance. We scale its independent transitory
shock to variance $T$ in the same way as for GKOS.

For GMR, we retain fitted loss incidence, the standardized log-loss distribution,
and $\widehat\psi$, and set persistence to $\rho$. Let
$\widehat S_c=\var(\widehat c_{it})$ and
$\widehat\tau_\epsilon^2=\var(\widehat\epsilon_{it})$. The loss
enters both the persistent innovation $\eta_{it}+\psi c_{it}$ and the
transitory component $(1-\psi)c_{it}+\epsilon_{it}$, so in general we
cannot preserve both standardized distributions. Scaling $\eta$ and $c$ by
a common factor preserves the first and matches $V$ at $s_P$; scaling $c$
and $\epsilon$ by a common factor preserves the second and matches $T$ at
$s_T$:
\begin{equation*}
 s_P^2=\frac{V}{(\widehat\sigma_\eta^2+\widehat\psi^2\widehat S_c)/(1-\rho^2)
   +\widehat\psi(1-\widehat\psi)\widehat S_c},
 \qquad
 s_T^2=\frac{T}{\widehat\tau_\epsilon^2+(1-\widehat\psi)\widehat S_c}.
\end{equation*}
We set $c_{it}^*=s_c\widehat c_{it}$ with $s_c=\min(s_P,s_T)$, and solve
for the other two variances:
\begin{equation*}
 (\sigma_\eta^*)^2
 =(1-\rho^2)\bigl[V-\widehat\psi(1-\widehat\psi)s_c^2\widehat S_c\bigr]
   -\widehat\psi^2s_c^2\widehat S_c,
 \qquad
 (\tau_\epsilon^*)^2=T-(1-\widehat\psi)s_c^2\widehat S_c,
\end{equation*}
scaling $\widehat\epsilon_{it}$ to variance $(\tau_\epsilon^*)^2$. The
smaller scale keeps both variances positive. The covariance formulas in
Appendix~\ref{subsec:diffusion-moments} then give
\eqref{eq:saving-covariance-target}. Under Gaussian transitory shocks,
$s_c=s_T=0.659$, so the transitory component keeps its standardized
distribution. Under mixture shocks, $s_c=s_P=0.791$, so the persistent
innovation keeps its standardized distribution, and $\epsilon$ is scaled
by 1.150. Scaling log losses
changes the implied distribution of fractional losses from the original
truncated-exponential family; it is a counterfactual transformation with
no additional estimated parameter.

\paragraph{Earnings transitions}
We use $N=65$ persistent states for the saving problem and compare
$N=33$ and $129$ for the fitted and covariance-matched processes. Every
model uses the same rule. The grid has $N$ evenly
spaced nodes on $\pm\sqrt{N-1}\sqrt{V_{X,m}}$, where $V_{X,m}$ is the
model's stationary persistent variance, and the transition probabilities
match the conditional mean and variance of each row by the method of
\citet{FarmerToda2017QE}. We use model-specific conditional moments and
reference weights within this common rule.
OU uses the Gaussian innovation density, GKOS the Gaussian-mixture
innovation density, and GMR the Gaussian density of the state given each
shared-loss node. For GMR, we retain the no-loss atom and integrate
continuous log losses with 16-point Gauss--Laguerre quadrature, correcting
its nodes to reproduce the exact first two loss moments. When the
variance target lies outside the grid's hull, a row matches only the
mean, or keeps its reference weights if the mean also lies outside. This
happens only in GMR rows with large losses.

For PLV, the reference weights are the one-year transition probabilities
of the generator of Appendix~\ref{subsec:diffusion-generator} on 2,001
cubic-spaced nodes over $[-40,40]$, augmented by the coarse nodes, which
we aggregate onto the coarse cells. The conditional variance comes from
the same fine transition. A common spatial rescaling then corrects the
stationary variance of the coarse chain without changing transition
probabilities or the linear conditional mean; the factor lies between
0.990 and 1.008 at $N=65$. Across the fitted and covariance-matched cases
and resolutions, the maximum
discrepancy between numerical and analytical annual covariances, relative
to the contemporaneous log-earnings variance, is below $3\times10^{-6}$,
except for GMR with mixture shocks. There, rows with large losses leave
discrepancies of up to $6\times10^{-4}$ at $N=65$ and $2\times10^{-5}$ at
$N=129$.

We integrate independent transitory shocks using seven Gauss--Hermite nodes
per Gaussian component. Let $w_{m,j}$ denote the stationary probability of
joint earnings outcome $j$, including the shared loss where applicable,
and let $z_{m,j}$ denote its centered log earnings. We set
\begin{equation*}
 y_{m,j}=\frac{\exp(z_{m,j})}{\sum_k w_{m,k}\exp(z_{m,k})},
 \qquad Y_{m,j}=y_{m,j}+b.
\end{equation*}
Thus mean raw earnings equal one at every numerical resolution, preserving
transition probabilities and log covariances.

\paragraph{Household solution and numerical checks}
We solve consumption policies by the endogenous grid method
\citep{Carroll2006}, using a 1,000-point exponential grid
for saving on $[0,10^6]$ with median grid point 10, as in
\citet{MaToda2022JME}.
Linear lotteries between adjacent asset nodes give the stationary joint
distribution of assets and persistent states. For GMR, we integrate the
lotteries and current utility using the joint law of the destination
state, shared loss, and independent transitory shock. We draw the previous
persistent state from stationarity before the first earnings transition
to obtain the stationary current earnings law at zero initial assets.

For the fitted and covariance-matched processes, at $N=65$, the 99th
percentile of off-grid relative Euler residuals is below $0.003\%$, and
Bellman, budget, and asset-bound errors are below $2\times10^{-6}$.
Truncating the saving grid at $10^5$ changes mean assets and CE by less
than $10^{-14}$ in every case.

Table~\ref{tab:saving-grid-convergence} compares persistent-state
resolutions under Gaussian-mixture transitory shocks. Changing $N$ from
65 to 129 changes mean assets by at most
$0.7\%$ and CE by at most $0.3\%$, except for fitted PLV, whose mean assets
rise by $1.7\%$ because a wider grid captures more of its Pareto upper
tail. $N=33$ is adequate except for fitted PLV, whose mean assets are
$2.7\%$ lower than at $N=65$. These changes are small relative to the differences across models.

\begin{table}[!htb]
 \centering
 \caption{Saving-model sensitivity to persistent-state resolution.}
 \label{tab:saving-grid-convergence}
 \begin{tabular}{lrrrrrr}
  \toprule
  & \multicolumn{2}{c}{$N=33$} & \multicolumn{2}{c}{$N=65$} & \multicolumn{2}{c}{$N=129$} \\
  \cmidrule(lr){2-3}\cmidrule(lr){4-5}\cmidrule(lr){6-7}
  Process & Assets & CE & Assets & CE & Assets & CE \\
  \midrule
  OU, fitted & $4.801$ & $0.793$ & $4.805$ & $0.793$ & $4.805$ & $0.793$ \\
  GKOS, fitted & $7.299$ & $0.746$ & $7.374$ & $0.746$ & $7.424$ & $0.746$ \\
  GKOS, matched & $4.617$ & $0.795$ & $4.636$ & $0.796$ & $4.662$ & $0.796$ \\
  GMR, fitted & $5.597$ & $0.802$ & $5.592$ & $0.802$ & $5.592$ & $0.802$ \\
  GMR, matched & $4.330$ & $0.805$ & $4.313$ & $0.805$ & $4.310$ & $0.805$ \\
  PLV, fitted & $10.227$ & $0.738$ & $10.508$ & $0.739$ & $10.683$ & $0.737$ \\
  PLV, matched & $4.231$ & $0.827$ & $4.193$ & $0.832$ & $4.165$ & $0.833$ \\
  \bottomrule
 \end{tabular}
 \caption*{\footnotesize\textit{Note:} Gaussian-mixture transitory shocks.
 Assets are stationary mean assets. We hold preferences, prices, mean raw
 earnings, the transfer, and the saving grid fixed.}
\end{table}

\begin{figure}[!htb]
 \centering
 \includegraphics[width=\textwidth]{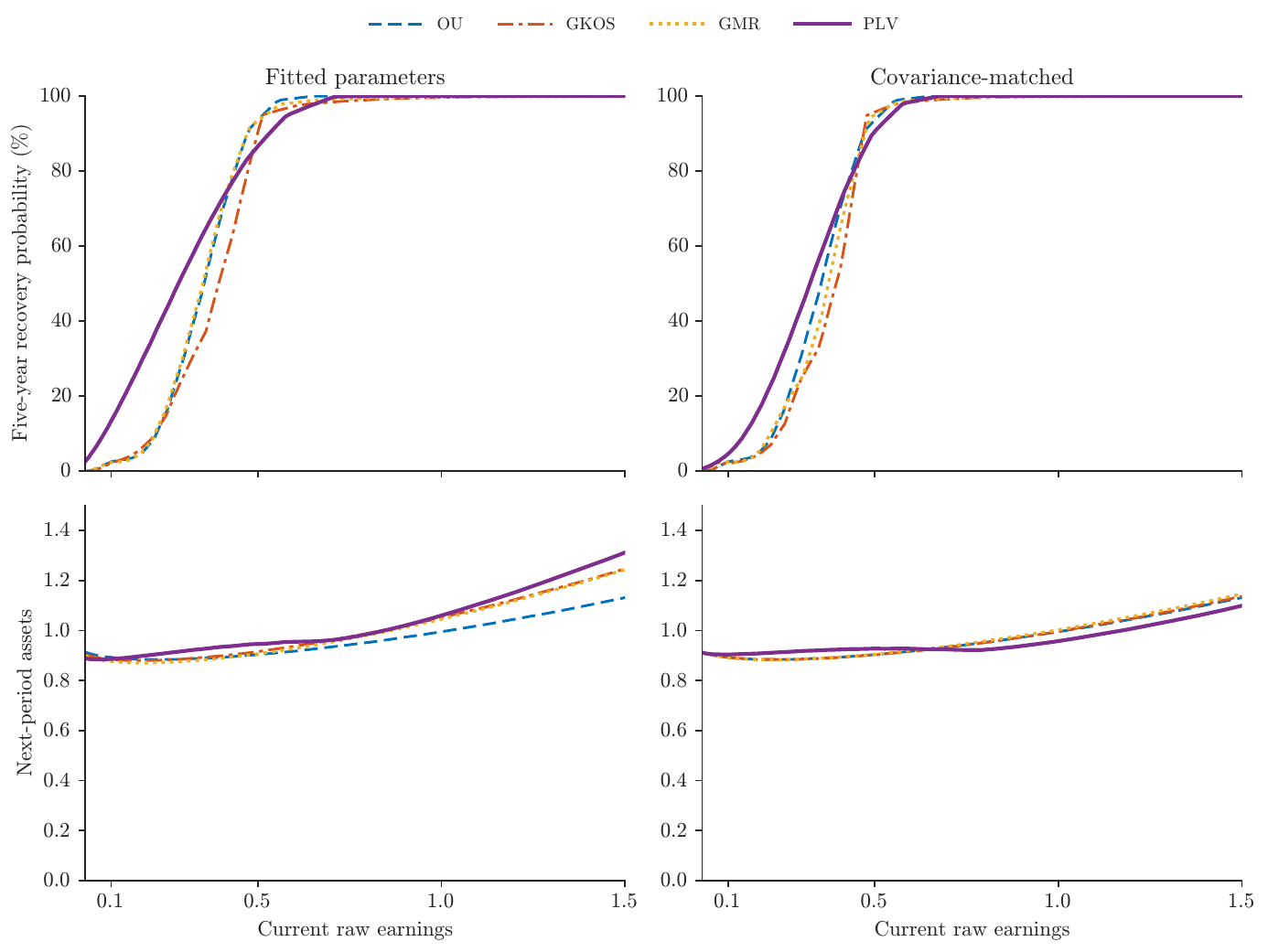}
 \caption{Recovery and saving under fitted and covariance-matched earnings dynamics.}
 \caption*{\footnotesize\textit{Note:} Gaussian-mixture transitory shocks.
 The left column uses fitted parameters; in the right column, each process
 matches the fitted OU log-earnings covariance function. We set current
 centered transitory shocks, including the shared GMR loss, to zero and
 integrate future shocks jointly with the persistent transition. Recovery
 means raw earnings of at least $0.5$ at one of the next five annual
 observations. The bottom row reports next-period assets at current assets
 of one and transfer $b=0.1$.}
 \label{fig:saving-recovery-policy}
\end{figure}

\paragraph{Conditional risk and saving policies}
Figure~\ref{fig:saving-recovery-policy} adds fitted recovery probabilities
and saving-policy profiles to the covariance-matched conditional-risk
comparison in Figure~\ref{fig:saving-recovery}. The left column uses
fitted dynamics. From current earnings of $0.2$, the probability of
earnings of at least $0.5$ within five years is 33 percent under PLV and
7 percent under OU and GMR.

The bottom row shows next-period assets at current assets of one. Around
and above mean earnings, households under all three fitted alternatives
save more than OU households. In the covariance-matched comparison,
PLV households save slightly more than OU households at low earnings but
less around and above mean earnings. We use these profiles to illustrate
conditional saving responses.

\FloatBarrier

\paragraph{Gaussian transitory shocks}

Table~\ref{tab:saving-gaussian} uses the Gaussian transitory-shock
estimates in Table~\ref{tab:model-comparison} and the same household
problem as Table~\ref{tab:saving-application}. We re-estimate each
earnings process when changing the transitory specification. The fitted
alternatives now imply less wealth than OU,
reversing the fitted comparison of Table~\ref{tab:saving-application}. In
the covariance-matched comparison, all alternatives again imply less
wealth and higher CE than OU, and PLV lowers mean assets by 47 percent.
With Gaussian transitory shocks, $N=33$ is too coarse for PLV; from
$N=65$ to 129, mean assets change by at most $1.1\%$.

\begin{table}[!htb]
 \centering
 \caption{Precautionary wealth and welfare with Gaussian transitory shocks.}
 \label{tab:saving-gaussian}
 \begin{tabular}{lrrrr}
   \toprule
   & OU & GKOS & GMR & PLV \\
   \midrule
   Mean assets, fitted & $6.936$ & $3.866$ & $5.855$ & $4.497$ \\
   Mean assets, covariance-matched & $6.936$ & $6.443$ & $6.668$ & $3.709$ \\
   \addlinespace
   CE, fitted & $0.841$ & $0.958$ & $0.821$ & $0.913$ \\
   CE, covariance-matched & $0.841$ & $0.850$ & $0.847$ & $0.904$ \\
   \bottomrule
  \end{tabular}
 \caption*{\footnotesize\textit{Note:} Conventions follow
 Table~\ref{tab:saving-application}, with the Gaussian-shock estimates of
 Table~\ref{tab:model-comparison}.}
\end{table}

\subsection{Calibration to moments of earnings changes and levels}
\label{subsec:calibration-exercise}

This appendix describes how we implement the calibration in
\S\ref{subsec:moment-selection}. We fixed the targets, weights, and
restrictions before running this calibration.

\paragraph{Targets}
Tables~\ref{tab:calibration-targets-1}--\ref{tab:calibration-targets-3}
list the 70 targets with their bootstrap standard errors and the
calibrated model moments. A change belongs to the age group of its
starting observation. The RE tercile of an observation is the tercile of
its residual within its age group, and a change inherits the RE tercile
of its starting observation. \citet{GuvenenKarahanOzkanSong2021} rank
workers by average earnings over the previous five years; our biennial
short panel does not allow this average for most workers. For the mean
subsequent changes in sets~3--4, we compute the deciles of the two-year
change among the changes whose follow-up change over the given horizon
exists.

We do not target medians, which are close to zero in the residuals and
make arc-percent deviations unstable. We also omit targets that our data
or models cannot address. The mean earnings of lifetime-earnings groups
in \citet{GuvenenKarahanOzkanSong2021} require long histories, and their
distribution of nonemployment requires the extensive margin. The slopes
of the tails of the log density in \citet{GuvenenMcKayRyan2026} are too
noisy with 22,619 two-year changes, and their cyclical moments have no
counterpart in our stationary models.

\paragraph{Objective and parameters}
Following \citet{GuvenenKarahanOzkanSong2021}, we minimize a weighted sum
of squared arc-percent deviations, $2(m_k-d_k)/(\abs{m_k}+\abs{d_k})$,
where $m_k$ and $d_k$ are the model and data values of moment $k$. We
assign weight $1/8$ to each set and equal weights within sets. We
calibrate the dynamic and transitory parameters together with $\sigma_0$
and $(a,b)$: 9 parameters for OU, 11 for PLV, and 12 for GKOS and GMR.

\paragraph{Restrictions}
We keep the restrictions of
Appendix~\ref{subsec:calibration-changes}: a lower bound of 0.025 on
the standard deviations of transitory components, GKOS innovation
components, and the GMR Gaussian innovation; $\rho\le0.99$; and, for
PLV, $2\kappa/c_+\ge1.1$ and $2\kappa/c_-\ge0.4$. We also restrict
$\sigma_0\in[0.025,1.5]$ and $a,b\in[1.1,20]$. These intervals contain
the likelihood estimates by a wide margin.

\paragraph{Model moments}
We compute all model moments from simulated panels with the observed
ages and observation gaps of all 4,683 men, as in
\S\ref{subsec:model-fit-diagnostics}, using the same formulas as in the
data. Each simulated individual starts from $\cN(0,\sigma_0^2)$ at age 30
and evolves annually through unobserved years. We hold all random draws
fixed across parameter values. We draw each permanent effect by the
inverse distribution function of the centered asymmetric Laplace law at
a fixed uniform draw, so that the permanent effects move continuously
with $(a,b)$.

OU, GKOS, and GMR evolve exactly at annual dates. For PLV, we simulate
the finite-state chain whose annual transition matrix is
$\exp(Q_\theta)$, where $Q_\theta$ is the generator of
Appendix~\ref{subsec:diffusion-generator} on cubic-spaced nodes over
$[-40,40]$, as in Appendix~\ref{subsec:saving-discretization}. This
matrix integrates over all movements within a year, so the simulation
requires no time step. We draw each annual transition by inverting the
cumulative probabilities of the current row at a fixed uniform draw.
We use 501 nodes in the search, 1,001 nodes in the refinement, and 2,001
nodes for every reported objective. At the PLV likelihood estimate, at
the PLV end point of Appendix~\ref{subsec:calibration-changes}, and at a
stress point with $c_-=0.83$, moments on 501 nodes differ from those on
2,001 nodes by at most 0.12, 0.17, and 0.48 bootstrap standard errors,
respectively, and moments on 1,001 nodes by at most 0.19.

\paragraph{Search and evaluation}
We evaluate the objective at the likelihood estimate, at the end point
of the calibration in Appendix~\ref{subsec:calibration-changes} (with
$\sigma_0$, $a$, and $b$ at their likelihood estimates), and at 64
quasi-random points. We then run the Nelder--Mead method from the
likelihood estimate, from that end point, and from the two best
quasi-random points, with at most 3,000 evaluations each. The search uses
40 simulated panels. We refine the best end point with 100 panels and at
most 500 evaluations. Optimizing over fixed draws can understate
discrepancies from the targets. We therefore evaluate every reported
objective on 200 new panels with draws independent of the search and
refinement.

\paragraph{Alternative calibrations}
To set the tolerance in \S\ref{subsec:moment-selection}, we sum the
squared ratios of the bootstrap standard errors to the absolute data
moments, using the calibration weights. The sum is 0.0092. This
first-order benchmark gives a scale for comparing objectives; it does
not define a joint test of target fit.

To construct the near-optimal ranges in Table~\ref{tab:calibration}, we
consider the end points of the four Nelder--Mead runs and nine search
evaluations spread over persistence, whose search objectives are within
0.0092 of the search minimum. We re-evaluate their objectives on the new
panels and retain candidates whose objectives are less than 0.0092 above
that of the reported calibration. Candidates with lower objectives also
qualify. The ranges cover the reported calibration and the retained
candidates; they are neither confidence intervals nor exhaustive bounds
over calibrations satisfying the tolerance.
Tables~\ref{tab:calibration-alternatives-ou}--\ref{tab:calibration-alternatives-plv}
show the reported calibration, the four run end points, and the
near-optimal candidates with the lowest and highest mean assets. Some
runs stop far from the best objective, for example two of the four PLV
runs.

\paragraph{Likelihood at the calibrated parameters}
We compute the log likelihood as in Table~\ref{tab:model-comparison}.
On a range twice as wide with the same spacing, as in
Table~\ref{tab:estimate-accuracy}, the log likelihood changes by less
than 0.001 for OU and PLV, by 3.6 for GMR, and by 11.9 for GKOS. These
changes are small relative to the differences in
Table~\ref{tab:calibration-moments}.

\begin{table}[!htb]
 \centering
 \caption{Targets in set 1: two-year changes by age and recent earnings.}
 \label{tab:calibration-targets-1}
 \begin{tabular}{lrrrrrr}
  \toprule
  Moment & Data & (SE) & OU & GKOS & GMR & PLV \\
  \midrule
  \multicolumn{7}{l}{\textit{Ages 30--39}} \\
  SD, bottom RE & $0.669$ & (0.019) & $0.628$ & $0.623$ & $0.613$ & $0.631$ \\
  Skewness, bottom RE & $0.49$ & (0.16) & $0.92$ & $0.64$ & $0.72$ & $0.90$ \\
  Excess kurtosis, bottom RE & $4.97$ & (0.64) & $6.63$ & $8.02$ & $7.56$ & $6.28$ \\
  SD, middle RE & $0.382$ & (0.014) & $0.408$ & $0.440$ & $0.418$ & $0.383$ \\
  Skewness, middle RE & $-1.77$ & (0.38) & $-2.23$ & $-2.19$ & $-2.16$ & $-2.03$ \\
  Excess kurtosis, middle RE & $14.55$ & (1.72) & $13.17$ & $13.97$ & $15.70$ & $17.76$ \\
  SD, top RE & $0.455$ & (0.016) & $0.430$ & $0.456$ & $0.467$ & $0.448$ \\
  Skewness, top RE & $-2.40$ & (0.27) & $-2.29$ & $-2.21$ & $-2.22$ & $-2.34$ \\
  Excess kurtosis, top RE & $13.53$ & (2.21) & $11.89$ & $13.00$ & $12.86$ & $13.61$ \\
  \addlinespace
  \multicolumn{7}{l}{\textit{Ages 40--49}} \\
  SD, bottom RE & $0.661$ & (0.022) & $0.629$ & $0.622$ & $0.610$ & $0.645$ \\
  Skewness, bottom RE & $0.81$ & (0.21) & $0.90$ & $0.61$ & $0.70$ & $0.82$ \\
  Excess kurtosis, bottom RE & $6.89$ & (0.95) & $6.55$ & $7.92$ & $7.74$ & $5.70$ \\
  SD, middle RE & $0.392$ & (0.017) & $0.410$ & $0.444$ & $0.426$ & $0.382$ \\
  Skewness, middle RE & $-2.88$ & (0.20) & $-2.25$ & $-2.17$ & $-2.16$ & $-2.00$ \\
  Excess kurtosis, middle RE & $15.33$ & (1.70) & $13.28$ & $14.05$ & $16.10$ & $17.99$ \\
  SD, top RE & $0.463$ & (0.018) & $0.429$ & $0.456$ & $0.465$ & $0.446$ \\
  Skewness, top RE & $-2.75$ & (0.30) & $-2.29$ & $-2.16$ & $-2.20$ & $-2.39$ \\
  Excess kurtosis, top RE & $15.87$ & (2.21) & $12.02$ & $12.55$ & $12.53$ & $13.95$ \\
  \addlinespace
  \multicolumn{7}{l}{\textit{Ages 50--60}} \\
  SD, bottom RE & $0.659$ & (0.024) & $0.631$ & $0.624$ & $0.609$ & $0.657$ \\
  Skewness, bottom RE & $0.28$ & (0.21) & $0.90$ & $0.61$ & $0.72$ & $0.75$ \\
  Excess kurtosis, bottom RE & $5.20$ & (0.51) & $6.59$ & $8.06$ & $7.44$ & $5.34$ \\
  SD, middle RE & $0.486$ & (0.022) & $0.409$ & $0.444$ & $0.425$ & $0.383$ \\
  Skewness, middle RE & $-2.94$ & (0.19) & $-2.25$ & $-2.21$ & $-2.10$ & $-2.03$ \\
  Excess kurtosis, middle RE & $13.92$ & (1.39) & $13.32$ & $13.94$ & $15.38$ & $17.71$ \\
  SD, top RE & $0.531$ & (0.025) & $0.429$ & $0.458$ & $0.461$ & $0.447$ \\
  Skewness, top RE & $-3.00$ & (0.22) & $-2.30$ & $-2.17$ & $-2.18$ & $-2.36$ \\
  Excess kurtosis, top RE & $14.28$ & (1.83) & $11.84$ & $12.90$ & $12.05$ & $13.56$ \\
  \bottomrule
 \end{tabular}
 \caption*{\footnotesize\textit{Note:} Gaussian-mixture transitory shocks. Standard errors come from 500 bootstrap samples of individuals. Model moments use 200 simulated panels with draws independent of the search. RE terciles are bottom, middle, and top.}
\end{table}

\begin{table}[!htb]
 \centering
 \caption{Targets in sets 2--6: changes.}
 \label{tab:calibration-targets-2}
 \begin{tabular}{lrrrrrr}
  \toprule
  Moment & Data & (SE) & OU & GKOS & GMR & PLV \\
  \midrule
  \multicolumn{7}{l}{\textit{Set 2: six-year changes}} \\
  SD, bottom RE & $0.766$ & (0.015) & $0.664$ & $0.689$ & $0.675$ & $0.741$ \\
  Skewness, bottom RE & $0.30$ & (0.12) & $0.65$ & $0.23$ & $0.29$ & $0.26$ \\
  Excess kurtosis, bottom RE & $4.03$ & (0.37) & $5.07$ & $5.90$ & $5.72$ & $3.65$ \\
  SD, middle RE & $0.514$ & (0.013) & $0.463$ & $0.530$ & $0.515$ & $0.454$ \\
  Skewness, middle RE & $-2.26$ & (0.14) & $-1.54$ & $-1.67$ & $-1.56$ & $-1.48$ \\
  Excess kurtosis, middle RE & $10.18$ & (0.83) & $7.83$ & $9.09$ & $9.08$ & $9.65$ \\
  SD, top RE & $0.620$ & (0.017) & $0.484$ & $0.548$ & $0.551$ & $0.493$ \\
  Skewness, top RE & $-1.97$ & (0.13) & $-1.59$ & $-1.63$ & $-1.60$ & $-1.82$ \\
  Excess kurtosis, top RE & $8.41$ & (0.68) & $7.35$ & $8.35$ & $7.81$ & $9.57$ \\
  \addlinespace
  \multicolumn{7}{l}{\textit{Sets 3--4: mean change after extreme two-year changes}} \\
  Next 2 years, bottom decile & $0.401$ & (0.021) & $0.544$ & $0.518$ & $0.504$ & $0.494$ \\
  Next 2 years, top decile & $-0.218$ & (0.014) & $-0.178$ & $-0.205$ & $-0.209$ & $-0.223$ \\
  Next 4 years, bottom decile & $0.438$ & (0.025) & $0.579$ & $0.535$ & $0.517$ & $0.505$ \\
  Next 4 years, top decile & $-0.221$ & (0.019) & $-0.214$ & $-0.210$ & $-0.215$ & $-0.223$ \\
  Next 6 years, bottom decile & $0.453$ & (0.026) & $0.597$ & $0.550$ & $0.530$ & $0.514$ \\
  Next 6 years, top decile & $-0.227$ & (0.022) & $-0.236$ & $-0.216$ & $-0.221$ & $-0.222$ \\
  Next 10 years, bottom decile & $0.531$ & (0.041) & $0.615$ & $0.573$ & $0.550$ & $0.533$ \\
  Next 10 years, top decile & $-0.253$ & (0.032) & $-0.257$ & $-0.224$ & $-0.234$ & $-0.220$ \\
  \addlinespace
  \multicolumn{7}{l}{\textit{Sets 5--6: percentiles and tails of changes}} \\
  P10, 2-year & $-0.424$ & (0.009) & $-0.407$ & $-0.416$ & $-0.393$ & $-0.349$ \\
  P90, 2-year & $0.418$ & (0.009) & $0.407$ & $0.424$ & $0.396$ & $0.327$ \\
  P10, 4-year & $-0.538$ & (0.013) & $-0.502$ & $-0.483$ & $-0.499$ & $-0.463$ \\
  P90, 4-year & $0.508$ & (0.010) & $0.502$ & $0.493$ & $0.501$ & $0.428$ \\
  P10, 6-year & $-0.613$ & (0.017) & $-0.548$ & $-0.547$ & $-0.578$ & $-0.544$ \\
  P90, 6-year & $0.557$ & (0.011) & $0.548$ & $0.553$ & $0.576$ & $0.499$ \\
  Share of two-year changes below $-1.2$ & $0.028$ & (0.001) & $0.023$ & $0.025$ & $0.025$ & $0.025$ \\
  Share of two-year changes above $1.2$ & $0.021$ & (0.001) & $0.023$ & $0.021$ & $0.023$ & $0.025$ \\
  \bottomrule
 \end{tabular}
 \caption*{\footnotesize\textit{Note:} Gaussian-mixture transitory shocks. Standard errors come from 500 bootstrap samples of individuals. Model moments use 200 simulated panels with draws independent of the search. Set 2 splits six-year changes by RE tercile.}
\end{table}

\begin{table}[!htb]
 \centering
 \caption{Targets in sets 7--8: levels by age.}
 \label{tab:calibration-targets-3}
 \begin{tabular}{lrrrrrr}
  \toprule
  Moment & Data & (SE) & OU & GKOS & GMR & PLV \\
  \midrule
  \multicolumn{7}{l}{\textit{Ages 30--39}} \\
  Variance & $0.465$ & (0.013) & $0.480$ & $0.432$ & $0.403$ & $0.369$ \\
  Autocovariance, lag 2 & $0.303$ & (0.011) & $0.348$ & $0.303$ & $0.281$ & $0.254$ \\
  Autocovariance, lag 4 & $0.277$ & (0.012) & $0.329$ & $0.287$ & $0.268$ & $0.250$ \\
  Autocovariance, lag 6 & $0.261$ & (0.014) & $0.318$ & $0.273$ & $0.257$ & $0.246$ \\
  P10 & $-0.753$ & (0.023) & $-0.855$ & $-0.730$ & $-0.711$ & $-0.718$ \\
  P90 & $0.723$ & (0.016) & $0.730$ & $0.716$ & $0.719$ & $0.645$ \\
  \addlinespace
  \multicolumn{7}{l}{\textit{Ages 40--49}} \\
  Variance & $0.532$ & (0.018) & $0.483$ & $0.517$ & $0.506$ & $0.516$ \\
  Autocovariance, lag 2 & $0.367$ & (0.017) & $0.351$ & $0.380$ & $0.377$ & $0.400$ \\
  Autocovariance, lag 4 & $0.329$ & (0.018) & $0.331$ & $0.355$ & $0.357$ & $0.393$ \\
  Autocovariance, lag 6 & $0.302$ & (0.018) & $0.319$ & $0.333$ & $0.338$ & $0.387$ \\
  P10 & $-0.804$ & (0.020) & $-0.859$ & $-0.827$ & $-0.822$ & $-0.875$ \\
  P90 & $0.758$ & (0.017) & $0.734$ & $0.772$ & $0.818$ & $0.744$ \\
  \addlinespace
  \multicolumn{7}{l}{\textit{Ages 50--60}} \\
  Variance & $0.675$ & (0.028) & $0.481$ & $0.542$ & $0.549$ & $0.661$ \\
  Autocovariance, lag 2 & $0.445$ & (0.026) & $0.348$ & $0.400$ & $0.415$ & $0.530$ \\
  Autocovariance, lag 4 & $0.368$ & (0.024) & $0.328$ & $0.372$ & $0.389$ & $0.507$ \\
  Autocovariance, lag 6 & $0.342$ & (0.025) & $0.316$ & $0.346$ & $0.366$ & $0.484$ \\
  P10 & $-0.924$ & (0.035) & $-0.856$ & $-0.861$ & $-0.871$ & $-0.973$ \\
  P90 & $0.849$ & (0.025) & $0.733$ & $0.794$ & $0.858$ & $0.817$ \\
  \bottomrule
 \end{tabular}
 \caption*{\footnotesize\textit{Note:} Gaussian-mixture transitory shocks. Standard errors come from 500 bootstrap samples of individuals. Model moments use 200 simulated panels with draws independent of the search. Autocovariances pair an observation in the age group with the observation the given number of years later.}
\end{table}

\begin{table}[!htb]
 \centering
 \caption{Alternative calibrations of OU.}
 \label{tab:calibration-alternatives-ou}
 \begin{tabular}{lrrrrrr}
  \toprule
  & (1) & (2) & (3) & (4) & (5) & (6) \\
  \midrule
  $\rho$ & $0.772$ & $0.774$ & $0.866$ & $0.940$ & $0.909$ & $0.688$ \\
  $\sigma$ & $0.207$ & $0.207$ & $0.157$ & $0.161$ & $0.169$ & $0.226$ \\
  $p_\epsilon$ & $0.066$ & $0.064$ & $0.064$ & $0.074$ & $0.075$ & $0.059$ \\
  $\delta_\epsilon$ & $0.736$ & $0.763$ & $0.794$ & $0.605$ & $0.607$ & $0.830$ \\
  $\sigma_{\epsilon,1}$ & $0.999$ & $1.000$ & $0.966$ & $1.095$ & $1.066$ & $1.032$ \\
  $\sigma_{\epsilon,2}$ & $0.026$ & $0.026$ & $0.116$ & $0.047$ & $0.025$ & $0.026$ \\
  $\sigma_0$ & $0.264$ & $0.242$ & $0.211$ & $0.277$ & $0.145$ & $0.205$ \\
  $a$ & $4.241$ & $3.967$ & $3.620$ & $4.068$ & $2.811$ & $2.140$ \\
  $b$ & $2.027$ & $2.059$ & $2.203$ & $2.757$ & $2.932$ & $2.668$ \\
  \addlinespace
  $V$ & $0.083$ & $0.083$ & $0.085$ & $0.209$ & $0.151$ & $0.068$ \\
  Objective & $0.0406$ & $0.0407$ & $0.0413$ & $0.0422$ & $0.0429$ & $0.0470$ \\
  Near-optimal & yes & yes & yes & yes & yes & yes \\
  \addlinespace
  Mean assets & $1.63$ & $1.63$ & $1.73$ & $3.65$ & $2.81$ & $1.37$ \\
  CE & $1.029$ & $1.029$ & $1.018$ & $0.924$ & $0.969$ & $1.041$ \\
  \bottomrule
 \end{tabular}
 \caption*{\footnotesize\textit{Note:} Gaussian-mixture transitory shocks. Column (1) is the reported calibration; columns (2)--(5) are the end points of the four Nelder--Mead runs, ordered by the final objective; the remaining columns are the near-optimal values with the lowest and highest mean assets, when they differ from earlier columns. $V$ is the stationary variance of the persistent state. Objectives use 200 simulated panels with draws independent of the search and refinement. A column is near-optimal if its objective is less than 0.0092 above that of column (1), including columns with lower objectives. Mean assets and CE follow Table~\ref{tab:saving-application}.}
\end{table}

\begin{table}[!htb]
 \centering
 \caption{Alternative calibrations of GKOS.}
 \label{tab:calibration-alternatives-gkos}
 \begin{tabular}{lrrrrrr}
  \toprule
  & (1) & (2) & (3) & (4) & (5) & (6) \\
  \midrule
  $\rho$ & $0.941$ & $0.939$ & $0.927$ & $0.849$ & $0.963$ & $0.980$ \\
  $p_\eta$ & $0.015$ & $0.013$ & $0.012$ & $0.014$ & $0.110$ & $0.013$ \\
  $\delta_\eta$ & $0.421$ & $0.389$ & $0.586$ & $1.255$ & $0.110$ & $0.297$ \\
  $\sigma_{\eta,1}$ & $1.191$ & $1.199$ & $0.980$ & $0.877$ & $0.380$ & $1.209$ \\
  $\sigma_{\eta,2}$ & $0.094$ & $0.098$ & $0.129$ & $0.052$ & $0.044$ & $0.060$ \\
  $p_\epsilon$ & $0.047$ & $0.049$ & $0.080$ & $0.045$ & $0.077$ & $0.053$ \\
  $\delta_\epsilon$ & $0.860$ & $0.834$ & $0.395$ & $0.341$ & $0.601$ & $0.777$ \\
  $\sigma_{\epsilon,1}$ & $1.024$ & $1.024$ & $1.013$ & $1.319$ & $0.984$ & $1.024$ \\
  $\sigma_{\epsilon,2}$ & $0.165$ & $0.161$ & $0.025$ & $0.180$ & $0.112$ & $0.176$ \\
  $\sigma_0$ & $0.286$ & $0.287$ & $0.288$ & $0.186$ & $0.325$ & $0.252$ \\
  $a$ & $3.615$ & $3.260$ & $4.211$ & $2.922$ & $3.951$ & $3.785$ \\
  $b$ & $3.397$ & $3.560$ & $2.728$ & $2.496$ & $3.214$ & $4.051$ \\
  \addlinespace
  $V$ & $0.283$ & $0.261$ & $0.233$ & $0.123$ & $0.260$ & $0.590$ \\
  Objective & $0.0206$ & $0.0213$ & $0.0250$ & $0.0281$ & $0.0374$ & $0.0291$ \\
  Near-optimal & yes & yes & yes & yes & no & yes \\
  \addlinespace
  Mean assets & $4.69$ & $4.44$ & $3.77$ & $2.20$ & $3.63$ & $8.16$ \\
  CE & $0.893$ & $0.903$ & $0.925$ & $1.000$ & $0.888$ & $0.704$ \\
  \bottomrule
 \end{tabular}
 \caption*{\footnotesize\textit{Note:} Conventions follow Table~\ref{tab:calibration-alternatives-ou}.}
\end{table}

\begin{table}[!htb]
 \centering
 \caption{Alternative calibrations of GMR.}
 \label{tab:calibration-alternatives-gmr}
 \begin{tabular}{lrrrrrrr}
  \toprule
  & (1) & (2) & (3) & (4) & (5) & (6) & (7) \\
  \midrule
  $\rho$ & $0.957$ & $0.960$ & $0.946$ & $0.967$ & $0.988$ & $0.873$ & $0.984$ \\
  $\sigma_\eta$ & $0.145$ & $0.140$ & $0.025$ & $0.121$ & $0.107$ & $0.029$ & $0.124$ \\
  $p_\zeta$ & $0.993$ & $0.992$ & $0.592$ & $0.973$ & $0.713$ & $0.591$ & $0.992$ \\
  $\lambda$ & $0.539$ & $0.542$ & $6.312$ & $1.941$ & $3.479$ & $6.016$ & $0.578$ \\
  $\psi$ & $0.848$ & $0.852$ & $0.890$ & $0.518$ & $0.166$ & $0.943$ & $0.836$ \\
  $p_\epsilon$ & $0.076$ & $0.075$ & $0.049$ & $0.103$ & $0.195$ & $0.047$ & $0.080$ \\
  $\delta_\epsilon$ & $0.499$ & $0.499$ & $0.665$ & $0.338$ & $0.048$ & $0.397$ & $0.402$ \\
  $\sigma_{\epsilon,1}$ & $1.025$ & $1.025$ & $1.213$ & $0.865$ & $0.468$ & $1.324$ & $1.025$ \\
  $\sigma_{\epsilon,2}$ & $0.089$ & $0.094$ & $0.149$ & $0.030$ & $0.077$ & $0.146$ & $0.079$ \\
  $\sigma_0$ & $0.296$ & $0.281$ & $0.271$ & $0.287$ & $0.397$ & $0.301$ & $0.242$ \\
  $a$ & $3.609$ & $3.757$ & $8.631$ & $3.138$ & $7.653$ & $7.092$ & $3.670$ \\
  $b$ & $4.391$ & $4.641$ & $2.687$ & $4.401$ & $4.509$ & $2.176$ & $4.501$ \\
  \addlinespace
  $V$ & $0.346$ & $0.365$ & $0.287$ & $0.321$ & $0.606$ & $0.163$ & $0.754$ \\
  Objective & $0.0197$ & $0.0198$ & $0.0223$ & $0.0259$ & $0.0482$ & $0.0286$ & $0.0278$ \\
  Near-optimal & yes & yes & yes & yes & no & yes & yes \\
  \addlinespace
  Mean assets & $4.98$ & $5.08$ & $4.39$ & $4.31$ & $5.15$ & $2.87$ & $7.22$ \\
  CE & $0.854$ & $0.844$ & $0.882$ & $0.855$ & $0.690$ & $0.974$ & $0.664$ \\
  \bottomrule
 \end{tabular}
 \caption*{\footnotesize\textit{Note:} Conventions follow Table~\ref{tab:calibration-alternatives-ou}. The loss probability is $1-p_\zeta$.}
\end{table}

\begin{table}[!htb]
 \centering
 \caption{Alternative calibrations of PLV.}
 \label{tab:calibration-alternatives-plv}
 \begin{tabular}{lrrrrrr}
  \toprule
  & (1) & (2) & (3) & (4) & (5) & (6) \\
  \midrule
  $\rho$ & $0.987$ & $0.987$ & $0.980$ & $0.588$ & $0.933$ & $0.986$ \\
  $\sigma$ & $0.101$ & $0.101$ & $0.059$ & $0.121$ & $0.089$ & $0.095$ \\
  $c_+$ & $0.000$ & $0.000$ & $0.030$ & $0.501$ & $0.000$ & $0.001$ \\
  $c_-$ & $0.066$ & $0.065$ & $0.082$ & $1.442$ & $0.347$ & $0.071$ \\
  $p_\epsilon$ & $0.084$ & $0.084$ & $0.095$ & $0.004$ & $0.095$ & $0.094$ \\
  $\delta_\epsilon$ & $0.400$ & $0.404$ & $0.409$ & $-0.125$ & $-0.041$ & $0.392$ \\
  $\sigma_{\epsilon,1}$ & $1.072$ & $1.072$ & $0.984$ & $0.100$ & $0.884$ & $0.985$ \\
  $\sigma_{\epsilon,2}$ & $0.025$ & $0.025$ & $0.025$ & $0.025$ & $0.025$ & $0.025$ \\
  $\sigma_0$ & $0.298$ & $0.284$ & $0.278$ & $0.853$ & $0.296$ & $0.197$ \\
  $a$ & $5.476$ & $5.560$ & $3.706$ & $6.070$ & $3.868$ & $6.202$ \\
  $b$ & $3.878$ & $3.604$ & $3.543$ & $1.879$ & $3.034$ & $2.744$ \\
  \addlinespace
  $V$ & $1.167$ & $1.186$ & $0.694$ & $0.220$ & $0.416$ & $1.037$ \\
  Objective & $0.0276$ & $0.0279$ & $0.0292$ & $0.0878$ & $0.1663$ & $0.0297$ \\
  Near-optimal & yes & yes & yes & no & no & yes \\
  \addlinespace
  Mean assets & $3.76$ & $3.80$ & $9.26$ & $2.29$ & $2.77$ & $3.14$ \\
  CE & $0.677$ & $0.673$ & $0.717$ & $1.009$ & $0.914$ & $0.709$ \\
  \bottomrule
 \end{tabular}
 \caption*{\footnotesize\textit{Note:} Conventions follow Table~\ref{tab:calibration-alternatives-ou}. PLV uses 2,001 nodes.}
\end{table}

\FloatBarrier

\subsection{Calibration to moments of earnings changes only}
\label{subsec:calibration-changes}

We first calibrated each model to 18 moments of earnings changes only. We
did this calibration first because it is computationally cheap:
permanent effects cancel in changes, so the targets do not involve
$\sigma_0$, $a$, or $b$, and the moments of a discretized PLV process have
closed forms. We then ran the calibration in
\S\ref{subsec:moment-selection} to follow the designs of
\citet{GuvenenKarahanOzkanSong2021} and \citet{GuvenenMcKayRyan2026} more
closely, including their moments of earnings levels. We report both
calibrations. For each calibration, we fixed the targets and weights
before computing its results. We added the parameter restrictions below
after pilot runs of this calibration, for the reasons we give there.

\paragraph{Targets}
We adapt the one-, three-, and five-year horizons in the two papers to
our biennial panel and target 18 moments of two-, four-, and six-year
changes in the earnings residuals in \eqref{eq:PSID-residualization}.
Table~\ref{tab:calibration-a-moments} groups the targets into five sets:
two-year moments, six-year moments, the 10th and 90th percentiles, tail
probabilities, and mean subsequent changes after extreme changes. Sets
(i)--(ii) follow \citet{GuvenenKarahanOzkanSong2021}, sets (iii)--(iv)
follow \citet{GuvenenMcKayRyan2026}, and set (v) approximates the impulse
responses in \citet{GuvenenKarahanOzkanSong2021}. We start simulated
histories from the age-30 law estimated by the HMM likelihood.

\paragraph{Objective and parameters}
The targets and their groups appear in Table~\ref{tab:calibration-a-moments}.
Following \citet{GuvenenKarahanOzkanSong2021}, we minimize a weighted sum
of squared arc-percent deviations, $2(m_k-d_k)/(\abs{m_k}+\abs{d_k})$,
where $m_k$ and $d_k$ are the model and data values of moment $k$. We
assign weight $1/5$ to each set and equal weights within sets.

Permanent effects cancel in changes, so the targets do not identify the
permanent-effect rates $(a,b)$. We retain their likelihood estimates in
Table~\ref{tab:model-comparison}. As a separate calibration choice, we
also hold the initial dispersion $\sigma_0$ at its likelihood estimate.
We calibrate the remaining dynamic and transitory parameters: 6 for OU,
8 for PLV, and 9 for GKOS and GMR.

\paragraph{Model moments}
For OU, GKOS, and GMR, we compute model moments from simulated panels with
the observed ages and observation gaps of all 4,683 men, as in
\S\ref{subsec:model-fit-diagnostics}. Each simulated individual starts from
the estimated Gaussian law at age 30 and evolves annually through
unobserved years. We compute each moment using the same formulas as in
the data. We use simulation because the coarse grids considered here
poorly resolve persistent innovation components much narrower than the
grid spacing. Such components arise in GKOS, whose innovation mixture
can place one component near the lower bound of 0.025.

For PLV, simulating the diffusion with fine time steps at every parameter
value is costly, so we compute exact moments of a discretized process
instead. We construct transitions from the fine generator as in
Appendix~\ref{subsec:saving-discretization}. For calibration, the grid spans
$\pm\sqrt{N-1}$ times the largest standard deviation of the state over
ages 30--60. We propagate the estimated age-30 law through this chain.
Each change is a finite mixture of Gaussian distributions. We evaluate
its moments, tail probabilities, and conditional means using closed
forms and find quantiles by numerically inverting the mixture
distribution function. We pool pairs and triples with the observed
counts by starting age.

These moment calculations use the age schedules in the earnings data.
For the saving experiment, we use each calibrated process's stationary
earnings law, as in \S\ref{sec:saving-application}.

\paragraph{Restrictions}
We impose a lower bound of 0.025 on transitory component standard
deviations, GKOS innovation component standard deviations, and the GMR
Gaussian innovation standard deviation. The bound matches the filtering
grid spacing used for Table~\ref{tab:model-comparison} and helps us
evaluate the likelihood at the calibrated parameters accurately.

We also restrict annual persistence to $\rho\le0.99$ and, for PLV,
require $2\kappa/c_+\ge1.1$. These restrictions limit persistence and
ensure that the stationary earnings distribution has a finite mean,
with a Pareto upper-tail exponent of at least 1.1 for PLV. Without these
restrictions, the pilot calibration pushes PLV toward a unit root
($\rho=0.995$) with an upper-tail exponent of 0.36. Stationary mean
earnings are then infinite, which prevents the normalization to mean
earnings of one used in \S\ref{sec:saving-application}.

For PLV, we also require $2\kappa/c_-\ge0.4$. This bound regularizes
extrapolation to the stationary lower tail beyond the observed ages
30--60. Without the bound, the calibrated exponent is 0.15, and the
stationary standard deviation of the persistent state is 1.9. For
context, the standard deviation of the observed log-earnings residual,
which also includes permanent effects and transitory shocks, is
0.68--0.82 at ages 30--60. We set the bound below the likelihood estimate,
0.49, so that the restriction retains that estimate.

\paragraph{Search and evaluation}
We evaluate the objective at the likelihood estimate and at 50
quasi-random starting values. We then run the Nelder--Mead method from
the likelihood estimate and from the three best of these values.
For OU, GKOS, and GMR, we hold random draws fixed across parameter values,
search with 40 simulated panels, and refine the best run end point with
100 panels. Optimizing over these draws can understate discrepancies
from the targets. We therefore evaluate every reported objective for
these models on 200 new panels with draws independent of the search and
refinement.

For PLV, we search with $N=129$ nodes and refine the best point with
$N=257$.\footnote{For PLV, the objective on 129 nodes differs from that on
257 nodes: at the best search point, it is 0.087 on 129 nodes and 0.065 on
257 nodes. The refinement on 257 nodes gives the reported calibration, but
the near-optimal points for PLV come from the coarser search. We therefore
re-evaluate them on 257 nodes.} At the calibrated PLV parameters, 40
simulated panels with 100 time steps per year give an objective of 0.065,
compared with 0.062 for the discretized process. We retain the refinement
result as the reported calibration for each model and use the final
evaluations to compare its objective with those of the other candidates.

\paragraph{Alternative calibrations}
To set the tolerance for this calibration, we sum the
squared ratios of the bootstrap standard errors in
Table~\ref{tab:calibration-a-moments} to the absolute data moments, using
the calibration weights. The sum is approximately 0.006. This first-order
benchmark gives a scale for comparing objectives; it does not define a
joint test of target fit.

To construct the near-optimal ranges in Table~\ref{tab:calibration-a}, we
consider the end points of the four Nelder--Mead runs and nine search
evaluations spread over persistence, whose search objectives are within
0.006 of the search minimum. This screening uses the search criterion.

We re-evaluate candidate objectives using the same procedure as for the
reported calibration: independent simulation for OU, GKOS, and GMR, and
the finer chain for PLV. We then retain candidates whose final objectives
are less than 0.006 above the reported calibration's final objective.
Candidates with lower final objectives also qualify. The ranges cover
the reported calibration and evaluated candidates with saving results;
they are neither confidence intervals nor exhaustive bounds over
calibrations satisfying the tolerance.
Tables~\ref{tab:calibration-a-alternatives-ou}--\ref{tab:calibration-a-alternatives-plv}
show the reported calibration, the four run end points, and the
near-optimal candidates with the lowest and highest mean assets.

\paragraph{Results}
Table~\ref{tab:calibration-a-moments} shows that calibration lowers the
moment objective in every model. GKOS and GMR have the lowest calibrated
objective values, 0.0010 and 0.0028, below the tolerance of 0.006. GKOS
closely matches the standard deviation, skewness, and excess kurtosis of
two-year changes. GMR closely matches skewness and excess kurtosis at the
two- and six-year horizons. Both models also generate larger mean
reversals after extreme losses than after extreme gains, as in the data.
OU and PLV leave larger discrepancies. OU matches the excess kurtosis of
two-year changes, but its symmetric changes prevent it from matching the
skewness targets. Calibrated PLV produces excess kurtosis that rises
between the two- and six-year horizons, whereas excess kurtosis falls in
the data.

The close fit to the targeted moments does not extend to earnings
histories. Calibration lowers the log likelihood by 700 to 1,500 points
relative to the likelihood estimates. Calibrated GKOS and GMR have log
likelihoods of $-14{,}361$ and $-14{,}087$, below every
likelihood-estimated specification with mixture shocks in
Table~\ref{tab:model-comparison}.

Table~\ref{tab:calibration-a} reports saving and welfare. Calibrated GKOS
and GMR imply mean assets of 18.1 and 13.2, more than twice their fitted
values, and CE of 0.44 and 0.49, compared with 0.75 and 0.80. Among the
near-optimal candidates, GKOS implies mean assets between 6.7 and 18.2,
and GMR between 4.8 and 13.5. The reported GKOS and GMR calibrations have
persistence near 0.99 and stationary persistent variances of about 1.5
(Tables~\ref{tab:calibration-a-alternatives-gkos}--\ref{tab:calibration-a-alternatives-gmr}).
Moments of changes leave these features weakly determined. In
\S\ref{subsec:moment-selection}, the moments of levels exclude such
values: no near-optimal calibration of GKOS or GMR has a stationary
persistent variance above 0.76. Both calibrations nevertheless lead to
the same conclusions. Calibrations with close fit to their targets have
log likelihoods below every likelihood-estimated specification with
mixture shocks, and calibrations with similar target fit imply mean
assets that differ by a factor of two or more.

\begin{table}[!htb]
 \centering
 \caption{Calibration to moments of changes only: targets, calibrated moments, and likelihoods.}
 \label{tab:calibration-a-moments}
 \begin{tabular}{lrrrrrr}
  \toprule
  Moment & Data & (SE) & OU & GKOS & GMR & PLV \\
  \midrule
  \multicolumn{7}{l}{\textit{(i) Two-year changes}} \\
  SD & $0.536$ & (0.008) & $0.516$ & $0.535$ & $0.517$ & $0.517$ \\
  Skewness & $-0.499$ & (0.096) & $0.005$ & $-0.490$ & $-0.507$ & $-0.233$ \\
  Excess kurtosis & $10.82$ & (0.60) & $10.93$ & $10.80$ & $11.64$ & $8.57$ \\
  \addlinespace
  \multicolumn{7}{l}{\textit{(ii) Six-year changes}} \\
  SD & $0.653$ & (0.011) & $0.592$ & $0.688$ & $0.633$ & $0.721$ \\
  Skewness & $-0.640$ & (0.102) & $-0.006$ & $-0.624$ & $-0.653$ & $-0.646$ \\
  Excess kurtosis & $7.22$ & (0.41) & $6.27$ & $6.93$ & $7.16$ & $10.77$ \\
  \addlinespace
  \multicolumn{7}{l}{\textit{(iii) Percentiles of changes}} \\
  P10, two-year & $-0.424$ & (0.008) & $-0.390$ & $-0.422$ & $-0.402$ & $-0.454$ \\
  P90, two-year & $0.418$ & (0.009) & $0.390$ & $0.432$ & $0.406$ & $0.448$ \\
  P10, four-year & $-0.538$ & (0.013) & $-0.506$ & $-0.493$ & $-0.530$ & $-0.585$ \\
  P90, four-year & $0.508$ & (0.010) & $0.505$ & $0.500$ & $0.531$ & $0.578$ \\
  P10, six-year & $-0.613$ & (0.017) & $-0.583$ & $-0.587$ & $-0.628$ & $-0.666$ \\
  P90, six-year & $0.557$ & (0.012) & $0.582$ & $0.570$ & $0.624$ & $0.661$ \\
  \addlinespace
  \multicolumn{7}{l}{\textit{(iv) Tails of two-year changes}} \\
  Share below $-1.2$ & $0.028$ & (0.001) & $0.024$ & $0.027$ & $0.025$ & $0.024$ \\
  Share above $1.2$ & $0.021$ & (0.001) & $0.024$ & $0.021$ & $0.022$ & $0.022$ \\
  \addlinespace
  \multicolumn{7}{l}{\textit{(v) Mean change after extreme two-year changes}} \\
  Next two years, bottom decile & $0.401$ & (0.020) & $0.470$ & $0.420$ & $0.437$ & $0.292$ \\
  Next two years, top decile & $-0.218$ & (0.015) & $-0.225$ & $-0.217$ & $-0.224$ & $-0.198$ \\
  Next four years, bottom decile & $0.438$ & (0.024) & $0.484$ & $0.430$ & $0.444$ & $0.378$ \\
  Next four years, top decile & $-0.221$ & (0.018) & $-0.241$ & $-0.222$ & $-0.228$ & $-0.182$ \\
  \midrule
  Objective, likelihood estimate & & & $0.736$ & $0.471$ & $0.088$ & $0.336$ \\
  Objective, calibrated & & & $0.534$ & $0.0010$ & $0.0028$ & $0.062$ \\
  \addlinespace
  LL, likelihood estimate & & & $-13{,}694$ & $-12{,}850$ & $-13{,}038$ & $-12{,}765$ \\
  LL, calibrated & & & $-14{,}476$ & $-14{,}361$ & $-14{,}087$ & $-13{,}538$ \\
  \bottomrule
 \end{tabular}
 \caption*{\footnotesize\textit{Note:} Gaussian-mixture transitory shocks.
 Changes are two-, four-, and
 six-year changes of the residuals in \eqref{eq:PSID-residualization}.
 Standard errors come from 500 bootstrap samples of individuals. Model
 moments for OU, GKOS, and GMR use 200 simulated panels with draws
 independent of the search; PLV moments are exact for the discretized
 process with $N=257$ nodes. The objective is the weighted sum of squared
 arc-percent deviations from the 18 targets. LL is the log likelihood, computed as in
 Table~\ref{tab:model-comparison}.}
\end{table}

\begin{table}[!htb]
 \centering
 \caption{Saving and welfare under the calibration to moments of changes only.}
 \label{tab:calibration-a}
 \begin{tabular}{lrrrr}
  \toprule
  & OU & GKOS & GMR & PLV \\
  \midrule
  Mean assets, fitted & $4.80$ & $7.37$ & $5.59$ & $10.51$ \\
  Mean assets, calibrated & $4.31$ & $18.05$ & $13.25$ & $5.30$ \\
  Mean assets, near-optimal & 2.4--4.3 & 6.7--18.2 & 4.8--13.5 & 5.3--5.9 \\
  \addlinespace
  CE, fitted & $0.793$ & $0.746$ & $0.802$ & $0.739$ \\
  CE, calibrated & $0.903$ & $0.443$ & $0.486$ & $0.825$ \\
  CE, near-optimal & 0.90--0.99 & 0.44--0.84 & 0.48--0.88 & 0.80--0.82 \\
  \bottomrule
 \end{tabular}
 \caption*{\footnotesize\textit{Note:} Gaussian-mixture transitory shocks.
 Mean assets and CE follow Table~\ref{tab:saving-application}. Near-optimal ranges
 cover the reported calibration and evaluated candidates with objectives
 less than 0.006 above it
 (Tables~\ref{tab:calibration-a-alternatives-ou}--\ref{tab:calibration-a-alternatives-plv}).}
\end{table}

\begin{table}[!htb]
 \centering
 \caption{Alternative calibrations of OU to moments of changes only.}
 \label{tab:calibration-a-alternatives-ou}
 \begin{tabular}{lrrrrrr}
  \toprule
  & (1) & (2) & (3) & (4) & (5) & (6) \\
  \midrule
  $\rho$ & $0.942$ & $0.887$ & $0.966$ & $0.976$ & $0.986$ & $0.872$ \\
  $\sigma$ & $0.170$ & $0.171$ & $0.141$ & $0.145$ & $0.154$ & $0.172$ \\
  $p_\epsilon$ & $0.075$ & $0.074$ & $0.090$ & $0.092$ & $0.085$ & $0.076$ \\
  $\delta_\epsilon$ & $0.431$ & $0.236$ & $0.530$ & $0.359$ & $-0.101$ & $0.242$ \\
  $\sigma_{\epsilon,1}$ & $1.104$ & $1.131$ & $0.930$ & $0.910$ & $1.056$ & $1.104$ \\
  $\sigma_{\epsilon,2}$ & $0.050$ & $0.077$ & $0.068$ & $0.025$ & $0.032$ & $0.070$ \\
  \addlinespace
  $V$ & $0.242$ & $0.122$ & $0.283$ & $0.436$ & $0.856$ & $0.109$ \\
  Objective & $0.5336$ & $0.5367$ & $0.5490$ & $0.5493$ & $0.5686$ & $0.5373$ \\
  Near-optimal & yes & yes & no & no & no & yes \\
  \addlinespace
  Mean assets & $4.31$ & $2.71$ & $4.03$ & $5.61$ & $9.82$ & $2.45$ \\
  CE & $0.903$ & $0.984$ & $0.871$ & $0.782$ & $0.594$ & $0.995$ \\
  \bottomrule
 \end{tabular}
 \caption*{\footnotesize\textit{Note:} Gaussian-mixture transitory shocks.
 Column (1) is the reported calibration; columns (2)--(5) are the end points
 of the four Nelder--Mead runs, ordered by the final objective; the remaining columns
 are the near-optimal values with the lowest and highest mean assets, when
 they differ from earlier columns. $V$ is the stationary variance of the
 persistent state. Objectives use 200 simulated panels with draws
 independent of the search and refinement. A column is near-optimal if its
 objective is less than 0.006 above that of column (1), including columns
 with lower objectives. For OU, the near-optimal candidate with the highest
 mean assets is the run end point in column (2). Mean assets and CE follow
 Table~\ref{tab:saving-application}.}
\end{table}

\begin{table}[!htb]
 \centering
 \caption{Alternative calibrations of GKOS to moments of changes only.}
 \label{tab:calibration-a-alternatives-gkos}
 \begin{tabular}{lrrrrrrr}
  \toprule
  & (1) & (2) & (3) & (4) & (5) & (6) & (7) \\
  \midrule
  $\rho$ & $0.983$ & $0.982$ & $0.959$ & $0.927$ & $0.982$ & $0.943$ & $0.983$ \\
  $p_\eta$ & $0.030$ & $0.031$ & $0.022$ & $0.027$ & $0.036$ & $0.032$ & $0.031$ \\
  $\delta_\eta$ & $0.292$ & $0.303$ & $0.374$ & $0.977$ & $0.961$ & $0.401$ & $0.303$ \\
  $\sigma_{\eta,1}$ & $1.220$ & $1.218$ & $1.316$ & $0.486$ & $0.286$ & $1.044$ & $1.213$ \\
  $\sigma_{\eta,2}$ & $0.063$ & $0.062$ & $0.097$ & $0.025$ & $0.114$ & $0.054$ & $0.061$ \\
  $p_\epsilon$ & $0.039$ & $0.039$ & $0.052$ & $0.044$ & $0.049$ & $0.046$ & $0.038$ \\
  $\delta_\epsilon$ & $0.727$ & $0.737$ & $0.529$ & $0.575$ & $0.452$ & $0.738$ & $0.758$ \\
  $\sigma_{\epsilon,1}$ & $1.024$ & $1.024$ & $1.025$ & $1.238$ & $1.380$ & $1.021$ & $1.027$ \\
  $\sigma_{\epsilon,2}$ & $0.181$ & $0.181$ & $0.156$ & $0.160$ & $0.027$ & $0.162$ & $0.183$ \\
  \addlinespace
  $V$ & $1.529$ & $1.487$ & $0.615$ & $0.231$ & $1.347$ & $0.385$ & $1.555$ \\
  Objective & $0.0010$ & $0.0009$ & $0.0015$ & $0.0092$ & $0.0249$ & $0.0048$ & $0.0009$ \\
  Near-optimal & yes & yes & yes & no & no & yes & yes \\
  \addlinespace
  Mean assets & $18.05$ & $17.94$ & $10.55$ & $5.27$ & $13.91$ & $6.66$ & $18.21$ \\
  CE & $0.443$ & $0.451$ & $0.727$ & $0.910$ & $0.515$ & $0.838$ & $0.438$ \\
  \bottomrule
 \end{tabular}
 \caption*{\footnotesize\textit{Note:} Conventions follow
 Table~\ref{tab:calibration-a-alternatives-ou}.}
\end{table}

\begin{table}[!htb]
 \centering
 \caption{Alternative calibrations of GMR to moments of changes only.}
 \label{tab:calibration-a-alternatives-gmr}
 \begin{tabular}{lrrrrrrr}
  \toprule
  & (1) & (2) & (3) & (4) & (5) & (6) & (7) \\
  \midrule
  $\rho$ & $0.987$ & $0.987$ & $0.977$ & $0.980$ & $0.989$ & $0.945$ & $0.987$ \\
  $\sigma_\eta$ & $0.157$ & $0.159$ & $0.171$ & $0.142$ & $0.156$ & $0.147$ & $0.160$ \\
  $p_\zeta$ & $0.990$ & $0.990$ & $0.990$ & $0.961$ & $0.966$ & $0.989$ & $0.990$ \\
  $\lambda$ & $0.585$ & $0.556$ & $0.091$ & $2.162$ & $0.533$ & $0.165$ & $0.581$ \\
  $\psi$ & $0.840$ & $0.819$ & $0.818$ & $0.525$ & $0.268$ & $0.754$ & $0.820$ \\
  $p_\epsilon$ & $0.078$ & $0.078$ & $0.073$ & $0.087$ & $0.101$ & $0.065$ & $0.078$ \\
  $\delta_\epsilon$ & $0.351$ & $0.338$ & $0.315$ & $0.254$ & $0.145$ & $0.407$ & $0.336$ \\
  $\sigma_{\epsilon,1}$ & $1.025$ & $1.025$ & $1.100$ & $0.947$ & $0.804$ & $1.059$ & $1.024$ \\
  $\sigma_{\epsilon,2}$ & $0.072$ & $0.072$ & $0.025$ & $0.042$ & $0.028$ & $0.138$ & $0.070$ \\
  \addlinespace
  $V$ & $1.409$ & $1.369$ & $0.900$ & $0.729$ & $1.307$ & $0.316$ & $1.407$ \\
  Objective & $0.0028$ & $0.0028$ & $0.0043$ & $0.0108$ & $0.0253$ & $0.0076$ & $0.0028$ \\
  Near-optimal & yes & yes & yes & no & no & yes & yes \\
  \addlinespace
  Mean assets & $13.25$ & $13.34$ & $11.54$ & $8.91$ & $12.68$ & $4.77$ & $13.52$ \\
  CE & $0.486$ & $0.493$ & $0.620$ & $0.661$ & $0.480$ & $0.876$ & $0.485$ \\
  \bottomrule
 \end{tabular}
 \caption*{\footnotesize\textit{Note:} Conventions follow
 Table~\ref{tab:calibration-a-alternatives-ou}. The loss probability is $1-p_\zeta$.}
\end{table}

\begin{table}[!htb]
 \centering
 \caption{Alternative calibrations of PLV to moments of changes only.}
 \label{tab:calibration-a-alternatives-plv}
 \begin{tabular}{lrrrrrrr}
  \toprule
  & (1) & (2) & (3) & (4) & (5) & (6) & (7) \\
  \midrule
  $\rho$ & $0.940$ & $0.946$ & $0.941$ & $0.915$ & $0.915$ & $0.946$ & $0.946$ \\
  $\sigma$ & $0.147$ & $0.151$ & $0.123$ & $0.140$ & $0.151$ & $0.151$ & $0.150$ \\
  $c_+$ & $0.000$ & $0.000$ & $0.074$ & $0.123$ & $0.081$ & $0.001$ & $0.003$ \\
  $c_-$ & $0.310$ & $0.276$ & $0.297$ & $0.439$ & $0.440$ & $0.278$ & $0.276$ \\
  $p_\epsilon$ & $0.074$ & $0.071$ & $0.048$ & $0.003$ & $0.633$ & $0.072$ & $0.071$ \\
  $\delta_\epsilon$ & $-0.042$ & $-0.050$ & $0.009$ & $-0.767$ & $0.381$ & $-0.041$ & $-0.030$ \\
  $\sigma_{\epsilon,1}$ & $0.860$ & $0.858$ & $1.038$ & $0.076$ & $0.147$ & $0.848$ & $0.858$ \\
  $\sigma_{\epsilon,2}$ & $0.025$ & $0.025$ & $0.039$ & $0.195$ & $0.042$ & $0.026$ & $0.030$ \\
  \addlinespace
  $V$ & $0.753$ & $0.830$ & $0.900$ & $0.908$ & $0.842$ & $0.829$ & $0.833$ \\
  Objective & $0.0623$ & $0.0650$ & $0.0882$ & $0.1240$ & $0.1240$ & $0.0652$ & $0.0657$ \\
  Near-optimal & yes & yes & no & no & no & yes & yes \\
  \addlinespace
  Mean assets & $5.30$ & $5.81$ & $11.40$ & $12.55$ & $8.30$ & $5.82$ & $5.89$ \\
  CE & $0.825$ & $0.801$ & $0.733$ & $0.741$ & $0.794$ & $0.801$ & $0.800$ \\
  \bottomrule
 \end{tabular}
 \caption*{\footnotesize\textit{Note:} Conventions follow
 Table~\ref{tab:calibration-a-alternatives-ou}, except that objectives use
 the discretized process with $N=257$ nodes.}
\end{table}

\clearpage
\printbibliography

\end{refsection}

\end{document}